\documentclass[%
reprint,
floatfix,
superscriptaddress,
amsmath,amssymb,
aps,
pra,,
]{revtex4-2}
\usepackage[utf8]{inputenc}

\usepackage{xcolor}
\usepackage{graphicx}
\usepackage{dcolumn}
\usepackage{bm}
\usepackage{appendix}
\usepackage{subfig}
\usepackage{hyperref}
\usepackage{float}

\begin{document}
\title{Asymmetric Tunneling of Bose-Einstein Condensates}

\author{Dusty R. Lindberg \href{https://orcid.org/0001-5335-7941}{\includegraphics[scale=0.05]{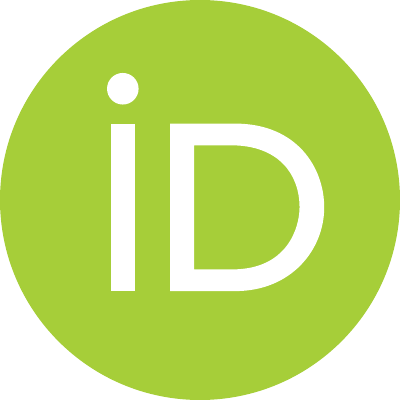}}}
\email{dlindberg@tulane.edu}
\affiliation{Tulane University, New Orleans, LA 70118, USA}

\author{Naceur Gaaloul
\href{https://orcid.org/0000-0001-8233-58481}{\includegraphics[scale=0.05]{orcidid.pdf}}}
\affiliation{Leibniz Universit\"{a}t Hannover, Institut f\"{u}r Quantenoptik, D-30167 Hannover, Germany}

\author{Lev Kaplan} 
\affiliation{Tulane University, New Orleans, LA 70118, USA}

\author{Jason R. Williams \href{https://orcid.org/0000-0002-3798-4424}{\includegraphics[scale=0.05]{orcidid.pdf}}}
\affiliation{Jet Propulsion Laboratory, California Institute of Technology, Pasadena, CA 91109, USA}

\author{Dennis Schlippert
\href{https://orcid.org/0000-0003-2168-1776}{\includegraphics[scale=0.05]{orcidid.pdf}}}
\affiliation{Leibniz Universit\"{a}t Hannover, Institut f\"{u}r Quantenoptik, D-30167 Hannover, Germany}
 
\author{Patrick Boegel \href{https://orcid.org/0000-0002-3606-2452}{\includegraphics[scale=0.05]{orcidid.pdf}}}
\affiliation{Institut f\"{u}r Quantenphysik and Center for Integrated Quantum Science and Technology (IQ$^{\rm ST}$), Ulm Universit\"{a}t, D-89069 Ulm, Germany}

\author{Ernst-Maria Rasel}
\affiliation{Leibniz Universit\"{a}t Hannover, Institut f\"{u}r Quantenoptik, D-30167 Hannover, Germany}

\author{Denys I. Bondar \href{https://orcid.org/0000-0002-3626-4804}{\includegraphics[scale=0.05]{orcidid.pdf}}}
\email{dbondar@tulane.edu}
\affiliation{Tulane University, New Orleans, LA 70118, USA}

\date{\today}

\begin{abstract}
In his celebrated textbook, \textit{Quantum Mechanics: Nonrelativistic Theory}, Landau argued that, for single particle systems in 1D, tunneling probability remains the same for a particle incident from the left or the right of a barrier. This left-right symmetry of tunneling probability holds regardless of the shape of the potential barrier. However, there are a variety of known cases that break this symmetry, e.g. when observing composite particles. We computationally (and analytically, in the simplest case) show this breaking of the left-right tunneling symmetry for Bose-Einstein condensates (BEC) in 1D, modelled by the Gross-Pitaevskii equation (GPE). By varying $g$, the parameter of inter-particle interaction in the BEC, we demonstrate that the transition from symmetric ($g=0$) to asymmetric tunneling  is a threshold phenomenon. Our computations employ experimentally feasible parameters such that these results may be experimentally demonstrated in the near future. We conclude by suggesting applications of the phenomena to design atomtronic diodes, synthetic gauge fields, Maxwell's demons, and black-hole analogues.
\end{abstract}

\maketitle


\section{\label{sec:intro}Introduction}
Tunneling is one of the most important quantum mechanical effects, underlying an extremely broad class of phenomena: nuclear fusion \cite{balantekin1998quantum}, ionization of atoms and molecules by strong laser fields \cite{amini2019symphony}, transport effects in condensed matter \cite{shi2013quantum}, single and two-proton tunneling in large molecules \cite{smedarchina2018entanglement}, semiconductor technology based on resonant tunneling \cite{guzun2013effect},  Josephson junctions \cite{albiez2005direct}, scanning tunneling microscopy \cite{bai2000scanning}, Hawking radiation from a black hole \cite{arzano2005hawking}, \emph{etc.} 
Furthermore, the role of tunneling in biochemical processes is one of the fundamental questions addressed in the burgeoning field of quantum biology \cite{lambert2013quantum}. 
Moreover, it is speculated in quantum cosmology that tunneling is the mechanism that created the universe~\cite{atkatz1994quantum}.

\begin{figure}
    \includegraphics[width=0.45\textwidth]{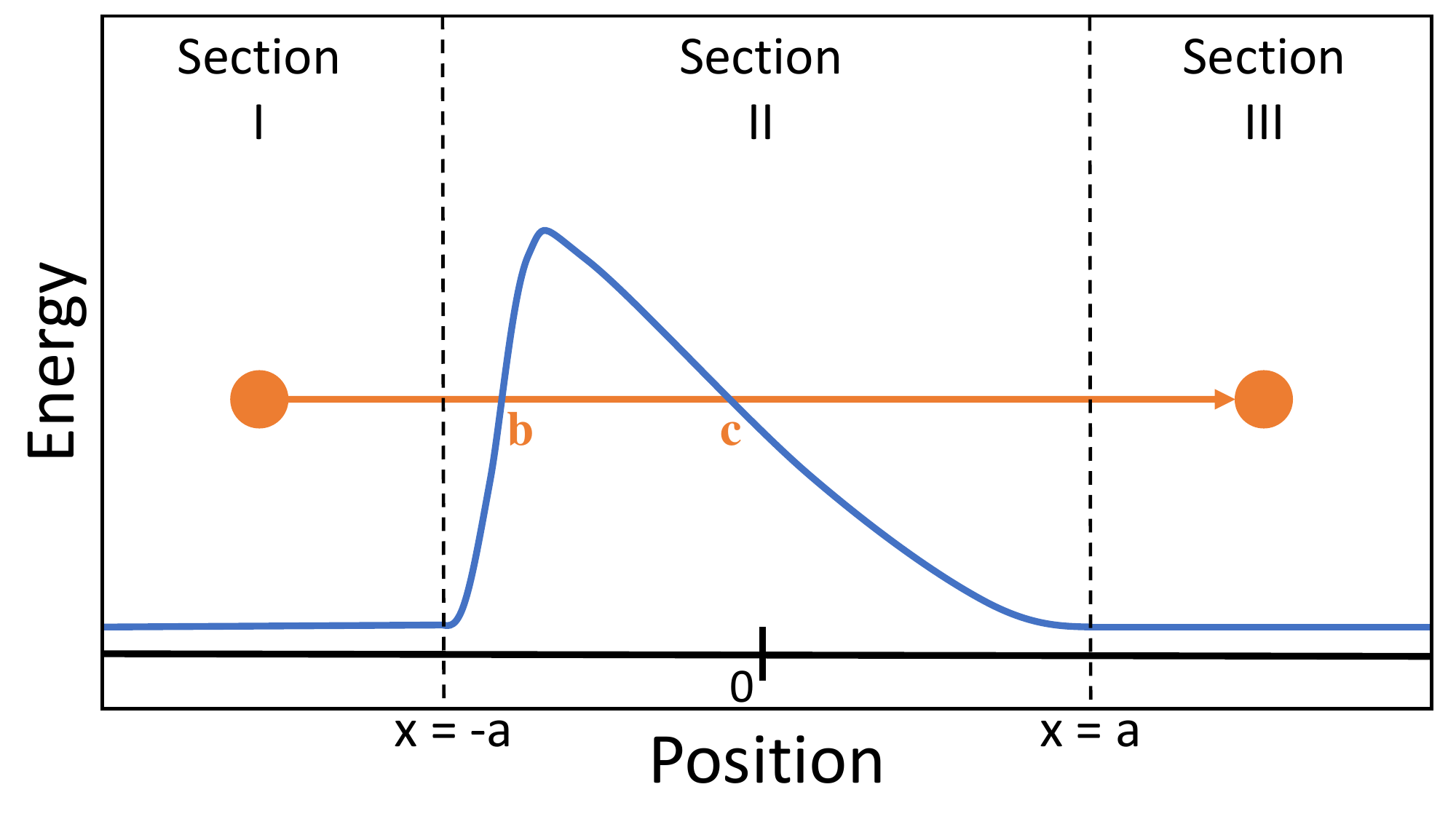}
    \caption{An arbitrary 1D potential, $V(x)$,  described in three sections for the purpose of proving symmetric tunneling. In sections I ($-\infty, -a$) and III ($a, \infty$), the potential is equal and constant, either 0 or reducible to 0 through a global shift. Section II [$-a, a$] contains the barrier, which the wavefunction will need to transmit across. Points $b$ and $c$ mark the classical turning points, where the energy of the particle is equal to the potential, $E(x) = V(x)$. Note that the barrier does not need to be symmetric, as only the points $x = \pm a$ matter for the purpose of proving symmetric tunneling.}
    \label{fig:potentialdemo}
\end{figure}

Given the many disciplines it touches on, the physics of tunneling is incredibly rich \cite{mohsen2013quantum}. Despite this, textbooks on quantum mechanics typically limit the discussion only to the semi-classical (i.e., WKB) approximation to the probability of 1D tunneling $\Gamma$ (see, e.g., Sec.~50 of Ref.~\cite{landau2013quantum}). For tunneling of a quantum particle with energy $E$ and mass $m$ through a potential barrier $V(x)$, the WKB approximation gives
\begin{align}\label{EqWKB1DTunneling}
    \Gamma \propto \exp\left[ -\frac{2}{\hbar} \int_b^c \sqrt{2m(V(x)-E)} dx \right],
\end{align}
see Fig.~\ref{fig:potentialdemo} for an illustration. This formula is responsible for a widely held belief that tunneling probability is exponentially small. It is important to note that the probability of tunneling exponentially depends on the area under the curve between the potential barrier and particle's energy.

Perhaps counterintuitively, Landau suggested (Sec. 25 of Ref.~\cite{landau2013quantum}) that transmission probabilities are symmetric in 1D according to the time-independent Schr\"odinger equation. In other words, a particle incident upon some potential barrier will have the same tunneling probability, regardless of whether it approaches the barrier from the right or the left. In the case when more than one degree of freedom (d.o.f.) is involved, however, this no longer works. For example, interactions between different d.o.f. in a complex system may enhance or suppress tunneling rates compared to the case when no interactions are present \cite{zakhariev1964intensified}. This can lead to an emergent dynamical asymmetry in the tunneling of a composite system \cite{Amirkhanov1966}, such that tunneling becomes more probable than flying above a barrier for a large class of potentials~\cite{bondar2010enhancement}. These works remain largely unknown and we are extending their insight to experimentally accessible systems. This is especially true given recent experimental advances in Bose-Einstein condensates (BECs), enabling researchers to directly observe the tunneling of complex systems for the first time \cite{albiez2005direct, potnis2017interaction, Ramos2020}. In addition, extensive computational research has been done with BEC tunneling, both using mean-field \cite{smerzi1997quantum, PhysRevA.61.023402, Salasnich2001, svidzinsky2003symmetric, jona2003asymmetric, jona2005nonlinear, carr2005macroscopic, PhysRevA.81.063638, manju2018quantum} and many-body \cite{PhysRevA.74.013605, glick2011macroscopic, Lode2014, PhysRevLett.118.210403} theories. Of particular relevance to the current work, Ref.~\cite{Haldar2019} uses many-body methods to study the suppression of oscillations in an asymmetric double well, which is itself a form of asymmetric tunneling.

In the current work, we study the breaking of tunneling symmetry with BECs modelled by the Gross-Pitaevskii Equation (GPE) and the ways in which BECs behave similarly to other systems that are known to break tunneling symmetry. The GPE is a non-linear equation used to describe weakly interacting BECs, where the non-linearity models the interaction between particles of the Bose gas. The Schr\"odinger equation, and resulting linearity, can be recovered simply by reducing the interaction parameter to zero.

Herein, we begin by introducing the intuition behind the breaking of tunneling symmetry in BECs (Sec.~\ref{sec:GPE}). Tunneling asymmetry for free BECs is then demonstrated numerically in two parts: First, in Sec.~\ref{sec:Modeling}.A the BEC is given a small momentum kick towards a single asymmetric barrier, which demonstrates asymmetry in single-pass tunneling probabilities; next, in Sec.~\ref{sec:Modeling}.B the BEC is kicked towards a double asymmetric barrier and the inter-particle interaction strength of the BEC is varied to examine its effect on the asymmetry of tunneling probability. Then, tunneling symmetry for trapping potentials is numerically illustrated for the Schr{\"o}dinger equation (Sec.~\ref{sec:TrapPot}.A), and we subsequently model a BEC confined to a trapping potential, demonstrating multi-pass asymmetric tunneling (Sec.~\ref{sec:TrapPot}.B). In Sec.~\ref{sec:analytic}, we show analytically how asymmetric tunneling in the GPE naturally emerges from symmetric tunneling in the Schr\"odinger equation when a small nonlinear interaction is added. We also offer a full proof of Landau's argument for symmetric tunneling in App.~\ref{sec:1dproof}.


\section{Principle of Asymmetric Tunneling Beyond the Single-Particle Case}\label{sec:GPE} 

In the overlooked Ref. \cite{Amirkhanov1966} (see also Eq.~(30) of Ref.~\cite{bondar2010enhancement}), it was shown that tunneling can become asymmetric for the case of a two-particle system. Let us give the intuition behind this phenomenon. First, recall that the dynamics generated by a time-independent Hamiltonian conserves the total energy. To understand the underlying physics of the two-particle system, it is convenient to represent it using the center of mass (C.M.) and inter-particle d.o.f. A tunneling barrier then couples these two d.o.f. 

\begin{figure}
\includegraphics[width=0.45\textwidth]{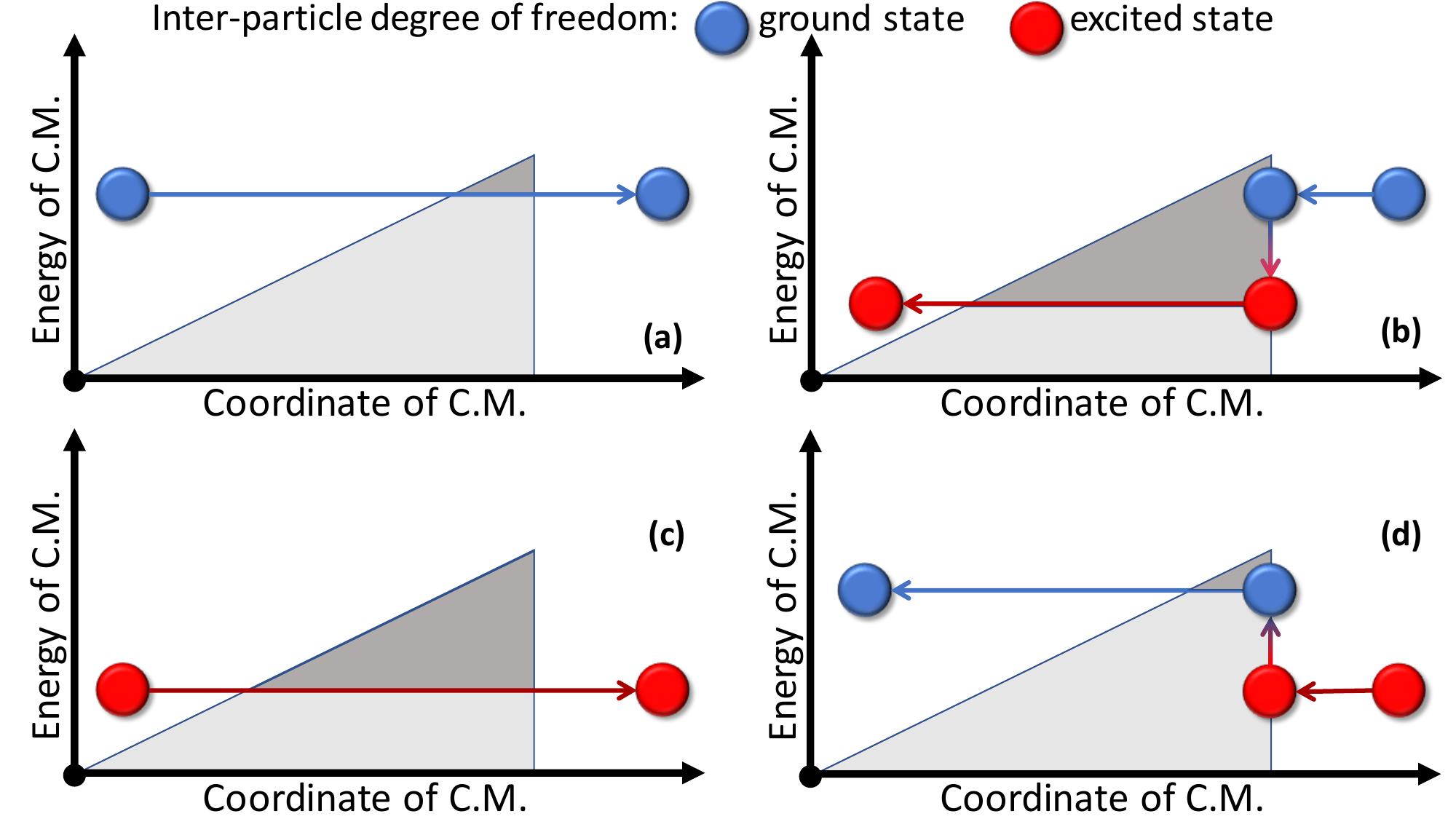}
\caption{The principle of asymmetric tunneling. A multi-particle system is incident upon an asymmetric barrier in either [(a), (b)] its ground state or [(c), (d)] an excited energy state.
In the cases (a) and (c), the wave packet experiences gradual change of the potential barrier, and it is unlikely to experience an excitation or de-excitation. In the cases (b) and (d), the wave packet experiences a sudden change of the barrier, and excitation or de-excitation is likely. In the case of the ground state wave packet (b), the excitation energy comes from the system's center of mass (C.M.) energy, making it less likely to transmit. In the case of an excited state (d), the system may relax to the ground state, adding to the C.M. energy and making it more likely to transmit.}
\label{fig:AsymmetrictunnelingMaxwellDemon}
\end{figure}

Consider a triangular potential barrier as depicted in Fig.~\ref{fig:AsymmetrictunnelingMaxwellDemon}. Assume that the inter-particle d.o.f. is initially in the lowest energy (i.e., ground) state as the wave packet approaches the angled side of the potential barrier, as shown Fig.~\ref{fig:AsymmetrictunnelingMaxwellDemon}(a). As the system moves from the left to right, the potential barrier gradually builds up. According to the adiabatic theorem, the inter-particle d.o.f. will not be excited; hence, tunneling dynamics of the two-particle system effectively resembles the 1D case (Fig.~\ref{fig:potentialdemo}) since the inter-particle d.o.f. is frozen. A very different dynamical process takes place when the wave packet approaches the vertical side of the triangular barrier, as shown in Fig.~\ref{fig:AsymmetrictunnelingMaxwellDemon}(b). Upon colliding with the edge of the barrier, the inter-particle d.o.f. experiences a sudden shakeup, which induces the change of state. Since the inter-particle d.o.f. is originally in the ground state, the only allowed transition is to the excited state. The total energy, which is the sum of the center of mass and inter-particle energies, is conserved, hence the excitation takes place by decreasing the center of mass kinetic energy. As a result, the center of mass d.o.f. effectively plunges deeper under the potential barrier, as depicted in Fig.~\ref{fig:AsymmetrictunnelingMaxwellDemon}(b). The gray shaded regions in Figs.~\ref{fig:AsymmetrictunnelingMaxwellDemon}(a, b) indicate the areas that enter into the  exponent in the tunneling probability as shown in Fig.~\ref{fig:potentialdemo} and Eq.~\eqref{EqWKB1DTunneling}. Thus, we achieve asymmetric tunneling since the probability of tunneling from left to right [Fig.~\ref{fig:AsymmetrictunnelingMaxwellDemon}(a)] is larger than the right-left probability  [Fig.~\ref{fig:AsymmetrictunnelingMaxwellDemon}(b)].

The preceding description of asymmetric tunneling has been done within the Schr{\"o}dinger equation for a two-particle system, where the asymmetric tunneling results from the coupling of the center of mass and inter-particle d.o.f. While we expect that broken tunneling symmetry is realizable in other multi-particle systems, here we examine mean-field predictions for the case of a BEC, which differs from the previous intuition as it behaves in many aspects like a single-particle system (since the majority of its $N$ particles occupy the same quantum state). The source of asymmetry comes from the BEC's self-interaction, which plays an analogous role to the excitation of internal d.o.f. in the previously established multi-particle case. 

\section{Numerical Demonstration of Asymmetric Tunneling of 1D BEC}\label{sec:Modeling}

To evaluate BEC dynamics in a mean-field approximation, we utilize the time-dependent Gross-Pitaevskii equation (GPE), which is written initially in 3D as
\begin{align}
    i\hbar \frac{\partial\Psi(\boldsymbol{r},t)}{\partial t} = 
    \left( -\frac{\hbar^2}{2m}\nabla^2+V(\boldsymbol{r})+g\left|\Psi(\boldsymbol{r},t)\right|^2 \right) \Psi(\boldsymbol{r},t), \label{eq:3DGPE}
\end{align}
or
\begin{align} \label{eq:3DTimeIndependentGPE}
    \mu \Psi(\boldsymbol{r}) = 
    \left( -\frac{\hbar^2}{2m}\nabla^2+V(\boldsymbol{r})+g\left|\Psi(\boldsymbol{r})\right|^2 \right) \Psi(\boldsymbol{r})
\end{align}
in its time-independent form, where $g$ is the interaction parameter that models the inter-particle interactions between bosons and $\mu$ is the chemical potential \cite{pethick2008bose}. The interaction parameter $g$ is the driving factor behind the expected asymmetry of tunneling, as reducing $g$ to 0 simply recovers the single-particle Schr{\"o}dinger equation and its accordingly symmetric dynamics. In order to better compare the transmission probabilities of the single-particle Schr\"odinger equation and the GPE, we choose to normalize the state to $\int \left| \Psi(x,t) \right|^2 dx = 1$. The full expression for the interaction parameter is
\begin{align}\label{eq:gparam}
    g =& 4 \pi N a_s \hbar^2 / m, 
\end{align}
where $a_s$ is the scattering length of the BEC, and $m$ is the mass of the individual bosons. We use the background scattering length value $a_{bg} = 100\,a_0$~\cite{Egorov_2013}, where $a_0 = 5.29 \times 10^{-11}\,\textrm{m}$, and the mass of $^{87}$Rb is $m = 1.44 \times 10^{-25}\,\textrm{kg}$.

For modelling the 1D behavior, we imagine the 3D BEC to be tightly confined in the $y-$ and $z-$ directions by a harmonic trap of angular frequency $\omega_y = \omega_z =1000\pi$\,Hz, and weakly trapped along the $x-$direction by a shallow trap of angular frequency $\omega_x = 100\pi$\,Hz. We use these frequencies to define the characteristic lengths,
\begin{align} \label{eq:charlength}
    L_i =& \sqrt{\frac{\hbar}{m \omega_i}}
\end{align}
for $i = x, y, z$. Utilizing these lengths, frequencies, and known constants, and noting that the physical potential $V(x)$ is measured in Kelvin, we introduce the dimensionless quantities:
\begin{align}
    \tau=&t\omega_x, \label{eq:dimlesst} \\
    \xi=&x/L_x, \label{eq:dimlessx}\\
    \tilde{V}(\xi)=&V(x) k_B / \hbar \omega_x, \label{eq:dimlessV}\\
    \tilde{\mu}=&\mu k_B / \hbar \omega_x, \label{eq:dimlessMu} \\
    \tilde{g}=& \frac{g}{\hbar \omega_x} \frac{1}{2\pi L_x L_y L_z}, \label{eq:dimlessg} \\
    \tilde{\Psi}(\xi,\tau) =& \Psi(x,t) \sqrt{L_x L_y L_z}, \label{eq:dimlessPsi} 
\end{align}
where $\hbar \omega_x / k_B \approx 2.4 \times 10^{-3}$\,\textmu K, 
    $L_x \approx 1.5$\,\textmu m, and $\quad L_y=L_z \approx 0.48$\,\textmu m, and arrive at the dimensionless quasi-1D GPE:
\begin{align}
    i \frac{\partial\tilde{\Psi}(\xi,\tau)}{\partial \tau} = 
    \left( -\frac{1}{2}\frac{\partial^2}{\partial\xi^2}+\tilde{V}(\xi)+\tilde g\left|\tilde{\Psi}(\xi,\tau)\right|^2 \right) \tilde{\Psi}(\xi,\tau), \label{eq:1DGPE}
\end{align}
or
\begin{align} \label{eq:1DTimeIndependentGPE}
    \tilde{\mu}\tilde{\Psi}(\xi) =& 
    \left( -\frac{1}{2}\frac{\partial^2}{\partial\xi^2}+\tilde{V}(\xi)+\tilde g\left|\tilde{\Psi}(\xi)\right|^2 \right) \tilde{\Psi}(\xi)
\end{align}
in its time-independent form.    
    
\begin{figure}
    \includegraphics[width=0.45\textwidth]{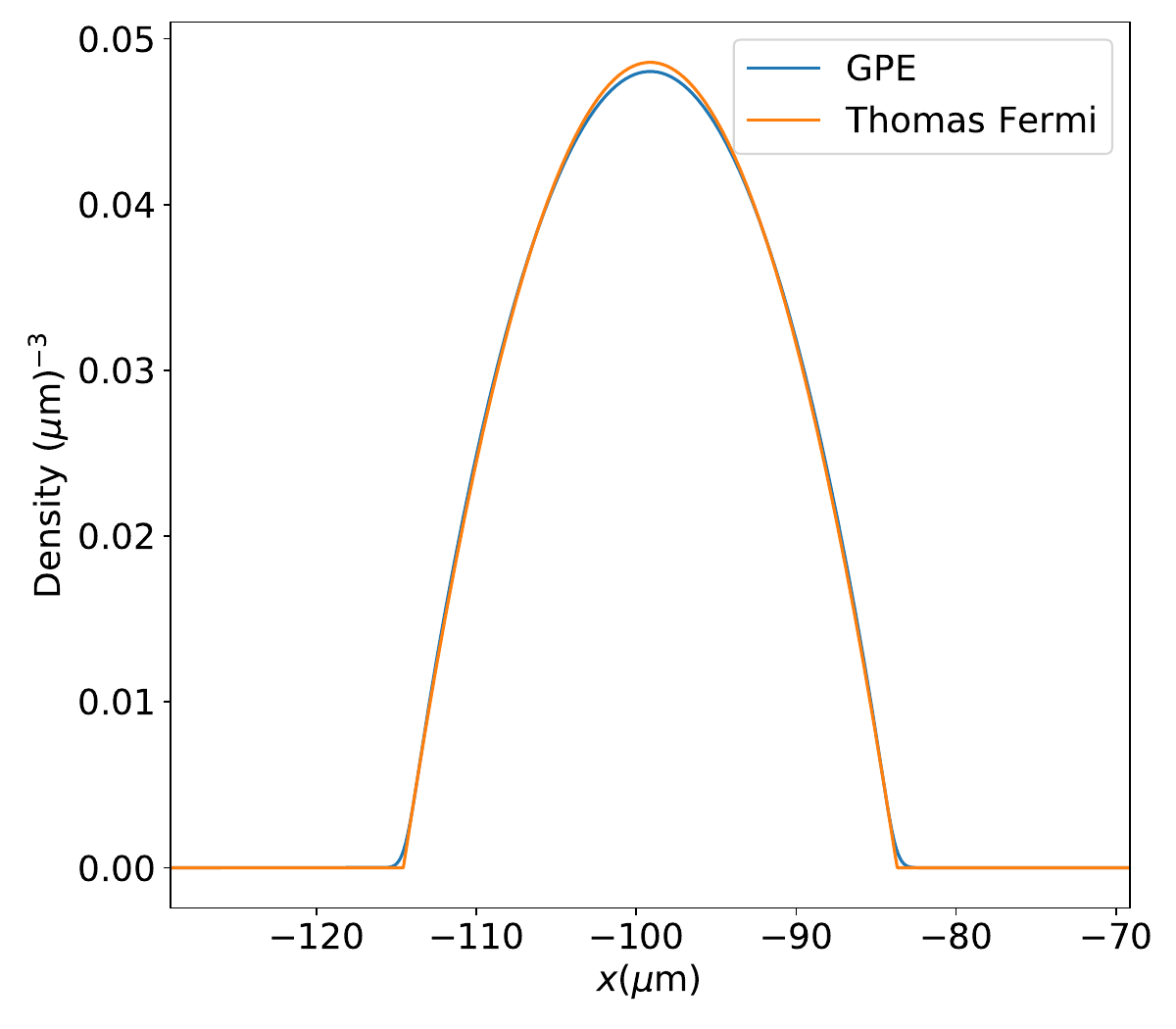}
    \caption{The BEC ground state for the trapping potential from Eq.~\eqref{eq:trapping} with parameters $v_0 = 0.5$ and $\xi_0 = -65$ (or $x = -99.1$ \textmu m) is compared with the Thomas-Fermi approximation. The normalized density of the BEC here is labelled as GPE, while the normalized right hand side of Eq.~\eqref{eq:TAapproximation} is labeled as Thomas-Fermi. Because the area under both curves is normalized to 1, the Thomas-Fermi approximation, which is low near the edges, compensates by being larger near the center of the trap.}
    \label{fig:TAapproximation}
\end{figure}

We consider a BEC of $N = 10^4$ particles trapped by the following quadratic potential centered around an offset $\xi_0$ 
\begin{align} \label{eq:trapping}
    \tilde{V}_{\textrm{trap}}(\xi) = v_0 (\xi - \xi_0)^2,
\end{align}
where $v_0$ is a scaling parameter for how tightly we set the trap, and $\xi_0$ is the position of the center of the trap (see Eq.~\eqref{eq:dimlessx} for unit conversion). The normalized ground state of Eq.~\eqref{eq:1DTimeIndependentGPE} in the trapping potential from Eq.~\eqref{eq:trapping} is obtained via imaginary time propagation and used as the initial condition at $t=0$ for the time evolution of the GPE. Despite the moderate number of atoms, the ground state reasonably matches the prediction of the Thomas-Fermi Approximation, 
\begin{align} \label{eq:TAapproximation}
    \left|\Psi(x)\right|^2 = \frac{\mu - V(x)}{g},
\end{align}
where the kinetic energy in Eq.~\eqref{eq:1DTimeIndependentGPE} is assumed to be negligibly small. In Eq.~\eqref{eq:TAapproximation},  $\mu$ is taken to be the numerically calculated chemical potential for the ground state. Plotting all positive values of $\left( \mu - V(x) \right) / g$, we see in Fig.~\ref{fig:TAapproximation} very good agreement of the calculated density of the ground state with the Thomas-Fermi approximation. The small disagreement in the center comes as a result of normalization of the wavefunction density. The Thomas-Fermi approximation is least accurate near the edges, where the kinetic energy is least likely to be negligible due to particle collisions with the trapping potential. As the exact wavefunction density, labeled GPE in Fig.~\ref{fig:TAapproximation}, includes the edge regions not included in the approximation, the density at the center, after normalization, is lowered accordingly. 

\subsection{Single-pass Asymmetric Tunneling}\label{sec:SinglePassAsymmetry}

In this section, the case of a 1D traveling BEC incident upon an asymmetric barrier is investigated. This approach will demonstrate the simplest asymmetric BEC tunneling behavior, since the BEC only interacts with the barrier once. The asymmetric barriers we implement for the single-pass case are sums of fixed-width Gaussians, which are employed due to their commonality with experimental methods. However, we are primarily concerned with the establishment of tunneling asymmetry for our initial discussion, and accordingly we prioritize simplified potentials and leave considerations of experimental implementation to Sec.~\ref{sec:TrapPot}, after the principle has been clearly demonstrated. The dimensionless form of the first potential is
\begin{align} \label{eq:m1V}
    \tilde{V}(\xi) =& 45 e^{-\left(\xi+15\right)^2/144.5} + 31.5 e^{-\xi^2/144.5} \nonumber \\
    &+ 13.5 e^{-\left(\xi-15\right)^2/144.5}.
\end{align}

\begin{figure}
    \includegraphics[width=0.45\textwidth]{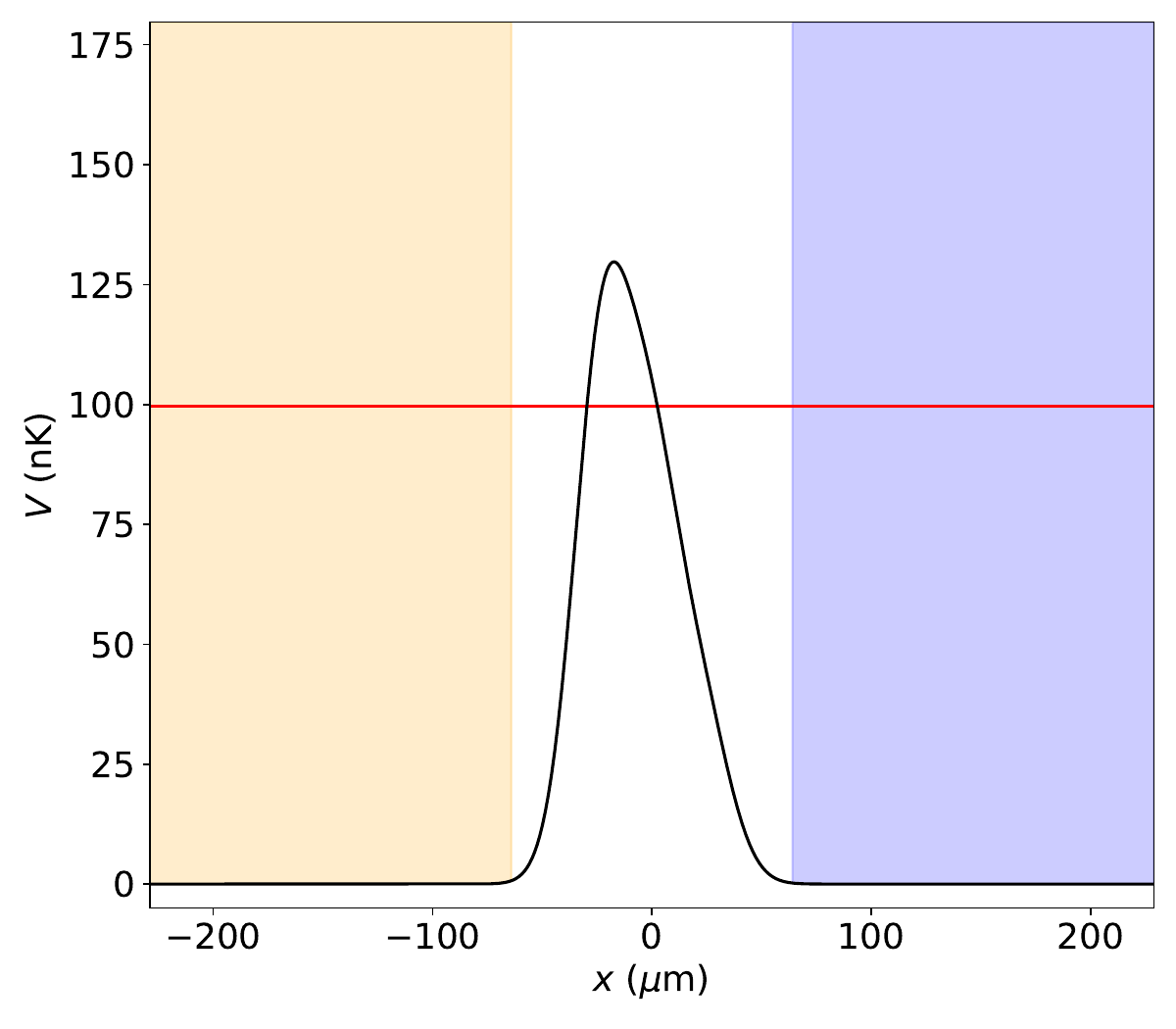}
    \caption{An asymmetric barrier used to demonstrate single-pass tunneling asymmetry, with the red line indicating the total energy of the BEC after it has been kicked (99.7\,nK). The shaded regions are the ones over which we integrate to obtain the probability of tunneling. The \textcolor[HTML]{5151cc}{blue shaded region} corresponds to the BEC prepared at $x=-99.1\,\textrm{\textmu m}$ and kicked such that it initially propagates to the right. The \textcolor[HTML]{FFA500}{orange shaded region} is the integration region for the flipped case, where we prepare the BEC at $x = 99.1\,\textrm{\textmu m}$ and kick it such that it propagates to the left.}
    \label{fig:method1Potential}
\end{figure}

\begin{figure}
    \subfloat[]{\includegraphics[width=0.45\textwidth]{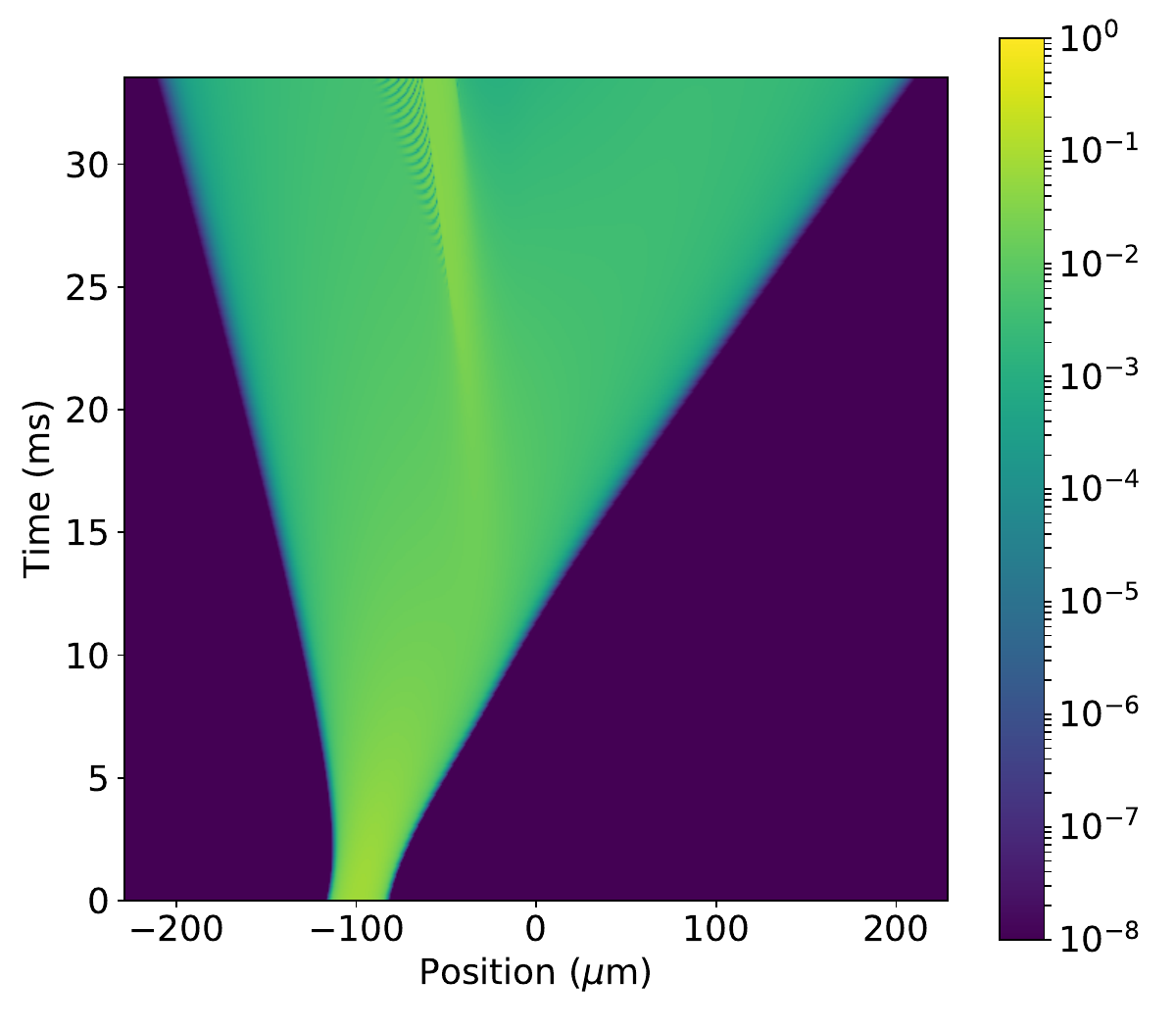}}
    
    \subfloat[]{\includegraphics[width=0.45\textwidth]{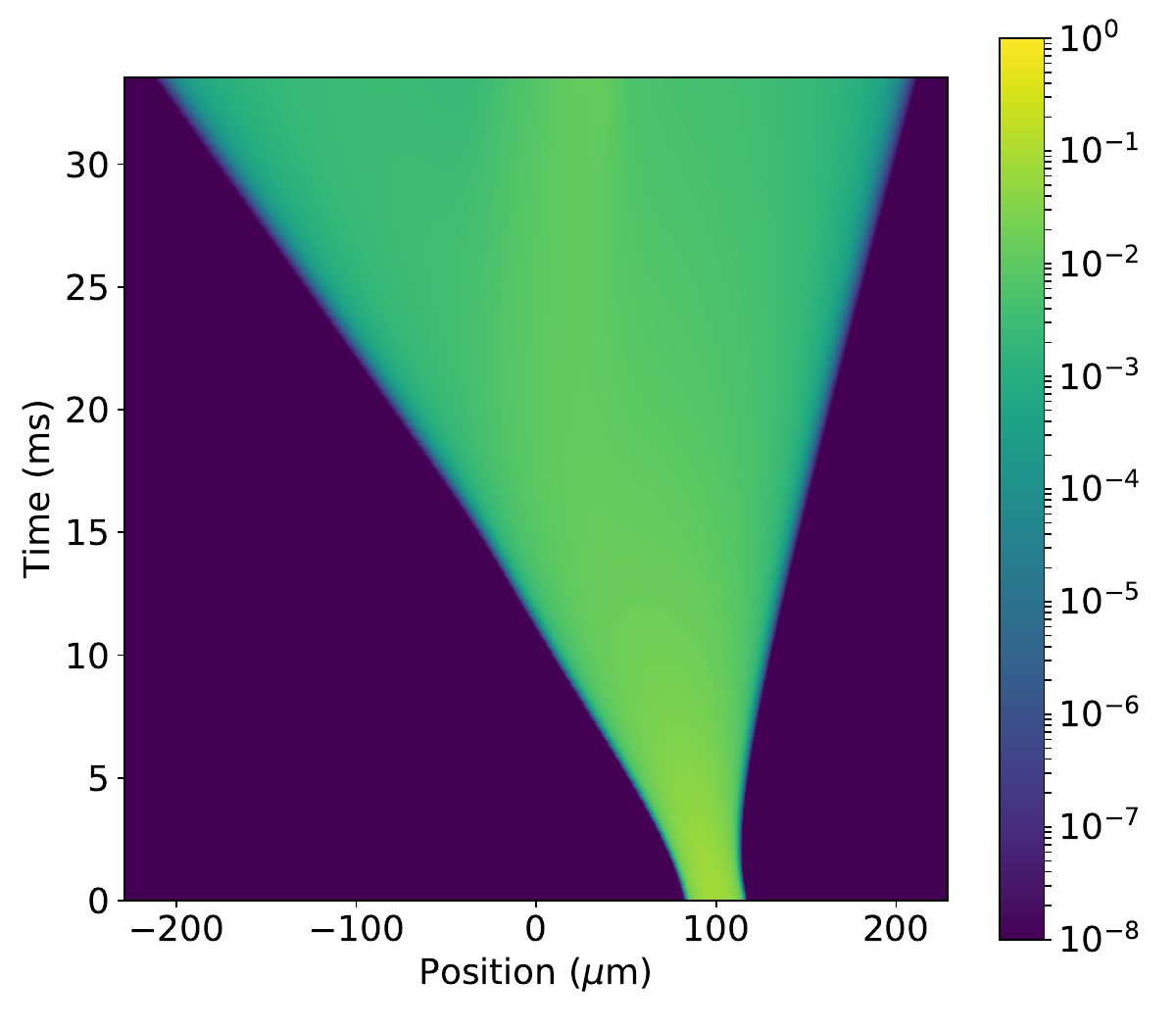}}
    \caption{(a)-(b)Evolution plots of a BEC modelled by the GPE, where (a) shows the BEC prepared on the left and kicked to the right and (b) shows the BEC initially on the right and kicked to the left. The color bar in each plot corresponds to $|\Psi(x, t)|^2$.
    Unlike the predictions of the single particle Schr\"odinger equation, we see significant spreading of the BEC as it propagates. This is the expected behavior of a self-interacting, multi-particle system.}
    \label{fig:method1GPEdensplot}
\end{figure}

\begin{figure}
    \subfloat[]{\includegraphics[width=0.45\textwidth]{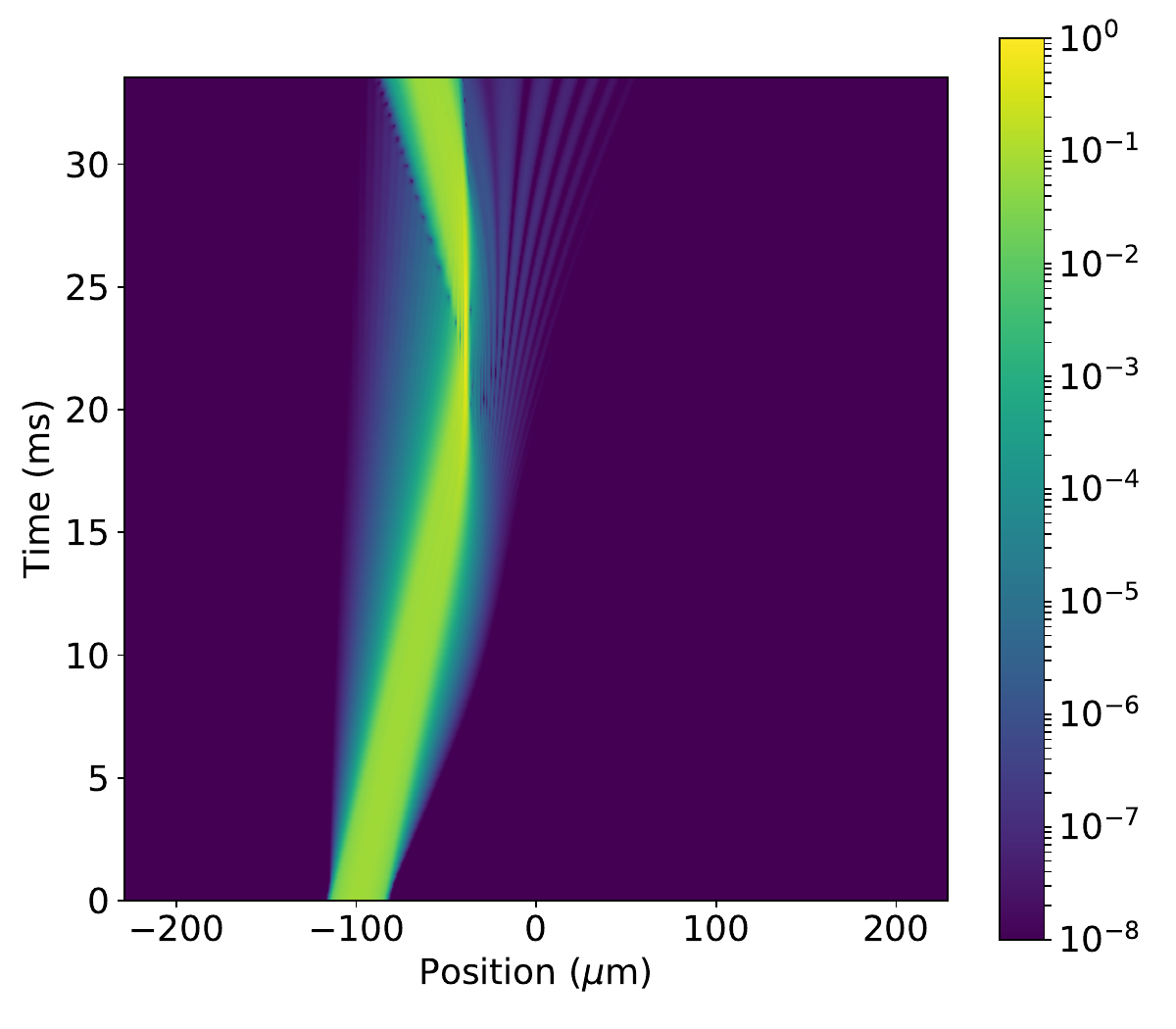}}

    \subfloat[]{\includegraphics[width=0.45\textwidth]{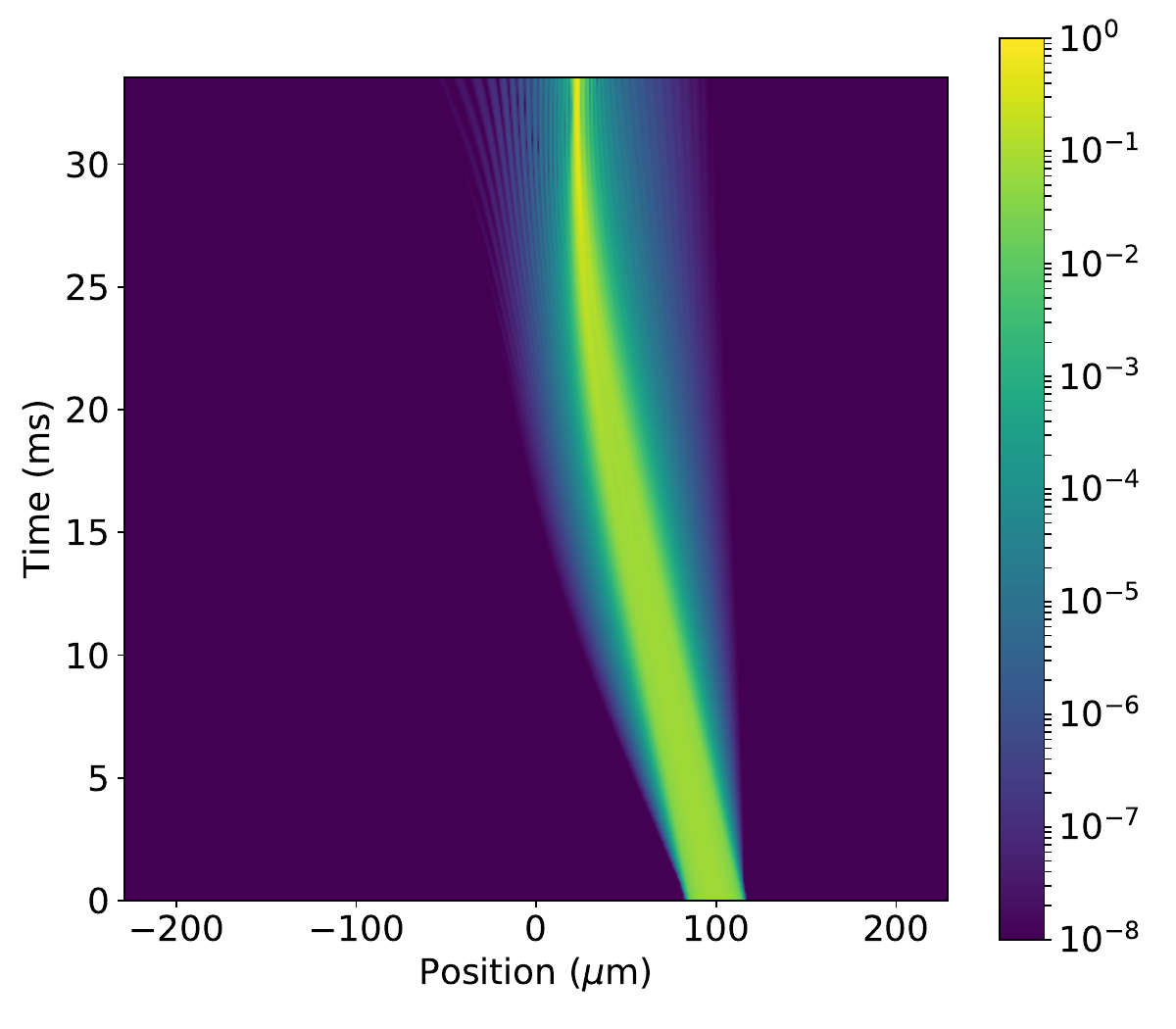}}
    \caption{(a)-(b) Evolution plots of a BEC with vanishing interactions, where (a) shows the BEC prepared on the left and kicked to the right and (b) shows the BEC initially on the right and kicked to the left. The color bar in each plot corresponds to $|\Psi(x, t)|^2$. Due to the low predicted energy, the BEC largely reflects off the barrier.}
    \label{fig:method1SCHRdensplot}
\end{figure}

In Fig.~\ref{fig:method1Potential}, the barrier is shown along with the regions for calculating tunneling probability. We  begin with a starting offset of  $\xi_0 = -65$, or $x = -99.1\,\textrm{\textmu m}$, and a scale of $v_0=0.5$ for the initial trap. The ground state of the BEC is then released by turning-off this trap and ``kicking" the condensate to give it an initial velocity towards the barrier. In this case, the initial wavefunction reads
\begin{align}
    \tilde{\Psi}(\xi, \tau=0) = e^{i \kappa \xi}\tilde{\Psi}_g(\xi),
\end{align}
where $\tilde{\Psi}_g$ is the BEC ground state, such that the state will now propagate to the right (for $\kappa>0$). With a kick parameter $\kappa = 6.5$, we allow the BEC to propagate for a time $\tau= 10.55$, or $t = 33.6\,\textrm{ms}$, using both the GPE (Fig.~\ref{fig:method1GPEdensplot}) and the Schr\"odinger equation (Fig.~\ref{fig:method1SCHRdensplot}) for comparison. The total elapsed time ensures that the BEC stays within the maximum bounds of propagation, but also allows the BEC sufficient time to either reflect or tunnel through the barrier. Because the method of simulated propagation has periodic boundary conditions due to use of the Fast Fourier Transform, the maximum bounds are chosen to be as large as possible while still maintaining a meaningful position- and time-grid resolution and avoiding significant numerical uncertainty constraints. While this is not ideal, it is sufficient for demonstrating the simplest implementation of asymmetric tunneling. We will examine the asymmetry without these constraints later in Sec.~\ref{sec:TrapPot}.

For now we concern ourselves entirely with the behavior of the single-pass case at hand. We see the behavior of the propagating BEC in Fig.~\ref{fig:method1GPEdensplot}, notably the expected spreading of the density due to positive inter-particle interaction parameter $g$, which corresponds to repulsive self-interactions. However, we have chosen our initial momentum ``kick''  parameter, $\kappa$, such that the BEC has a net initial motion towards the barrier even as it spreads in both directions. We expect this repulsive BEC self-interaction to drive the tunneling asymmetry in a manner similar to the inter-particle interaction in the multi-particle case considered in Fig.~\ref{fig:AsymmetrictunnelingMaxwellDemon}. In the case of Fig.~\ref{fig:method1GPEdensplot} (a), the BEC meets the steeper side of the barrier first. We remember for the multi-particle case seen in Fig.~\ref{fig:AsymmetrictunnelingMaxwellDemon}(b) that approaching from the steeper side is more likely to lead to a sudden excitation and a lowering of the transmission probability as a result. We expect that if the BEC inter-particle interaction term is a proper analogue to the multi-particle intuition, the sharp increase will cause the BEC to also experience a decrease in transmission probability.
We repeat this process separately for the flipped case [Fig.~\ref{fig:method1GPEdensplot}(b)], preparing the BEC at $\xi_0 = 65$, or $x = 99.1 \textrm{\textmu m}$, and then applying the same magnitude kick in the opposite direction. This time, the BEC sees a more adiabatic increase in the potential barrier, and should experience a higher transmission probability, similar to the case presented in Fig.~\ref{fig:AsymmetrictunnelingMaxwellDemon}(a). We repeat the same numerical calculations in Fig.~\ref{fig:method1SCHRdensplot} for a BEC with vanishing interaction ($g=0$), which reduces to the Schr\"odinger case. This will be used for contrast with the interacting case and to highlight the role of the interaction in the asymmetry of tunneling probability.

\begin{figure}
    \subfloat[]{\includegraphics[width=0.45\textwidth]{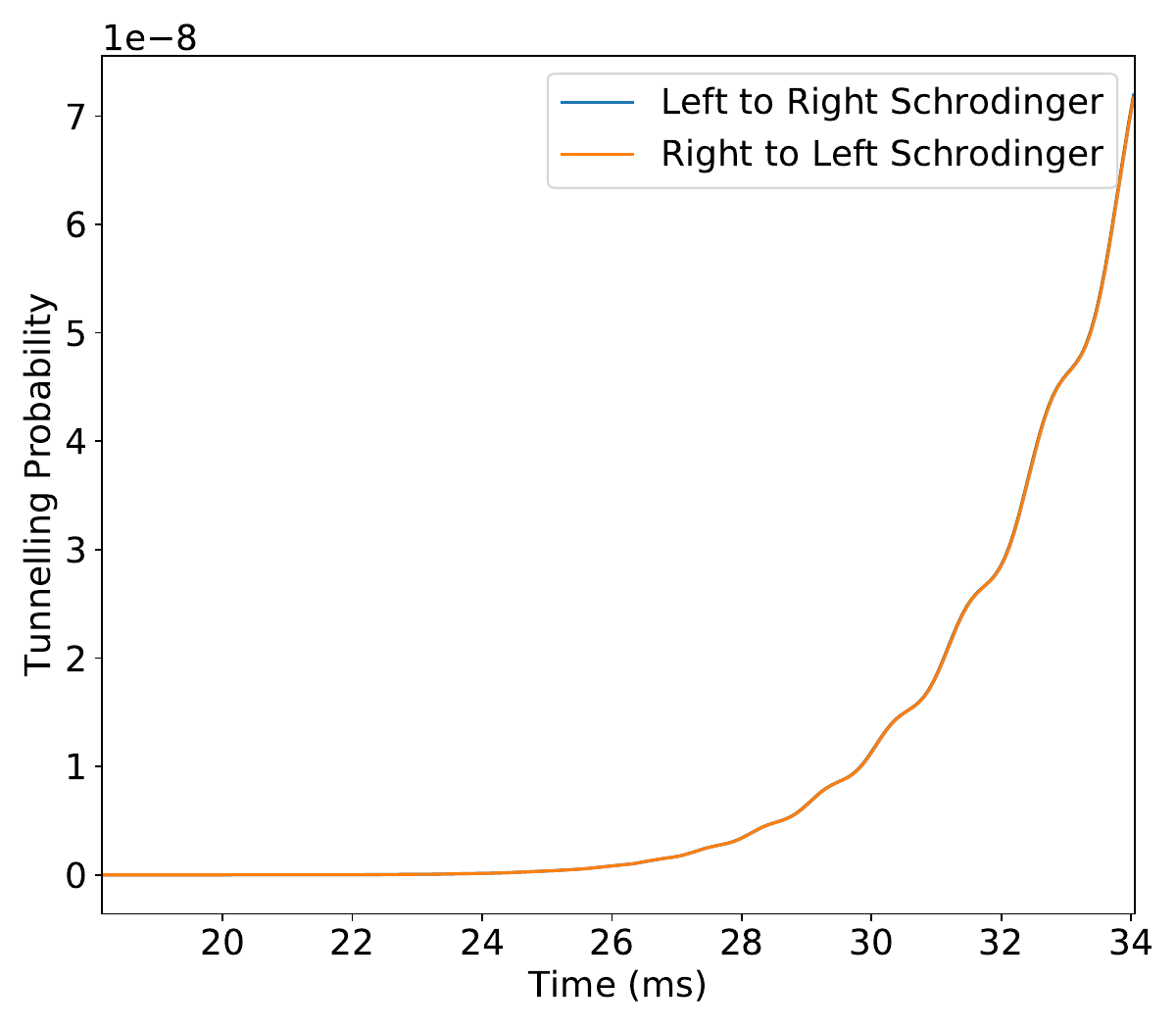}}
    
    \subfloat[]{\includegraphics[width=0.45\textwidth]{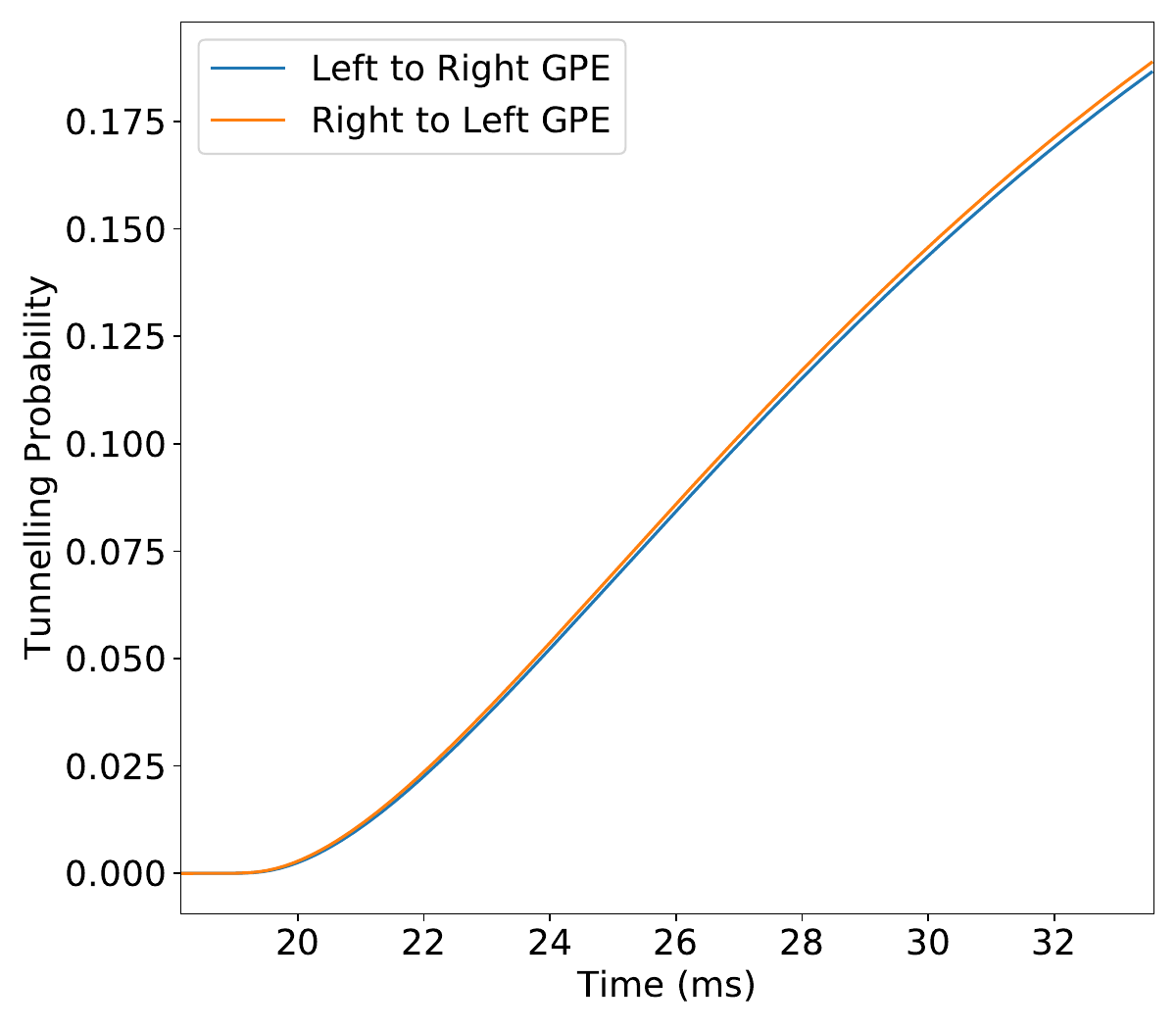}}
    \caption{(a) Tunneling probabilities for a BEC modelled by the Schr\"odinger equation (SE), which overlap completely. The size of the tunneling probability is very low due to the lower predicted energy of the SE.
    (b) Tunneling probability for a BEC modelled by the GPE. While the difference is small, there is a clear split between the left-to-right probability and the right-to-left. The intuition from Fig.~\ref{fig:AsymmetrictunnelingMaxwellDemon} holds, as the more gradual side of the barrier has  higher tunneling probability.
    Note the colors are synchronized to Fig.~\ref{fig:method1Potential}, with the \textcolor[HTML]{5151cc}{left to right BEC} tunneling corresponding to integration over the \textcolor[HTML]{5151cc}{blue shaded region} and the \textcolor[HTML]{FFA500}{right to left BEC} tunneling corresponding to integration over the \textcolor[HTML]{FFA500}{orange shaded region}.}
    \label{fig:method1TP}
\end{figure}

To quantify tunneling asymmetry in the interacting BEC case [Fig.~\ref{fig:method1TP}(b)], we introduce the following metric
\begin{align} \label{eq:asymmetrymeasure}
    d_r = \frac{|T_L - T_R|}{\left(T_L+T_R\right)/2} \times 100\%,
\end{align}
where $T_R$ is the tunneling probability for the BEC incident from the left to tunnel through to the right, and $T_L$ is the tunneling probability for the BEC incident from the right to tunnel through to the left. These probabilities are calculated as: 
\begin{align}
    T_R &=  \int_{\textcolor[HTML]{5151cc}{blue}} |\Psi (x, t)| ^2 dx \textrm{ and}  \label{eq:TR} \\
    T_L &= \int_{\textcolor[HTML]{FFA500}{orange}} |\Psi (x, t)| ^2 dx , \label{eq:TL}
\end{align}

\begin{figure}
    \includegraphics[width=0.45\textwidth]{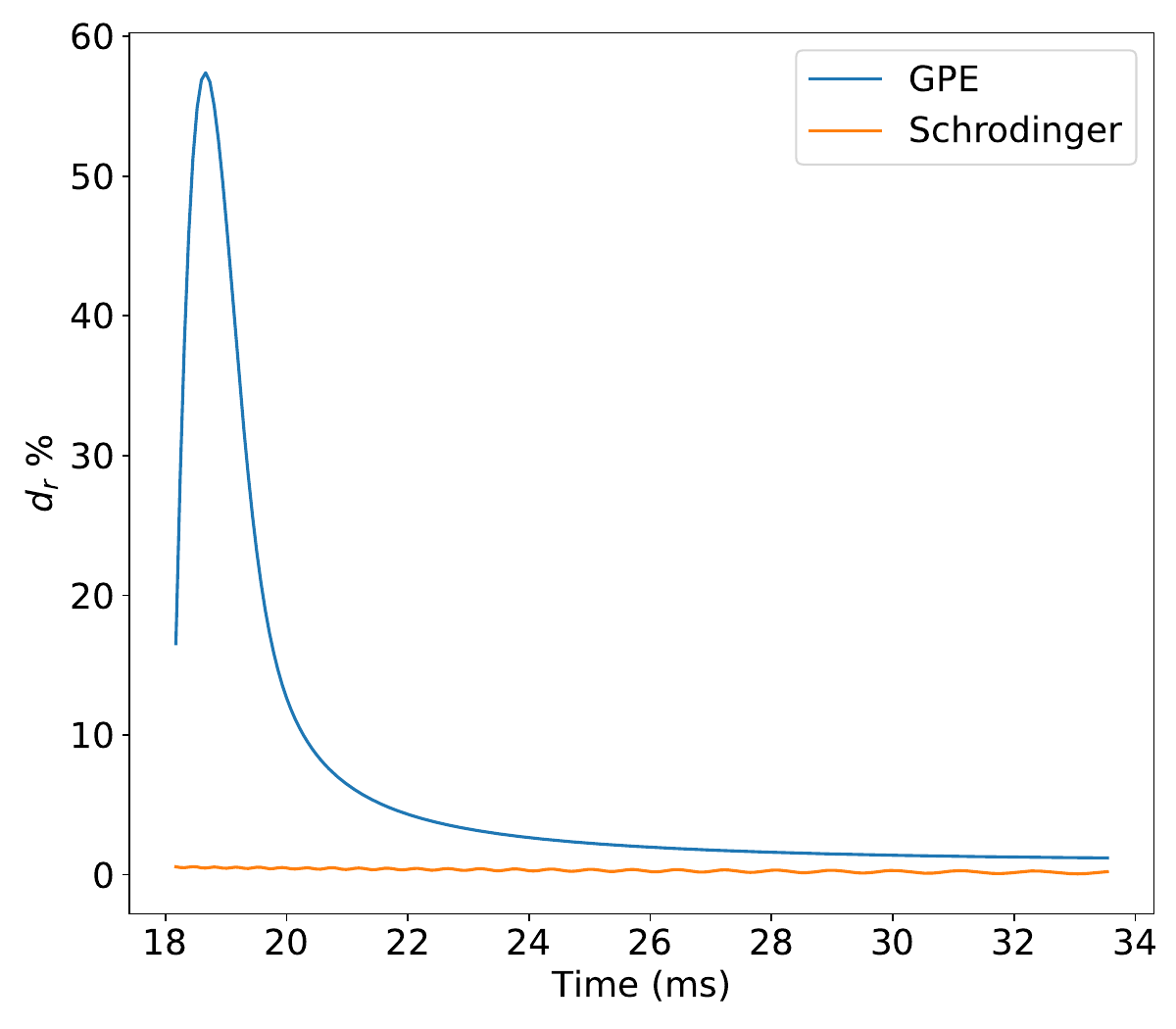} 
    \caption{ Relative difference between tunneling probabilities modelled by the Schr\"odinger equation and the GPE. Note that, for the Schr\"odinger case, the order of tunneling probability is $10^{-8}$. We do see minor fluctuation on the order of $\approx 0.1 \%$ difference in tunneling probability, consistent with numerical noise.
    For the case of the GPE, we see a spike of  $d_r = 57\% $ at the peak, observed soon after the tunneling probabilities first increase above $\approx 0$ and immediately begin to diverge. The asymmetry declines to $d_r = 1\% $ by the cutoff time, showing that while the tunneling probabilities are asymmetric, a better method is needed to achieve higher asymmetries.}
    \label{fig:method1dr}
\end{figure}

 The colors in the integration of $|\Psi (x, t)| ^2$  correspond to the shaded regions seen in Fig.~\ref{fig:method1Potential}. Accordingly, $T_R$ corresponds to the case where the BEC is cooled on the left of the barrier, tunnels through the barrier, and enters into the \textcolor[HTML]{5151cc}{blue} shaded region on the right of the potential, whereas $T_L$ corresponds to the case that the BEC is cooled on the right of the barrier, tunnels through the barrier, and enters the \textcolor[HTML]{FFA500}{orange} region to the left of the potential. These integration regions are chosen away from the barrier, so that particles still traveling through the barrier are not counted as having tunneled. Additionally, these regions are equally sized, located at equal distance from the initial positions of the cooled BEC, and the magnitude of the initial kick towards the barrier is the same for the left-to-right and right-to-left cases. 
 In these single-pass examples, shown in Fig.~\ref{fig:method1TP}(b), we see that $T_L$ and $T_R$ are already different immediately after the initial time at which either the left-to-right or right-to-left propagating BEC has tunneled through the barrier and entered the \textcolor[HTML]{5151cc}{blue} or \textcolor[HTML]{FFA500}{orange} shaded regions respectively. Seeing that $T_L$ and $T_R$ begin increasing at identical time, albeit at different rates, we are confident that the asymmetry is not a result of time delay between the two cases. Furthermore, the difference between the two probabilities persists until the simulations are stopped, reflecting the asymmetry in Fig~.\ref{fig:method1GPEdensplot}(a) and (b). While it would be optimal to verify this by allowing the BEC to completely evacuate the region of the barrier, the previously mentioned grid size and resolution constraints currently do not allow this at a scale where the barrier may be replicated in experiment. For reference, the Schrödinger case is shown in Fig.~\ref{fig:method1TP}(a).

According to  Fig.~\ref{fig:method1TP}(a), the Schr\"odinger equation shows identical tunneling probabilities from either side. This is reinforced by the relative difference maximum of $d_r = 0.5\%$ seen in Fig.~\ref{fig:method1dr}, corresponding to tunneling probabilities on the order of $10^{-8}$ and numerical fluctuations on the order of $10^{-10}$. The GPE, however, predicts that the two probabilities diverge, as seen in Fig.~\ref{fig:method1TP}(b). We see that these probabilities continue to rise, while $d_r$ decreases as the overall tunneling probability increases. At this point, however, we stop the simulation due to the previously mentioned grid-size constraints. We do note that in Fig.~\ref{fig:method1TP}(a), the Schr\"odinger equation predicts overall an exponentially smaller magnitude of tunneling probability. The lowered magnitude of tunneling probability is due to the much lower wave packet energy predicted by the single-particle Schr\"odinger equation, which we recall from Eq.~\eqref{EqWKB1DTunneling} corresponds to exponentially smaller tunneling probability. Indeed, exponentially large enhancement in tunneling rate in a BEC due to a mean-field energy shift has been well studied in many contexts~\cite{Lee_2007,Schlagheck2007,paul2007,Zenesini_2008}, and even in a linear system, BEC tunneling rates can be dramatically  enhanced due to resonances~\cite{paul2005,sias2007}.
Our focus, however, is specifically on tunneling {\it asymmetry} rather than on an overall increase or decrease in tunneling rate due to nonlinear coupling.

Overall, the GPE predicts a peak asymmetry of $d_r = 57\% $ soon after significant tunneling probabilities first appear and a final asymmetry of $d_r = 1.2 \%$ at $t=33.6$ ms when the simulation is stopped (Fig.~\ref{fig:method1dr}). Ultimately this gives a confirmation of asymmetry, but does not show a clear method to obtaining high asymmetries. In order to better examine the asymmetric behavior, a more optimized case is considered in the following sections.


\subsection{Transition from Symmetric to Asymmetric Tunneling}

We now directly examine the interaction parameter $g$ and its impact on overall tunneling asymmetry. To this end, we vary $g$ by scaling the scattering length, the effect of which can be seen in Eq.~\eqref{eq:gparam}. This may be accomplished experimentally using Feshbach resonance to scale the scattering length, $a_s$ \cite{RevModPhys.82.1225}:
\begin{align} \label{eq:FRscaling}
    a_s(B) = \gamma\,a_{bg} \textrm{, where } \gamma = 1 - \frac{\Delta}{B -B_0}.
\end{align}
The scaling factor $\gamma$ is introduced in Eq.~\eqref{eq:FRscaling} to simplify notation; $\gamma$ could, in principle, range from $\approx 0$ to arbitrarily high numbers for repulsive interactions. 
Alternatively, Eq.~\eqref{eq:gparam} also shows that changing the number of particles accomplishes the same effect. While we consider both cases, it should be noted that manipulation of $g$ is more practically feasible through Feshbach resonance, as the scaling may be carried out without changing the sample BEC. As for the asymmetric barrier, we utilize the previous potential, but modify it such that it becomes two identical asymmetric barriers:
\begin{align} \label{m3V}
    \tilde{V}(\xi)=& 300e^{-\left(\xi+15.0\right)^2/4.5} + 240 e^{-\left(\xi+12.4\right)^2/4.5} \nonumber \\
    &+ 180 e^{-\left(\xi+9.7\right)^2/4.5} + 120 e^{-\left(\xi+7.1\right)^2/4.5} \nonumber \\
    &+ 60 e^{-\left(\xi+4.4\right)^2/4.5} + 300e^{-\left(\xi-4.4\right)^2/4.5} \nonumber \\
    &+ 240 e^{-\left(\xi-7.1\right)^2/4.5} + 180 e^{-\left(\xi-9.7\right)^2/4.5}  \nonumber \\
    &+ 120 e^{-\left(\xi-12.4\right)^2/4.5} + 60 e^{-\left(\xi-15.0\right)^2/4.5}.
\end{align}

\begin{figure}[h]
    \includegraphics[width=0.45\textwidth]{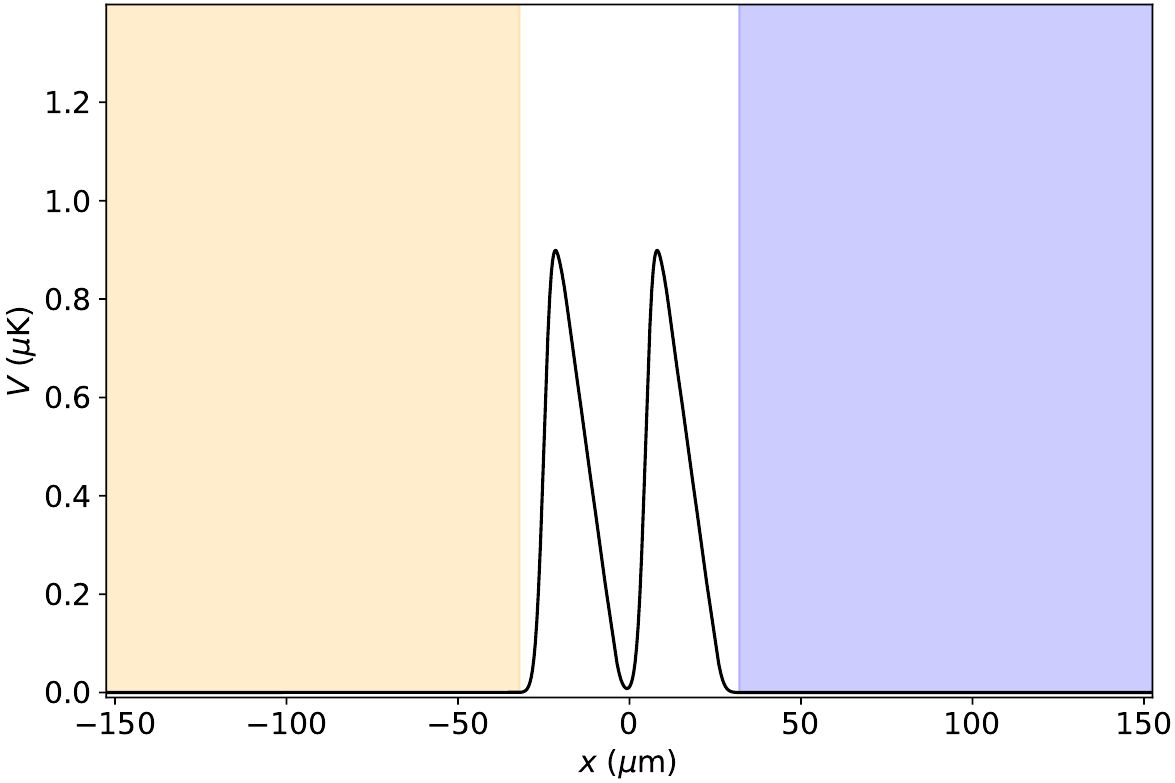}
    \caption{A two-peak asymmetric barrier, which offers multiple chances for transmission and reflection. As with the calculations above (see Fig.~\ref{fig:method1Potential}), the \textcolor[HTML]{5151cc}{blue shaded region} corresponds to the case in which we cool the BEC on the left of the asymmetric barrier at $x=-76.3\,\textrm{\textmu m}$ and kick the BEC such that it propagates to the right. The \textcolor[HTML]{FFA500}{orange shaded region} corresponds to the probability of the flipped case, where we instead cool the BEC at $x = 76.3\,\textrm{\textmu m}$ and kick it such that it propagates to the left.}
    \label{fig:method3potential}
\end{figure}

\begin{figure}
    \subfloat[]{\includegraphics[width=0.45\textwidth]{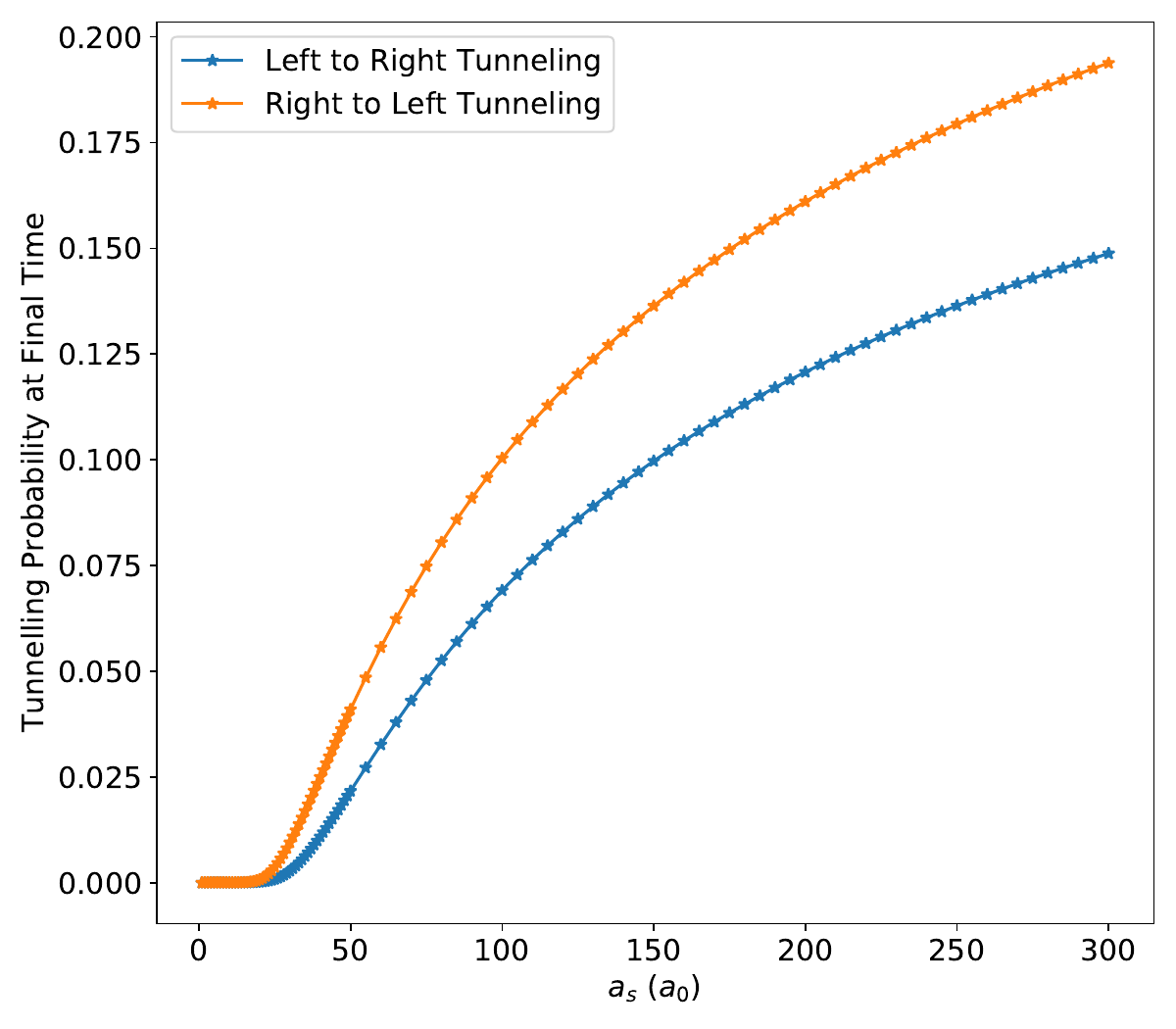}}

    \subfloat[]{\includegraphics[width=0.45\textwidth]{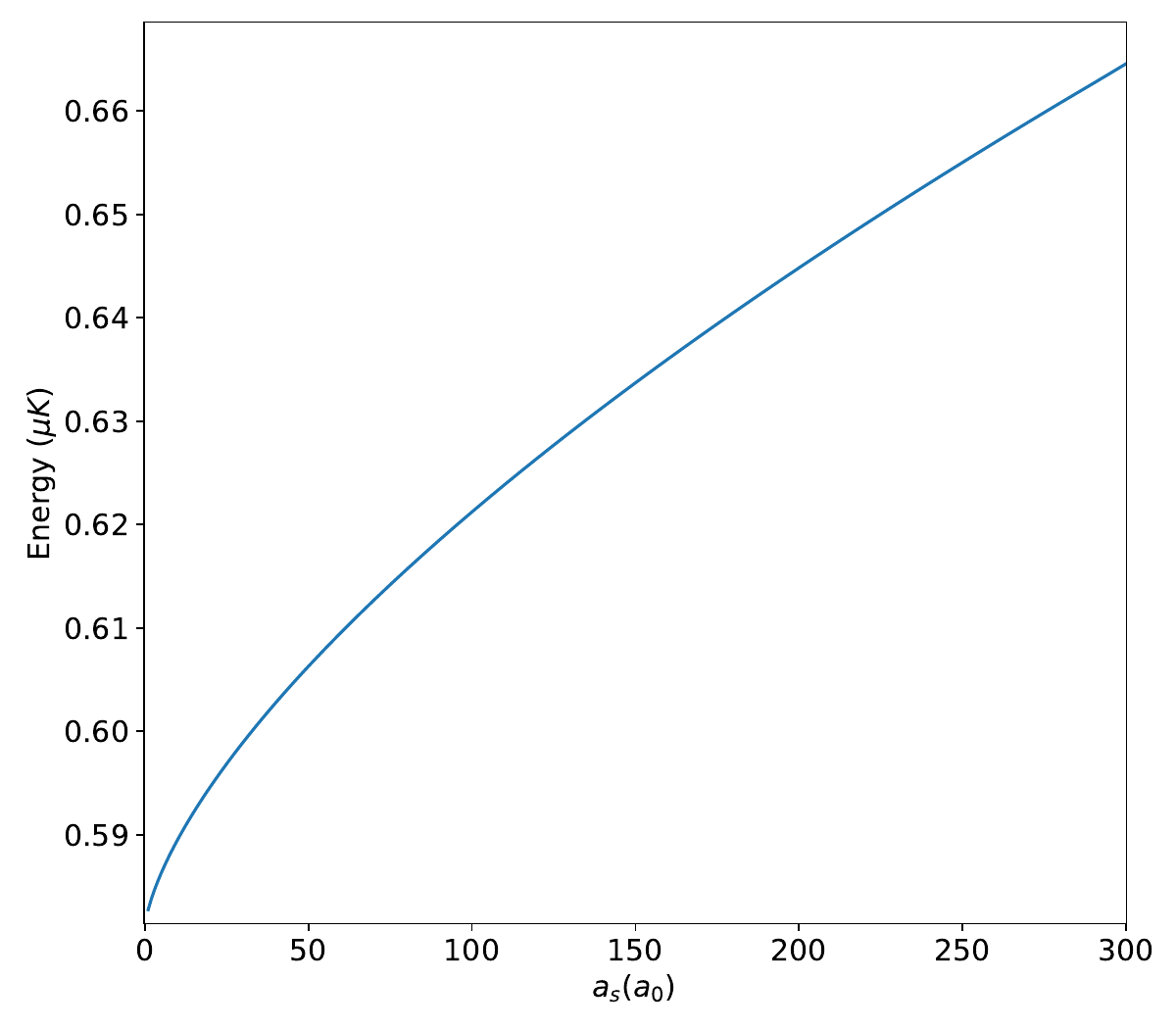}}
    \caption{For initial wavepackets prepared at the same location, this plot shows the change in tunneling probabilities and energy as the scattering length is varied. Here $a_s=0$ corresponds to $g=0$, the Schr{\"o}dinger case.
    (a) The threshold of tunneling is clearly visible as  $a_s$ is increased. As in the previous figures, the plot is color coded to match Fig.~\ref{fig:method3potential}.
    (b) The energy of the wavepacket is affected by tuning $a_s$ and correspondingly $g$. The resulting change in energy, as we recall from Eq.~\eqref{EqWKB1DTunneling}, corresponds to an exponential change in tunneling probability. However, we also note that the apparent square-root dependence of energy on scattering length does not completely explain the seemingly threshold-like behavior in tunneling probability, pointing to a contribution of $g$ that goes beyond simple changes in energy.
    }
    \label{fig:method3scalingprobdiff}
\end{figure}

This creates a total asymmetric barrier that allows the BEC multiple chances for tunneling or for reflection, as seen in Fig.~\ref{fig:method3potential}. As each chance to tunnel is asymmetric, increasing the number of barriers should further highlight the asymmetry. In future work, this may eventually be extended such that a larger number of barriers may act as an atomtronic diode, allowing a BEC to transmit from, say, left to right, while effectively eliminating the possibility that it travels right to left.

We perform 100 simulations, increasing $\gamma$ in steps of 0.01 over the interval $0 < \gamma \leq 0.50$ and in steps of 0.05 over the interval $0.50 < \gamma \leq 3.0$. Because of this number of individual simulations as well as the more complex nature of the barrier, we reduce the distance between the initial state's offset and the center of the asymmetric barrier, with each state being prepared in a potential with $v_0 = 0.5$ at $x_0 = \pm 76.3$\,\textmu m. As compared to previous simulations, we reduce the time of propagation to $t=11.14\, \textrm{ms}$ and the range of $x$ is taken to be  $\pm 152.5$\,\textmu m. Accordingly, we increase the initial kick applied to the BEC to $\kappa = 22$ such that it propagates faster, and with much higher kinetic energy. This provides adequate resolution to demonstrate the relationship between tunneling probability at the final time, $t=11.14$\,ms, and $a_s$, as shown in Fig.~\ref{fig:method3scalingprobdiff}. In the case of $g = a_s = 0$, symmetric tunneling is restored as expected.

Additionally, Fig.~\ref{fig:method3scalingprobdiff} can also be visually compared to Fig.~\ref{fig:method1TP}, as $a_s=100a_0$ is the background scattering length used in Sec.~\ref{sec:Modeling}.A. As previously discussed, the tunneling asymmetry of the BEC in Fig.~\ref{fig:method1TP} is quite small at the final time. We can see from Fig.~\ref{fig:method3scalingprobdiff} that, with the same scattering length $a_s = 100a_0$, the splitting is much more pronounced, showing a clear example of barrier modification that enhances the tunneling asymmetry. It is most surprising, though, that the relationship between the tunneling probability and $a_s$ exhibits a threshold-like behavior, where the tunneling probabilities remain symmetric and close to 0 for the lowest values of the scattering length $a_s$ but increase and become asymmetric above $a_s \approx 20 a_0$. As increasing $g$ by scaling $a_s$ changes the energy and correspondingly affects tunneling probability, we must ensure that this threshold-like behavior is not simply a result of rapid changes in total energy. In Fig.~\ref{fig:method3scalingprobdiff}(b), we illustrate the change in total energy as a function of scaling $a_s$ and see that, relative to the magnitude of total energy, the contribution of $g$ is small due to the increased kinetic energy from the larger kick parameter $\kappa$ in the current example. The change in energy is also smooth and monotonic throughout, even within the region centered about the threshold-like behavior, $15 a_0 \leq a_s \leq 25 a_0$. The relatively small and well-behaved changes in energy due to scaling $a_s$ within that region cannot alone explain the threshold-like behavior of the tunneling probability seen in Fig.~\ref{fig:method3scalingprobdiff}(a). Thus, we are confident that there is an inter-particle interaction $g$ contribution predicted by the GPE that goes beyond a simple increase in energy.

The effects presented here have been indirectly noted in previous many-body simulations~\cite{Haldar2019}, albeit for the purpose of studying different aspects of tunneling. Ref.~\cite{Haldar2019} utilizes many-body multi-configurational time-dependent Hartree methods~\cite{AlonStreltsov2008} to investigate the effect of varying the inter-particle BEC interaction strength in double-well potentials. The modified oscillation of tunneling probability~\footnote{In Ref.~\cite{Haldar2019}, the notion of survival probability is used, which is just 1 minus the tunneling probability.} depending on which well the BEC starts in, can be seen in Figs.~1(c-f) of Ref.~\cite{Haldar2019}. 
Furthermore, an increase in the asymmetry of tunneling was observed in that Figure as inter-particle interactions were increased. Of particular note is the recognition that very small inter-particle interactions do not lead to noticeable tunneling asymmetry. Rather, the asymmetry of tunneling emerges as inter-particle interactions are increased beyond a certain levelg, which the single-pass results seen here in Fig.~\ref{fig:method3scalingprobdiff} also demonstrate.

Additionally, Ref.~\cite{Haldar2019} also explores the effects of barrier asymmetry by adding a scalable slope to the central barrier of an idealized double-well potential. This is implemented as a simple and tunable control of barrier asymmetry, and the mechanisms behind its effect on tunneling asymmetry are not directly examined. However, we note that the results certainly correspond to the intuition presented in Fig.~\ref{fig:AsymmetrictunnelingMaxwellDemon} of the current work, where a more gradual slope is more conducive to tunneling than a steeper slope (when starting in a multi-particle ground state).
Finally, Ref.~\cite{Haldar2019} suggests that, for weak, repulsive interactions, the mean-field approach very closely matches many-body predictions. Accordingly, we move beyond the case of a free BEC experiencing single-pass asymmetric tunneling to examine the behavior of confined BECs.


\section{Going Beyond Landau: Tunneling in a Trapping Potential}\label{sec:TrapPot}

Landau's prediction of symmetric tunneling (see Sec. 25 of Ref.~\cite{landau2013quantum}) relies upon the requirement that the potential vanishes on either side of the barrier (see Fig.~\ref{fig:potentialdemo}); as a result, a solution to the time-independent Schr{\"o}dinger equation in these outside regions is a linear combination of incoming and outgoing plane waves (see App.~\ref{sec:1dproof} for the detailed proof). The examples seen in Sec.~\ref{sec:Modeling} above were carried out to test BEC tunneling under these same conditions. Landau's argument does not apply to the case of trapping potentials because the condition of asymptotically vanishing potential is violated. Because trapping potentials are better suited for experimentally studying asymmetric tunneling of a BEC by utilizing painted potentials~\cite{ryu2015integrated, Roy2016, Albers2021alloptical}, we ought first to confirm the symmetry of tunneling for the Schr{\"o}dinger equation under trapped conditions before contrasting the results with the GPE.

\subsection{Symmetric Tunneling for the Schr{\"o}dinger Equation in a Trap}

Unlike in the cases presented in Sec.~\ref{sec:Modeling}, the solution to the Schr{\"o}dinger equation on either side of the barrier depends on the form of the trapping potential, and a rigorous proof of left-right tunneling symmetry that accounts for an arbitrary trapping potential is non-trivial. While we expect that this may be accomplished in future work, here we present numerical evidence for symmetric tunneling. 

\begin{figure}
    \centering
    \includegraphics[width=0.4\textwidth]{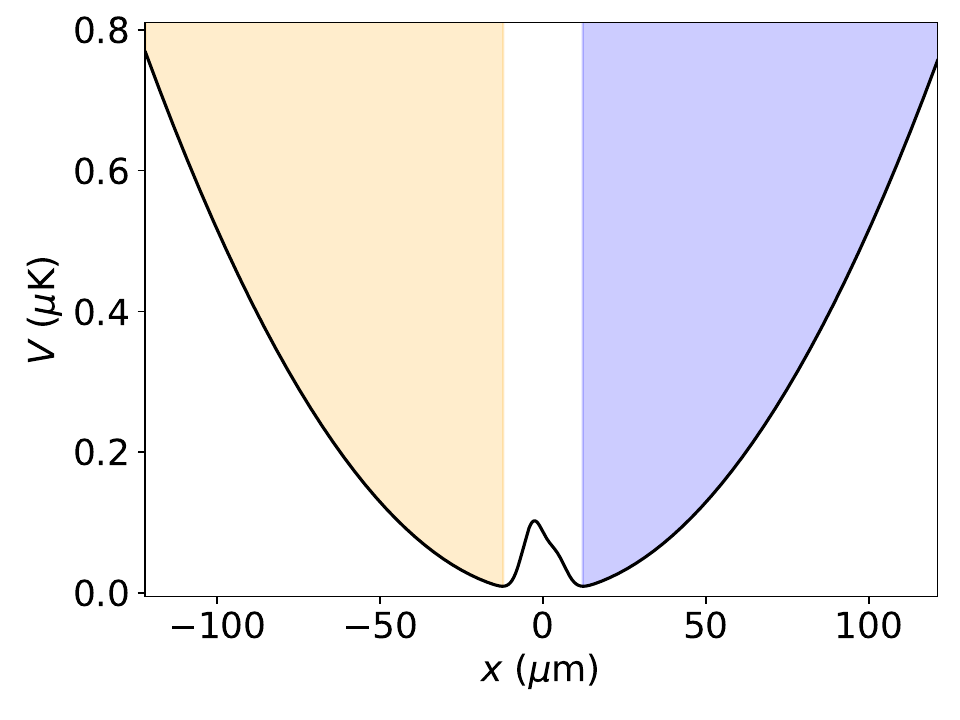}
    \caption{The asymmetric barrier from Eq.~\eqref{eq:symschroV1} used for examining tunneling symmetry in the Schr{\"o}dinger equation. As this potential is not used for GPE modelling, we do not include the GPE ground state energy as in previous potential plots. The shaded regions  are the ones over which we integrate to obtain the probability of tunneling, with the \textcolor[HTML]{5151cc}{blue shaded region} corresponding to the BEC cooled on the left tunneling to the right and the \textcolor[HTML]{FFA500}{orange shaded region} corresponding to the probability of the flipped case.}
    \label{fig:symschroV1}
\end{figure}

For an initial test, we consider a simple asymmetric potential composed of two fixed-width Gaussians and an added quadratic trapping term of the dimensionless form:

\begin{align}\label{eq:symschroV1}
     \tilde{V}(\xi) = 0.05 \xi^2 + 40e^{-\left(\xi +2\right)^2/9} + 20.82e^{-\left(\xi -2.5\right)^2/9}.
\end{align}

\begin{figure}
    \centering
    \subfloat[]{\includegraphics[width=0.45\textwidth]{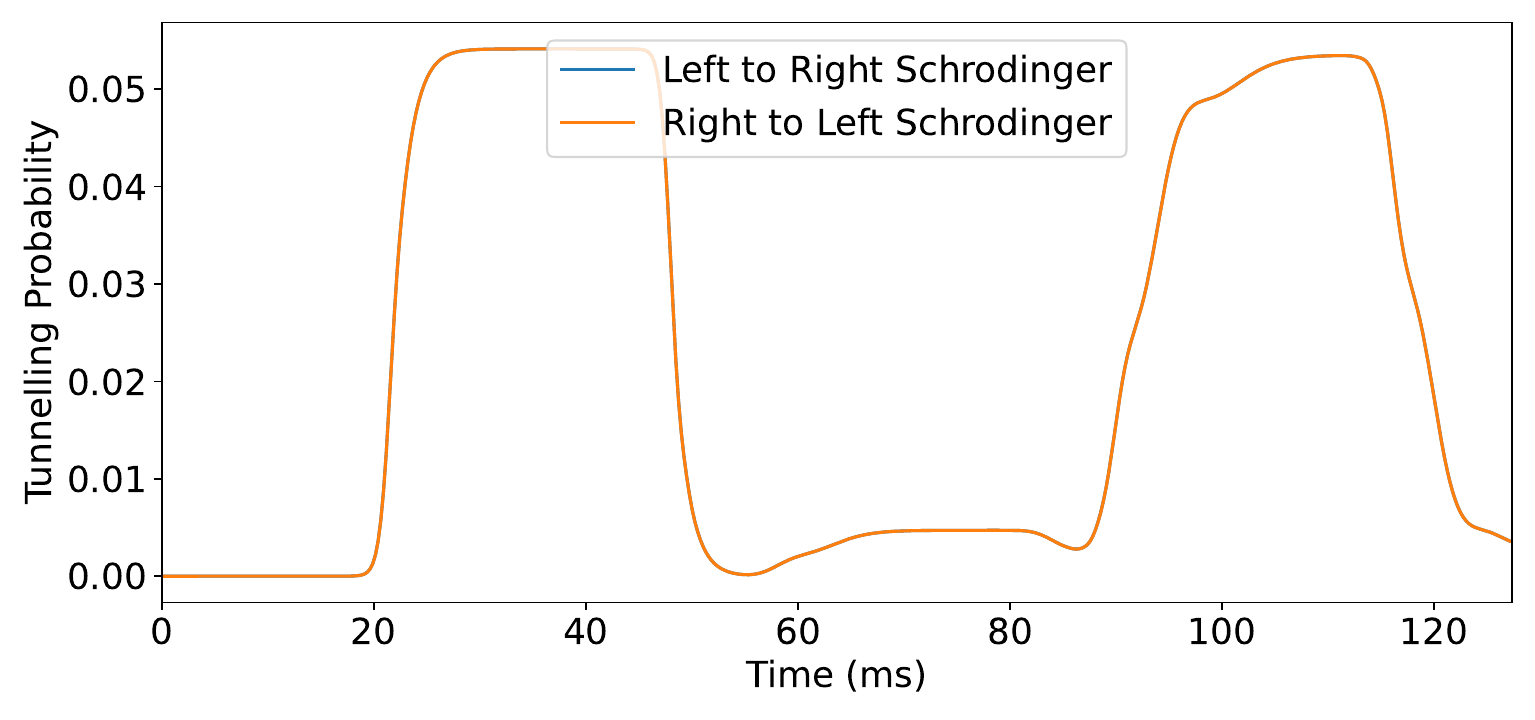}}
    
    \subfloat[]{\includegraphics[width=0.45\textwidth]{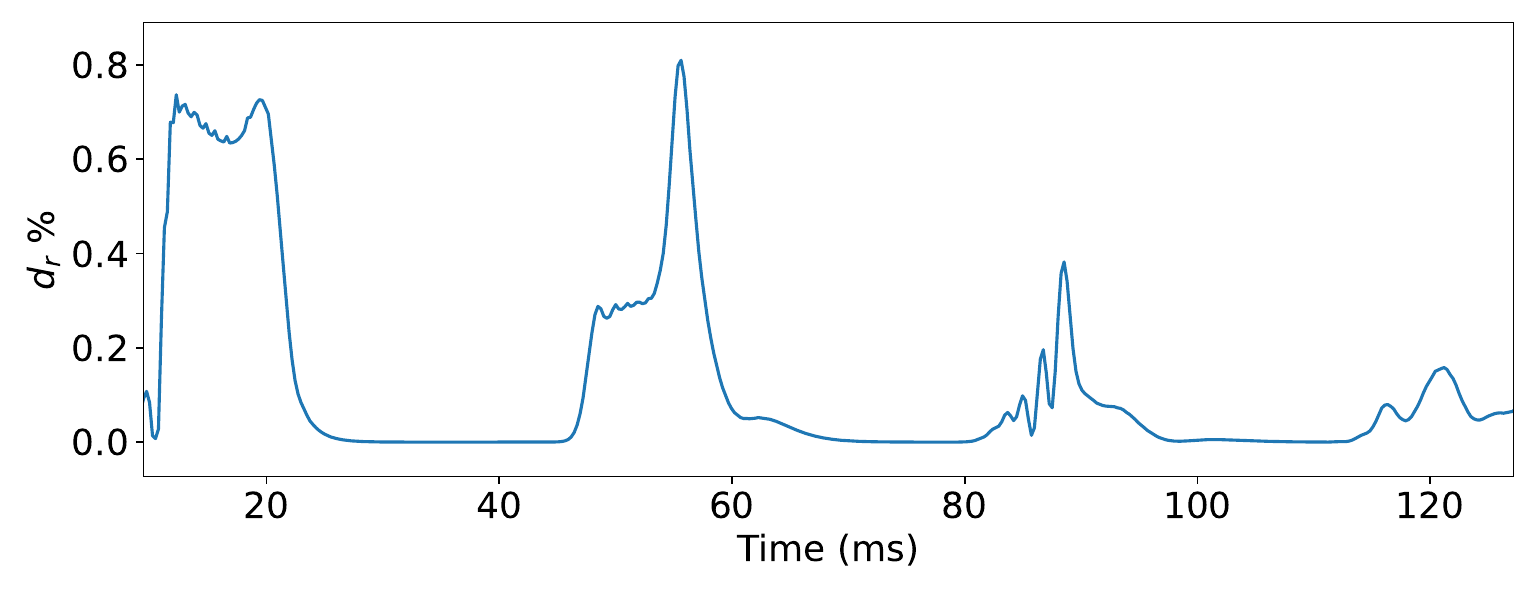}}
    \caption{(a) Tunneling probability for a BEC modelled by the Schr\"odinger equation. Because the BEC is trapped, it can be propagated for $t \approx 127$ ms, which is much longer than in the Landau case. Note that the colors are synchronized to the potential in Fig.~\ref{fig:symschroV1}, with the \textcolor[HTML]{5151cc}{left to right Schr\"odinger}  tunneling corresponding to integration over the \textcolor[HTML]{5151cc}{blue shaded region} and the \textcolor[HTML]{FFA500}{right to left Schr\"odinger} tunneling corresponding to integration over the \textcolor[HTML]{FFA500}{orange shaded region}
    (b) Relative difference between Schr{\"o}dinger tunneling probabilities using Eq.~\eqref{eq:asymmetrymeasure}. There are small spikes as the probabilities approach and leave $0$, but $d_r$ remains on the order of $0.1\%$}
    \label{fig:symschroTP1}
\end{figure} 

This potential (see Fig.~\ref{fig:symschroV1}) is a natural extension of those in Sec.~\ref{sec:Modeling} but with an added quadratic trap. We begin, as in the previous section, by simulating the cooling of a BEC at $x=\pm 30.5$ \textmu m with a cooling trap tightness parameter of $v_0 = 0.25$ (see Eq.~\eqref{eq:trapping}), and then propagate the BEC for $t \approx 127$ ms using the Schr\"{o}dinger equation. The results can be seen in Fig.~\ref{fig:symschroTP1}. Even with the addition of a quadratic trapping term, the tunneling probabilities are identical [Fig.~\ref{fig:symschroTP1}(a)], and the relative difference between the left and right cases remains on the order of $0.1\%$ [Fig.~\ref{fig:symschroTP1}(b)], which is consistent with numerical noise and is of the same order as in Fig.~\ref{fig:method1dr}, the rigorously proven case of Landau.

\begin{figure}
    \includegraphics[width=0.45\textwidth]{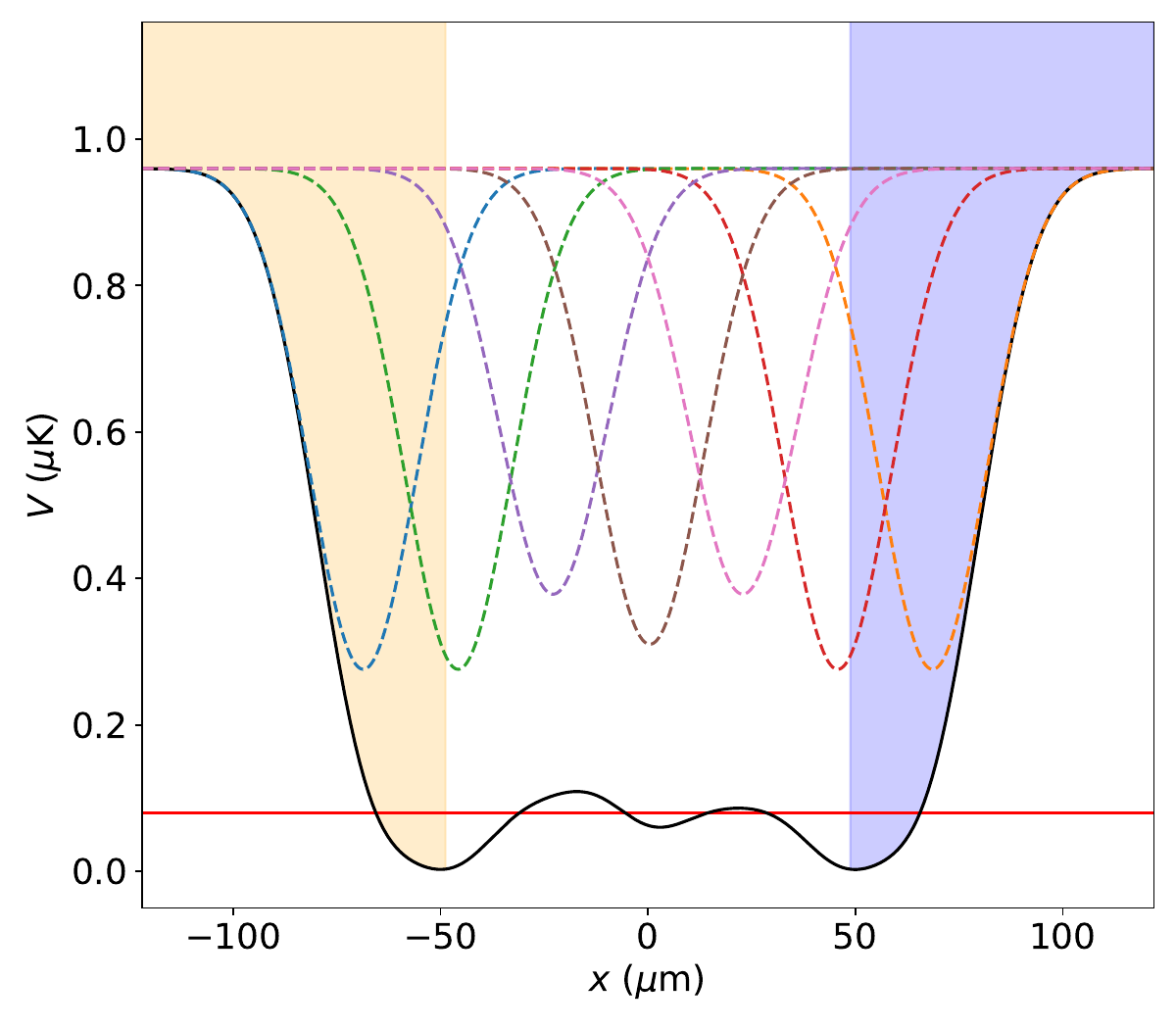}
    \caption{The asymmetric barrier in Eq.~\eqref{eq:m2V} used for  Schr{\"o}dinger and GPE tunneling simulations. The dotted lines indicate the seven fixed-width Gaussians used to create the potential, and the red line indicates the GPE ground state energy of the BEC ($\approx 80$ nK). The shaded regions, again, are the ones over which we integrate to obtain the probability of tunneling, with the \textcolor[HTML]{5151cc}{blue shaded region} corresponding to the BEC cooled on the left tunneling to the right and the \textcolor[HTML]{FFA500}{orange shaded region} corresponding to the flipped case.}
    \label{fig:method2Potential}
\end{figure}

To continue our investigation, we have chosen parameters such that our trapping asymmetric potential [see Eq.~\eqref{eq:m2V}] may be entirely created by a series of fixed-width Gaussians of varying intensity, which could be realized by a time-averaged experimental technique~\cite{ryu2015integrated, Roy2016, Albers2021alloptical}. To best match future experimental parameters, the potential is comprised of Gaussians corresponding to a waist size of $\approx 17$ \textmu m (Fig.~\ref{fig:method2Potential}):
\begin{align} \label{eq:m2V}
    \tilde{V}(\xi)=& 400 - 285 \Big(e^{-\left(\xi+45\right)^2/144.5} + e^{-\left(\xi-45\right)^2/144.5} \nonumber \\
    &+ e^{-\left(\xi+30\right)^2/144.5} + e^{-\left(\xi-30\right)^2/144.5} \Big) \nonumber \\
    &- 242.25 \left(e^{-\left(\xi+15\right)^2/144.5} + e^{-\left(\xi-15\right)^2/144.5} \right) \nonumber \\
    &- 270.25e^{-\left(\xi-0.3\right)^2/144.5}.
\end{align}

\begin{figure}[h]
    \centering
    \subfloat[]{\includegraphics[width=0.45\textwidth]{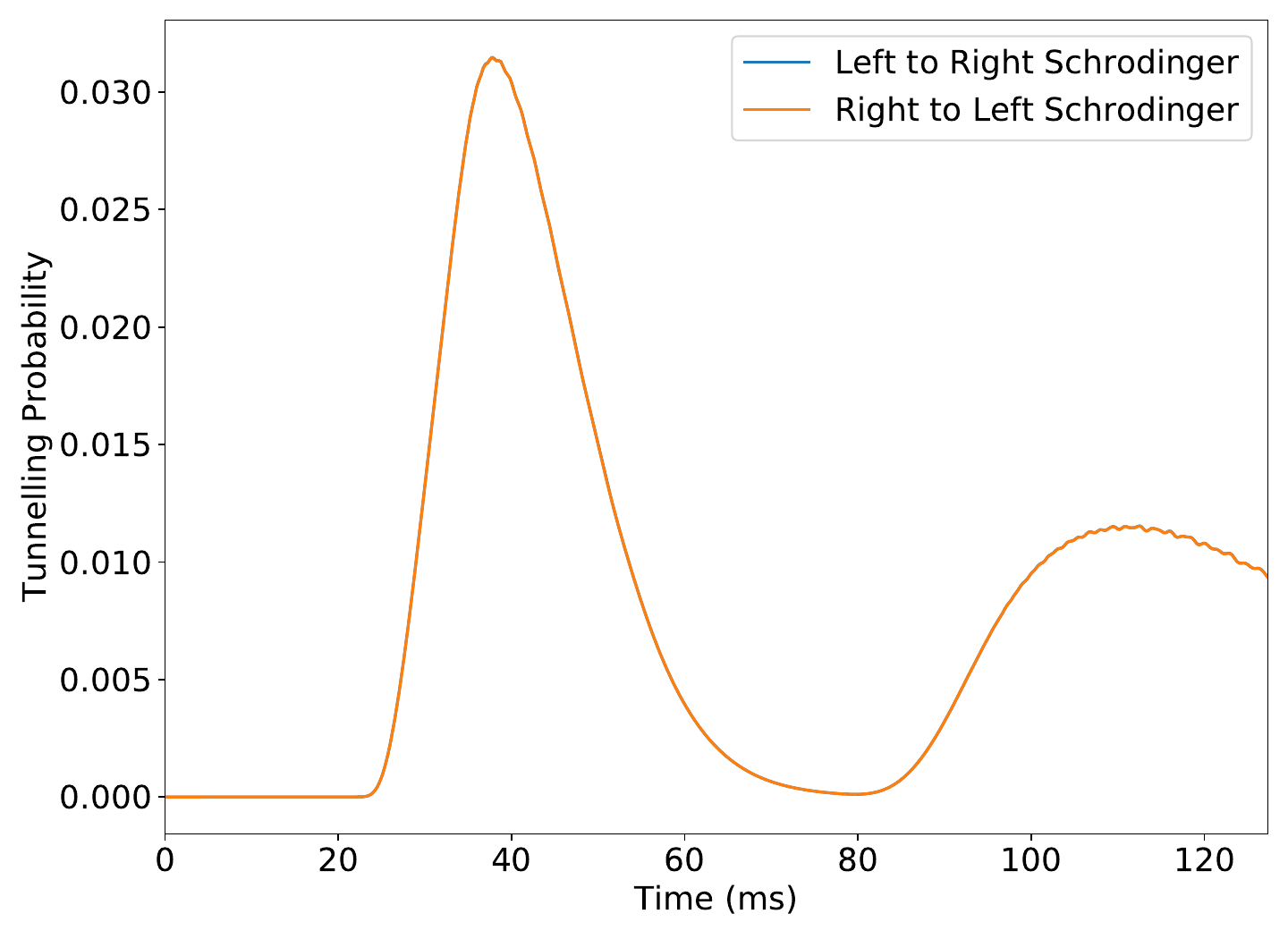}}
    
    \subfloat[]{\includegraphics[width=0.45\textwidth]{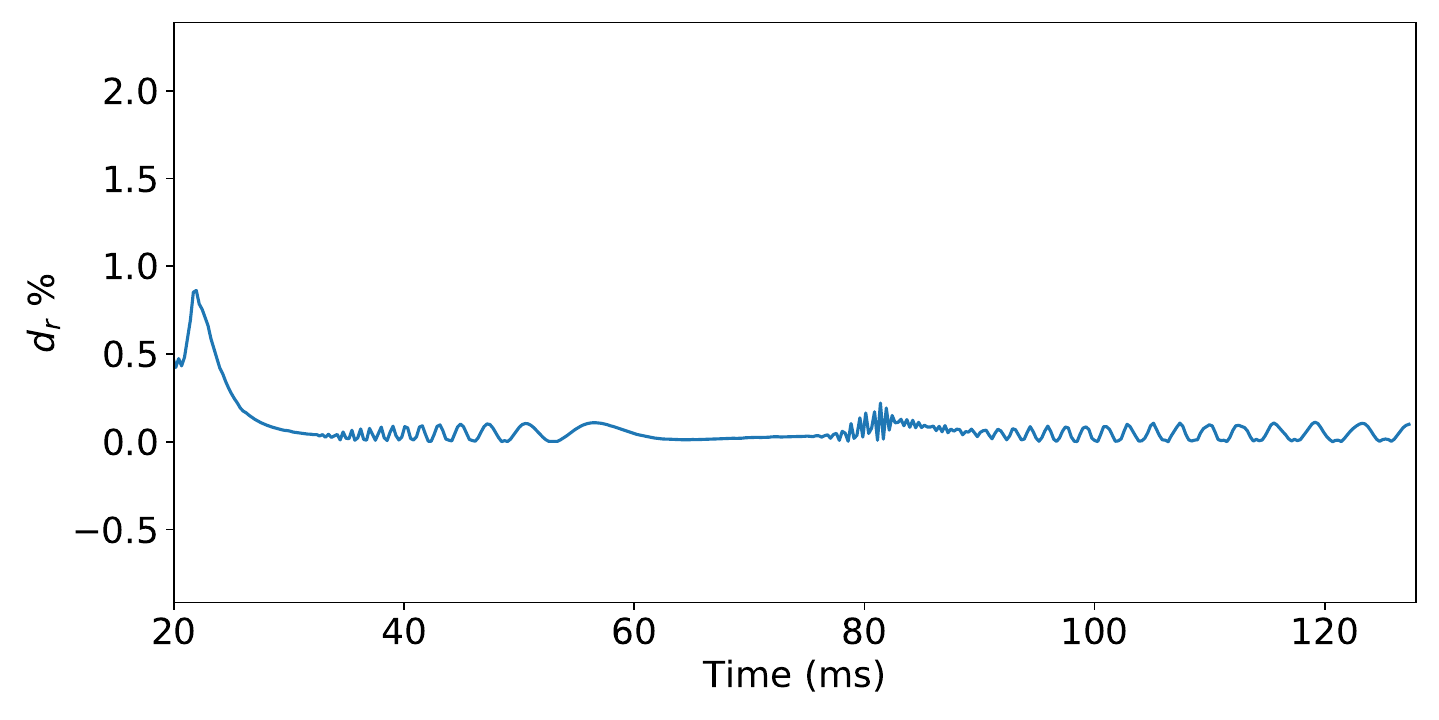}}
    \caption{(a) Tunneling probability for a BEC modelled by the Schr\"odinger equation. Because the BEC is trapped, we are able to propagate for $t \approx 127$ ms, which is much longer than in the Landau case. Note that the colors are synchronized to the potential in Fig.~\ref{fig:method2Potential}, with the \textcolor[HTML]{5151cc}{left to right Schr\"odinger}  tunneling corresponding to integration over the \textcolor[HTML]{5151cc}{blue shaded region} and the \textcolor[HTML]{FFA500}{right to left Schr\"odinger} tunneling corresponding to integration over the \textcolor[HTML]{FFA500}{orange shaded region}.
    (b) Relative difference between Schr{\"o}dinger tunneling probabilities using Eq.~\eqref{eq:asymmetrymeasure}. There is a small spike around $t\approx 20$ ms which is just before the tunneling probability rises above the level of machine error.}
    \label{fig:symschroTP2}
\end{figure}

The BEC is prepared further off-center to accommodate the wider asymmetric potential barrier and is cooled with a slightly tighter trapping potential, choosing $v_0 = 0.5$ for the tightness parameter and  $\xi_0 = \pm 37$, or $x = \pm 56$\,\textmu m, as the offset in Eq.~\eqref{eq:trapping}. As with the previous potential, we perform the Schr\"{o}dinger  propagation for $t=127.3$ ms. As we see in Fig.~\ref{fig:symschroTP2}, the tunneling probabilities overlap, and the relative difference between the left and right tunneling probabilities oscillates around $0.1\%$, which is again on the same order as in Fig.~\ref{fig:method1dr}.


\subsection{Recurrent Asymmetric Tunneling of BECs}

For GPE modelled BEC tunneling, we use the previous potential from Eq.~\eqref{eq:m2V} as it utilizes experimentally realizable parameters.
\begin{figure}
    \subfloat[]{\includegraphics[width=0.45\textwidth]{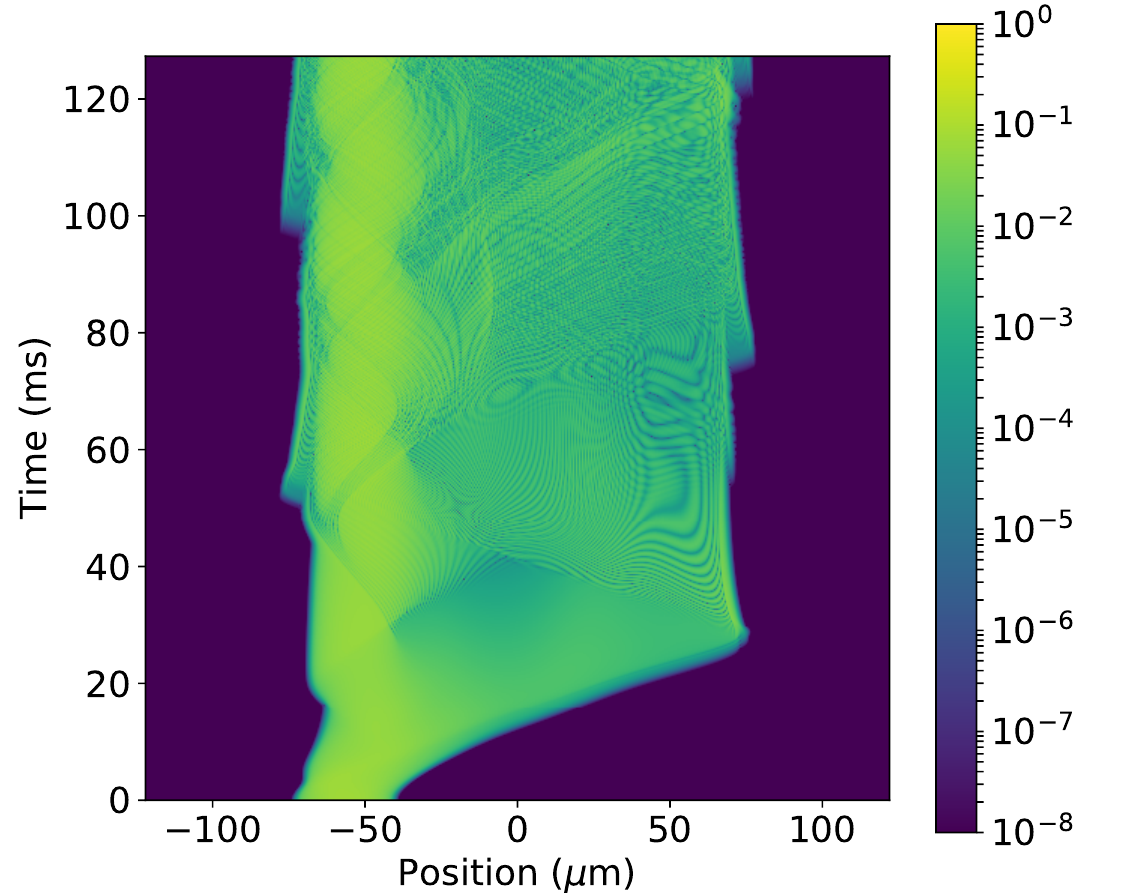}}
    
    \subfloat[]{\includegraphics[width=0.45\textwidth]{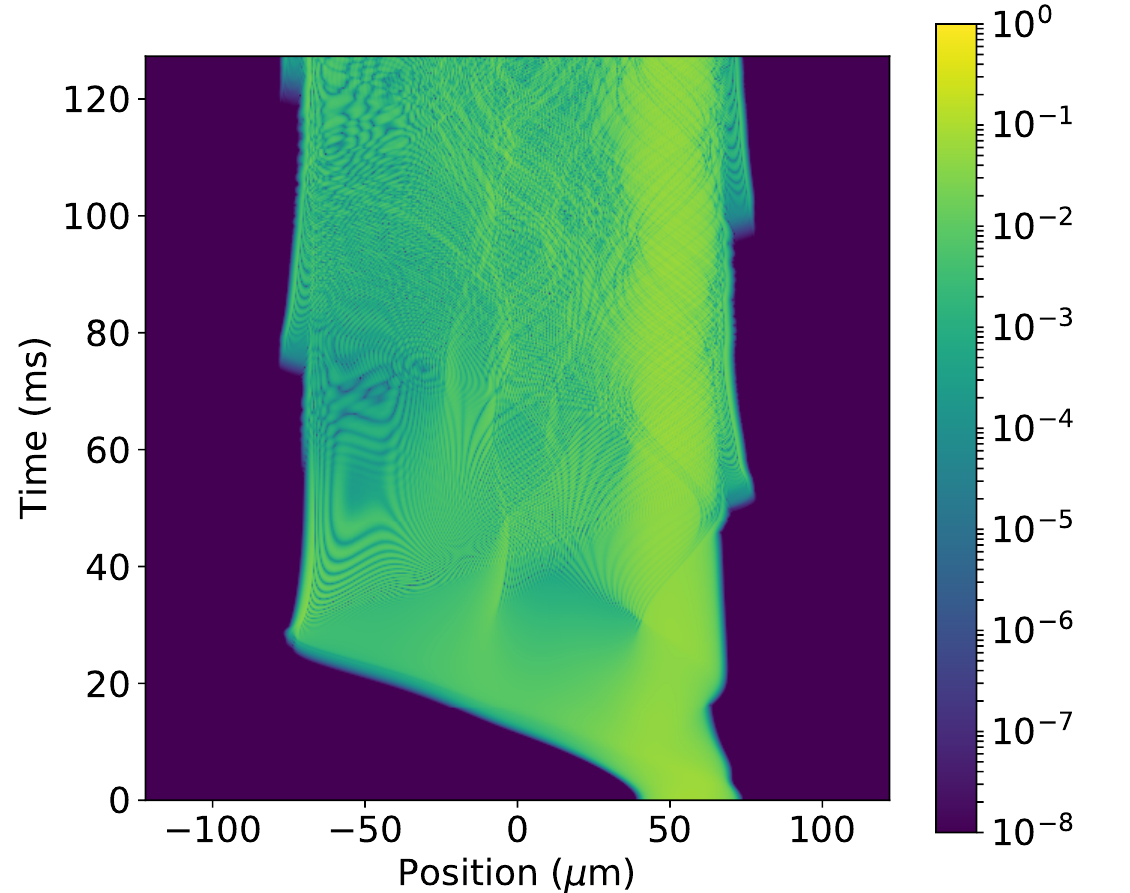}}
    \caption{Time evolution of a BEC modelled by the GPE, where (a) shows the BEC prepared on the left side of the trap and (b) shows the BEC initially on the right side of the trap's central barrier. The color bar in each plot corresponds to $|\Psi(x, t)|^2$. Because particles of the BEC may tunnel from one side of the barrier to the other and then back, we notice that the tunneling probability may increase or decrease over time.}
    \label{fig:method2GPEdensplot}
\end{figure}
This potential (see Fig.~\ref{fig:method2Potential}) confines the BEC in a smooth well such that it will have multiple chances to reflect or tunnel through the asymmetric barrier in the center. Additionally, once tunneled through, particles on the other side of the barrier will then have a chance to tunnel back to their original side. Because the BEC is confined within a trapping potential (Fig.~\ref{fig:method2Potential}), it can be propagated over a long time, $\tau=40$, or $t \approx 127$\,ms, allowing multiple chances for tunneling. This allows larger disparities between left-to-right and right-to-left tunneling probabilities to form. The BEC is prepared closer to the barrier than in the previous calculation, with a much tighter trapping potential. Here the tightness parameter is $v_0 = 0.5$,  and the offset in Eq.~\eqref{eq:trapping} is $\xi_0 = \pm 37$, or $x \pm 56$\,\textmu m. Examining the density plots in Fig.~\ref{fig:method2GPEdensplot}, we see the oscillatory behavior expected for a trapped matterwave, but with added complexity arising from the repeated tunneling behavior.

\begin{figure}[H]
    \includegraphics[width=0.45\textwidth]{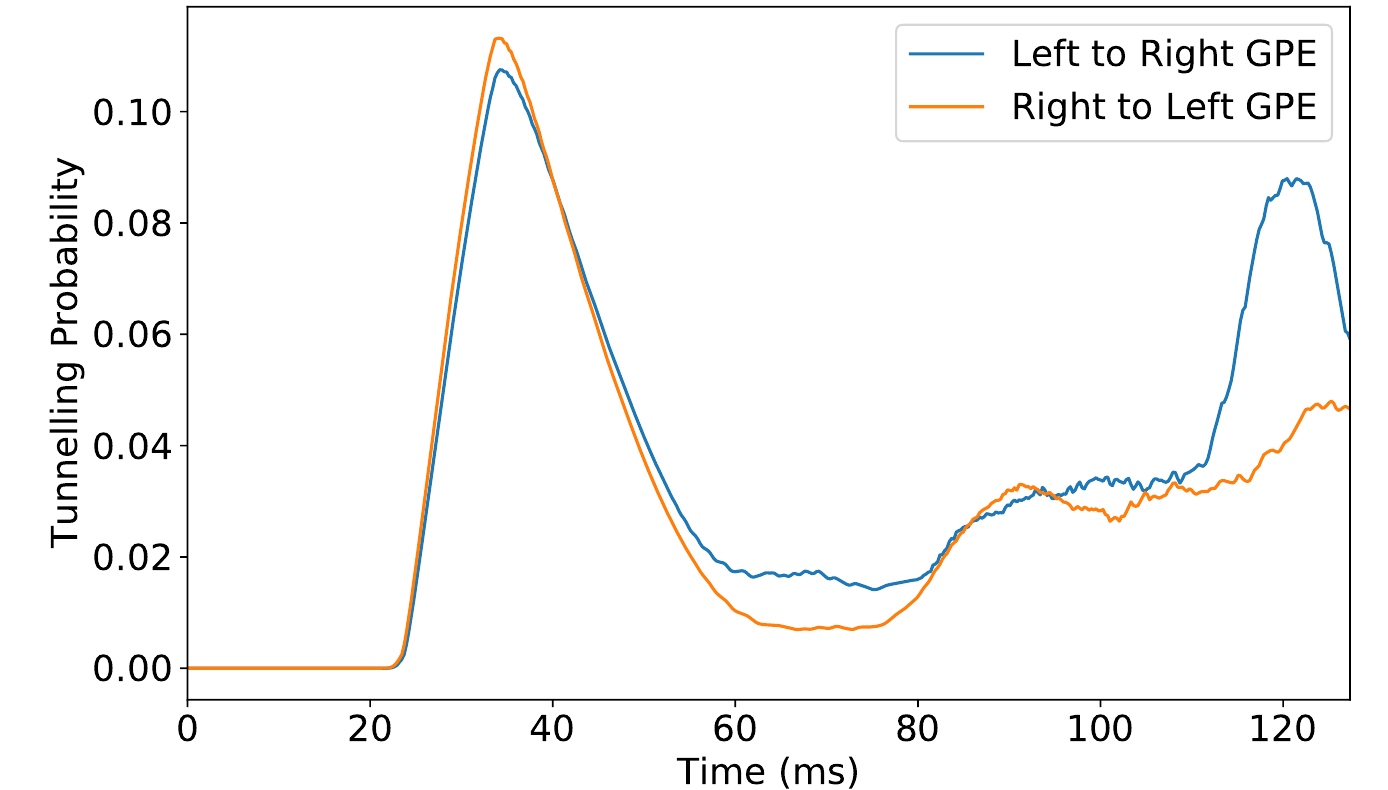}
    \caption{Tunneling probability for a BEC modelled by the GPE.
    Because the BEC is trapped, it can be propagated for $t \approx 127$ ms. Note that the colors are synchronized to the potential in Fig.~\ref{fig:method2Potential}, with the \textcolor[HTML]{5151cc}{left to right GPE} tunneling corresponding to the \textcolor[HTML]{5151cc}{blue shaded region} and the \textcolor[HTML]{FFA500}{right to left GPE} tunneling corresponding to the \textcolor[HTML]{FFA500}{orange shaded region}.}
    \label{fig:method2TP}
\end{figure}

Looking at the tunneling probabilities predicted by the GPE [Fig.~\ref{fig:method2TP}], we see an initial divergence similar to that in the single-pass calculation. However, after multiple passes, we see clear and strong asymmetry resulting at $t \approx~70 \textrm{ ms and } t \approx~120$ ms arising as the trapped BEC experiences subsequent chances at reflection and transmission. Astoundingly $d_r$ reaches values $\approx 85\%$ and $\approx 75\%$ respectively at these times. This clear tunneling asymmetry is nearly two orders of magnitude higher than the single-pass asymmetry seen in  Sec.~\ref{sec:SinglePassAsymmetry}.

\section{\label{sec:analytic} Analytic Results for Weak BEC Interaction}

Although the most spectacular examples of asymmetric BEC tunneling, as seen in Secs.~\ref{sec:Modeling} and \ref{sec:TrapPot}, are naturally observed for large BEC interaction strength $g$, it is instructive to consider also the case of especially small interaction, where the tunneling asymmetry may be calculated explicitly in perturbation theory. We begin, for simplicity, with a quantum asymmetric double-well potential, where the tunneling is dominated by one pair of stationary states -- one on the left side of the barrier and one on the right. The basis states here are eigenstates of the Hamiltonian in the limit of an infinitely high barrier. The actual Hamiltonian in this basis is given by
\begin{equation}
    H = \begin{pmatrix} \Delta & \Omega \\
    \Omega & -\Delta \end{pmatrix}\,,
\end{equation}
where $2 \Delta$ is the energy difference between the states on the left and right (related to the potential asymmetry) and $\Omega$ is the tunneling coupling. We emphasize that any asymmetry between the wells will generically give rise to an energy difference $2 \Delta$ between the quantum levels, even where the bottom of each well has the same classical energy (e.g. the state in the steeper or narrower well will have a higher energy). The GPE may then be written as
\begin{align}
\frac{d\psi_1}{dt}&= -i\left(\Delta \psi_1 +\Omega \psi_2 +\epsilon |\psi_1|^2 \psi_1\right) \nonumber \\
\frac{d\psi_2}{dt}&= -i\left(-\Delta \psi_2 +\Omega \psi_1 +\epsilon |\psi_2|^2 \psi_2\right)\,,
\label{GPE2state}
\end{align}
where $\psi_1$ and $\psi_2$ are the amplitudes for being on the left and right sides of the barrier, respectively, $\epsilon$ is a measure of nonlinearity proportional to $g$, and time units have been chosen such that $\hbar=1$. This minimal model of BEC tunneling in an asymmetric double well has appeared, for example, in Refs.~\cite{smerzi1997quantum,raghavan}. A more sophisticated model along these lines, where the overlap between the left and right wavefunctions is included in the nonlinear coupling, giving rise to additional terms such as $\epsilon |\psi_2|^2 \psi_1$, has been analyzed in Ref.~\cite{cataldojezek}, see also Refs.~\cite{ananikian,jia}. In a scenario where the nonlinearity and the overlap between the wavefunctions are both non-negligible, these additional terms can give rise to further tunneling asymmetry, but the minimal toy model of Eq.~\eqref{GPE2state} is sufficient to see the effect. We also note that, in the limit of an infinitely high barrier, the coupling $\Omega$ between the two states reduces to 0.

For the Schr\"odinger case ($\epsilon=0$) the equations are of course trivial. If we start initially on the left ($\psi_1(0)=1$, $\psi_2(0)=0$), we have
\begin{equation}
    \psi_2^{Sch}(t)=
    -\frac{i \Omega  \sin \left(t \sqrt{\Delta ^2+\Omega ^2}\right)}{\sqrt{\Delta ^2+\Omega ^2}}
\end{equation}
and the tunneling probability (from left to right) is
\begin{equation}
    T_R=|\psi_2^{Sch}(t)|^2=
    \frac{\Omega ^2 \sin ^2\left(t \sqrt{\Delta ^2+\Omega ^2}\right)}{\Delta ^2+\Omega ^2}\,.
\end{equation}

Since the left and right wells differ only by the sign of $\Delta$, the right-to-left tunneling probability $T_L$ may be obtained by changing the sign of $\Delta$. Of course, in the Schr\"odinger case we have $T_L=T_R$ as expected.

Now we may solve Eq.~\eqref{GPE2state} to first order in the nonlinearity parameter $\epsilon$. With the same initial conditions $\psi_1(0)=1$, $\psi_2(0)=0$, we obtain for small $\epsilon$,
\begin{widetext}
\begin{align}
    \psi_2(t)=&\psi_2^{Sch}(t)
    - \frac{\epsilon \Omega}{16 \left(\Delta ^2+\Omega ^2\right)^3} 
    \left[\Omega ^2 \left(\Delta ^2+\Omega ^2\right) \cos \left(3 t \sqrt{\Delta ^2+\Omega
   ^2}\right) \right.
   \nonumber \\
    +&2 \sqrt{\Delta ^2+\Omega ^2} \sin \left(t \sqrt{\Delta ^2+\Omega ^2}\right)
   \left(-i \Delta  \Omega ^2 \cos \left(2 t \sqrt{\Delta ^2+\Omega ^2}\right)+4 \Delta ^3 (\Delta  t+i)+\Delta  \Omega^2 (10 \Delta  t-i)+6 t \Omega ^4\right) \\
   +& \left. i \left(\Delta ^2+\Omega ^2\right) \left(-8 \Delta ^3 t+\Omega ^2 (4 \Delta  t+i)\right) \cos \left(t
   \sqrt{\Delta ^2+\Omega ^2}\right) \right] + O(\epsilon^2) \,.\nonumber
\end{align}

We note that, unlike $\psi_2^{Sch}(t)$, $\psi_2(t)$ is not an even function of $\Delta$, implying an asymmetry in tunneling between left and right for nonzero $\epsilon$. Indeed, squaring $\psi_2(t)$ to obtain the tunneling probability and then extracting the part odd in $\Delta$, we obtain an explicit expression for the tunneling asymmetry to leading order in $\epsilon$,
\begin{align}
    T_R-T_L =& \frac{\epsilon \Delta \Omega ^2}{8 \left(\Delta ^2+\Omega ^2\right)^3}
    \left[-4 t \left(\Omega ^2-2 \Delta ^2\right) \sqrt{\Delta ^2+\Omega ^2}
   \sin \left(2 t \sqrt{\Delta ^2+\Omega ^2}\right)-16 \Delta ^2 \sin^2 \left( t \sqrt{\Delta ^2+\Omega ^2}\right)\right. \nonumber \\
   +&2\Omega^2 \left. \sin^2 \left(2 t \sqrt{\Delta^2+\Omega^2}\right) 
   \right] +O(\epsilon^2)  \,. \label{tdiff}
\end{align}
\end{widetext}
The strength of the effect is proportional to $\epsilon\Delta$. Indeed, that can be seen directly from the form of Eq.~(\ref{GPE2state}), where the difference between time evolution in the right and left wells is that in one case the $\Delta$ and $\epsilon$ terms contribute to the phase with the same sign, while in the other case they contribute with opposite signs. Thus, nonlinearity acts differently on the left (higher energy) and right (lower energy) sides of the barrier. 

Notably the first term in the brackets in Eq.~(\ref{tdiff}) grows linearly with $t$, confirming in perturbation theory that tunneling asymmetry accumulates for multiple passes through the barrier. Specifically, for $\Delta,\epsilon<\Omega$, the asymmetry $T_R-T_L$ reaches values of order $\epsilon \Delta/\Omega^2$ after times comparable to one traversal through the barrier, and values $T_R-T_L \sim 1$ are naturally achieved after $\sim \Omega^2/\epsilon \Delta$ traversals through the barrier.

We notice in this very simple model a very nontrivial time-dependence of the tunneling probability, due to the fact that the time-dependent density drives a time dependence in the rate of flow~\cite{paul2005,Schlagheck2007}. Nevertheless, 
for short times ($t \sqrt{\Delta^2+\Omega^2}\ll 1$, i.e., on time scales shorter than the tunneling time), we have the simple form
\begin{equation}
    T_R-T_L=\frac{2}{3} \epsilon \Delta  \Omega ^2 t^4 +O(t^6)
\end{equation}
and the convenient measure of tunneling asymmetry defined in Eq.~(\ref{eq:asymmetrymeasure})
yields
\begin{equation}\label{eq:dranalytic}
    d_r = \frac{|T_L - T_R|}{\left(T_L+T_R\right)/2}=
    \frac{2}{3} |\epsilon \Delta|   t^2 +O(t^4)\,.
\end{equation}
Thus, we see that even at short times a tunneling asymmetry is always present in principle, as long as two conditions are satisfied: (i) the BEC interaction $\epsilon$ is nonzero and (ii) the potential has an asymmetry, $\Delta \ne 0$.

Of course in the general case of an asymmetric double well, there will be multiple pairs of states located on either side of the barrier, and depending on the preparation of the initial wave packet, many of these may contribute to the tunneling probability. In this more general situation, $T_R-T_L$ will naturally be given by a weighted sum of terms of the form of Eq.~\eqref{tdiff}. The overall conclusion remains unchanged: As long as asymmetry is present in the potential, even a small BEC interaction $g$ will give rise to a tunneling asymmetry proportional to $g$ (and also proportional to potential asymmetry $\Delta$ for small $\Delta$, and to $t^2$ for short times $t$).


\section{\label{conclusion} Conclusion}

Through several examples, we have demonstrated the asymmetric tunneling of a BEC through asymmetric potentials in 1D. This is in contrast to the case of non-interacting particles described by the Schr\"{o}dinger equation, where the tunneling probability respects left-right symmetry even for asymmetric barriers, as has been rigorously proven.

For future work, there is much room to optimize barrier shapes and scattering length to boost tunneling asymmetry in order to steer the dynamics of a BEC; this is a natural and direct extension of efforts to optimize scattering length to control transmission \cite{manju2018quantum}. This approach may enable atomtronic diodes \cite{daley2015towards}, facilitate BEC driven interferometers \cite{manju2020atomic}, and gravimetry~\cite{schach_tunneling_2022}.
The analytic treatment in Sec.~\ref{sec:analytic} suggests at least three strategies for obtaining large tunneling asymmetry already in the weak-interaction regime: (a) large tunneling asymmetry may naturally be obtained when a relatively small number of quantum states participate in the dynamics, (b) tunneling asymmetry is enhanced when the tunneling rate, potential asymmetry, and BEC interaction strength are all comparable, in appropriate units, and (c) tunneling asymmetry is enhanced for fast, repeated tunneling back and forth through a barrier as compared to a slow single pass. 

Asymmetric tunneling also suggests a new approach for experimental implementation of synthetic gauge fields \cite{galitski_artificial_2019, aidelsburger_artificial_2018}. Gauge fields are the building blocks of the Standard Model. Their unique physics have opened new horizons in topological quantum matter and technology, attracting a significant effort to engineer gauge fields in the laboratory. The complexity of the current experimental setups results from the assumption that tunneling is symmetric \cite{aidelsburger_artificial_2018, galitski_artificial_2019}. The presented asymmetric tunneling dynamics suggests a simpler implementation: At the bottom of a deep ring-shaped trap, arrange several asymmetric barriers while preserving their orientation, such that the vertical side of one triangular barrier abuts the angled side of the next triangle (similar to Fig.~\ref{fig:method3potential}). This trap should exhibit a strong left-right asymmetry in the transmission probability, thereby inducing a chiral motion into the BEC. This is a signature of synthetic gauge fields, with the chiral current appearing as if it were induced by a magnetic field.

A black hole analogue in BEC may also be created by asymmetric tunneling overlying self-trapping. Recall that when a wave packet is placed on one side of a symmetric double well potential, the Schr\"{o}dinger equation predicts that the wave packet undergoes recurrent oscillatory tunneling between the two wells. However, for sufficiently strong inter-particle interactions, a BEC placed on one side of the symmetric double well potential will remain trapped. This phenomenon is known as self-trapping \cite{smerzi1997quantum, albiez2005direct, julia2010bose}. Our findings suggest that it should be possible to find a value of the scattering length and an asymmetric double well barrier such that when the condensate is placed in one well it will be able to tunnel to the other but not be able to tunnel back, remaining self-trapped. 

We want to point out that self-trapping and asymmetric tunneling are two complementary phenomena. The self-trapping reveals asymmetric dynamics of BEC even in a symmetric potential; hence, it is natural to expect asymmetric tunneling through a non-symmetric barrier. 

A Maxwell’s demon may also be implemented via asymmetric tunneling since the latter is sensitive to the state of the inter-particle d.o.f. As discussed in Sec.~\ref{sec:GPE}, the left-to-right tunneling probability [Fig.~\ref{fig:AsymmetrictunnelingMaxwellDemon}(a)] is smaller than the right-to-left probability [Fig.~\ref{fig:AsymmetrictunnelingMaxwellDemon}(b)] when the inter-particle d.o.f. is initially in the ground state. This asymmetry is reversed when the system is initially in the excited state. According to Fig.~\ref{fig:AsymmetrictunnelingMaxwellDemon}(b), the angled side of the triangular barrier builds adiabatically, leaving the inter-particle state unchanged, and hence the two-particle system behaves effectively as a 1D system. If the excited inter-particle d.o.f. hits the vertical side of the barrier [Fig.~\ref{fig:AsymmetrictunnelingMaxwellDemon}(d)], the resulting shakeup forces a state transition. One possibility is for the inter-particle d.o.f. to drop to the ground state. The energy difference will be transferred to the kinetic energy of the center of mass in order to preserve the total energy, hence the system will be able to fly above the barrier as shown in Fig.~\ref{fig:AsymmetrictunnelingMaxwellDemon}(d). Comparing Figs.~\ref{fig:AsymmetrictunnelingMaxwellDemon}(a) and \ref{fig:AsymmetrictunnelingMaxwellDemon}(c), we conclude that if the wavepacket is initially placed on the left side of the barrier, then the transport rate across the barrier is insensitive to the initial state of the inter-particle d.o.f. Conversely, the transport rates shown in Figs.~\ref{fig:AsymmetrictunnelingMaxwellDemon}(b) and \ref{fig:AsymmetrictunnelingMaxwellDemon}(d) are very sensitive to the state of the inter-particle d.o.f. In this case, the system initially in the excited state is much more likely to cross the barrier than the system initially the ground state. Hence, the triangular barrier with the vertical side facing the wavepacket should act as a Maxwell’s demon.

\appendix


\section{\label{sec:1dproof} Proof of Symmetric Tunneling With the 1D Schr\"odinger Equation}

It is relatively easy to show within the time-independent 1D Schr{\"o}dinger equation that the probability of tunneling from left to right equals the probability of tunneling in the opposite direction if the barrier is symmetric, i.e., an even function of the coordinate \cite{flugge2012practical}. Generalising this, Landau gave an intuition as to why tunneling remains symmetric for an arbitrary potential that remains bounded as $x\to\pm\infty$ (see Sec. 25 of Ref.~\cite{landau2013quantum}). The tunneling symmetry in a similar case has also been derived from flux conservation in Ref.~\cite{shegelski_equal_2020} as well as in Sec. 7.3.3 of Ref.~\cite{tannor2007introduction}. In Sec. V of Ref.~\cite{schach_tunneling_2022} it was shown that symmetry of tunneling is violated for a potential $V(x) \to \text{const} \cdot x$ as $x\to\pm\infty$, which is asymptotically an unbounded linear ramp.

In this Appendix, we present another proof for tunneling symmetry.

Before proving Landau's statement, we reiterate that because we only examine the dynamics generated by a time-independent Hamiltonian, any solution of the time-dependent Schr\"odinger equation is a linear combination of the eigen-solutions to the time-independent Schr\"odinger equation. Therefore, it suffices to examine only the time-independent case.

We begin the proof by considering the wavefunction $\Psi (x)$, a solution to the 1D time-independent Schr\"odinger equation with energy $E$, incident upon a potential barrier $U(x)$.  The form of $U(x)$ consists of three separate sections (see Fig.~\ref{fig:potentialdemo}): $U(x) = 0$ within sections (I) and (III); whereas in section (II), $U(x)$ takes the form of some arbitrary barrier, $U_{\textrm{II}} (x)$. We will define our origin such that $U_{\textrm{II}} (x)$ spans the range of $-a\leq x\leq a$. Within section (II), the solution $\psi (x)$ of the time-independent Schr\"odinger equation can be written as a linear combination of two linearly independent solutions, $\psi_1(x)$ and $\psi_2(x)$, since the Schr\"odinger equation is second order. For a wave incident from the left (negative $x$) and moving to the right (positive $x$), the complete solution to the time-independent Schr\"odinger equation then takes the following form:
\begin{align}
  \Psi_L(x)  = 
                \begin{cases} \label{eq:psileft}
                  e^{ikx}+B_Le^{-ikx}           & \textrm{for} \quad -\infty<x\leq -a\\
                  C_1\psi_1(x)+C_2\psi_2(x)     & \textrm{for} \quad -a\leq x\leq a\\
                  D_Le^{ikx}                    & \textrm{for} \quad a\leq x<\infty, \\
                \end{cases} \\
    \textrm{where } k=\frac{1}{\hbar}\sqrt{2m(E)}. \nonumber 
\end{align}
For a wave incident from the right, the solution reads
\begin{align}
    \Psi_R(x)  = 
        \begin{cases} \label{eq:psiright}
                  D_Re^{-ikx}                   & \textrm{for} \quad -\infty<x\leq -a\\
                  C_1\psi_1(x)+C_2\psi_2(x)     & \textrm{for} \quad -a\leq x\leq a\\
                  e^{-ikx}+B_Re^{ikx}           & \textrm{for} \quad a\leq x<\infty. \\
        \end{cases}
\end{align}
From this we define the reflection and transmission probabilities as $R_L=\left|B_L\right|^2$ and $T_L=\left|D_L\right|^2$ respectively for the wave incident from the left, and similarly $R_R=\left|B_R\right|^2$ and $T_R=\left|D_R\right|^2$ for the wave incident from the right. In order to calculate the probabilities, we use the condition that the solutions of the Schr\"odinger equation must be  continuously differentiable. Using the relations from Eq.~\eqref{eq:psileft} and their first derivatives at the boundary $x=\pm a$, we generate the following system of four equations and four unknowns for the case of a wave incident from the left:
\begin{align} \label{eq:systemLeft}
    e^{-ika}+B_Le^{ika} &= C_1\psi_1(-a)+C_2\psi_2(-a), \nonumber \\
    ik( e^{-ika}-B_Le^{ika}) &= C_1\psi'_1(-a)+C_2\psi'_2(-a), \nonumber \\
    D_Le^{ika} &= C_1\psi_1(a)+C_2\psi_2(a), \nonumber \\
    ikD_Le^{ika} &= C_1\psi'_1(a)+C_2\psi'_2(a). 
\end{align}
By rearranging Eq.~\eqref{eq:systemLeft} and applying Cramer's rule, we obtain
\begin{align}
    B_L =& num_{B_L}/den_L, \quad D_L = num_{D_L}/den_L, \\
     \label{denLeft}
    den_L =&e^{2ika}\bigg[ ik\Big(\psi'_1(-a)\psi_2(a)-\psi_1(-a)\psi'_2(a) \nonumber \\
    &+ \psi_2(-a)\psi'_1(a)-\psi'_2(-a)\psi_1(a)\Big) \nonumber \\
    &+  \Big(\psi'_1(a)\psi'_2(-a)-\psi'_1(-a)\psi'_2(a) \nonumber \\
    &+ k^2\psi_2(-a)\psi_1(a)-k^2\psi_1(-a)\psi_2(a)\Big)\bigg], \\
    num_{B_L} =& ik\Big(-\psi'_1(-a)\psi_2(a)+\psi'_2(-a)\psi_1(a) \nonumber \\
        &-\psi_1(-a)\psi'_2(a)+\psi_2(-a)\psi'_1(a)\Big) \nonumber \\
        &+ \Big( \psi'_1(-a)\psi'_2(a)-\psi'_2(-a)\psi'_1(a) \nonumber \\
        &-k^2\psi_1(-a)\psi_2(a)+k^2\psi_2(-a)\psi_1(a)\Big), \label{eq:numBL} \\
    num_{D_L} =& -2ik\Big(\psi'_2(a)\psi_1(a)-\psi_2(a)\psi'_1(a)\Big). \label{eq:numDL}
\end{align}
Note that Eq.~\eqref{eq:numDL} contains the Wronskian, $\psi'_2(x)\psi_1(x)-\psi_2(x)\psi'_1(x)$. By repeating the same process for $\Psi (x)$ incident from the right, we obtain solutions in a similar form
\begin{align}
    B_R =& num_{B_R}/den_R, \quad D_R = num_{D_R}/den_R, \\
    den_R=&-e^{2ika}\bigg[ik\Big(\psi'_1(-a)\psi_2(a)-\psi_1(-a)\psi'_2(a) \nonumber \\
        &+\psi_2(-a)\psi'_1(a)-\psi'_2(-a)\psi_1(a)\Big) \nonumber \\
        &+\Big(\psi'_1(a)\psi'_2(-a)-\psi'_1(-a)\psi'_2(a) \nonumber \\
        &+k^2\psi_2(-a)\psi_1(a)-k^2\psi_1(-a)\psi_2(a)\Big)\bigg], \label{eq:denRight} \\
    num_{B_R} =&ik\Big(-\psi'_1(-a)\psi_2(a)+\psi'_2(-a)\psi_1(a) \nonumber \\
        & -\psi_1(-a)\psi'_2(a)+ \psi_2(-a)\psi'_1(a)\Big) \nonumber \\
        & - \Big( \psi'_1(-a)\psi'_2(a)-\psi'_2(-a)\psi'_1(a) \nonumber \\
        & -k^2\psi_1(-a)\psi_2(a)+k^2\psi_2(-a)\psi_1(a)\Big), \label{numBR} \\
    num_{D_R} =& 2ik\left(\psi'_2(-a)\psi_1(-a)-\psi_2(-a)\psi'_1(-a)\right). \label{eq:numDR}
\end{align}
From these results, we see $\left|den_L\right|^2 = \left|den_R\right|^2$ and  $\left|num_{B_L}\right|^2 = \left|num_{B_R}\right|^2$, and accordingly $R_L = R_R$; moreover, 
\begin{align}
    \left|num_{D_L}\right|^2 =& 4k^2\Big(\psi'_2(a)\psi_1(a)-\psi_2(a)\psi'_1(a)\Big)^2, \label{eq:num2DL} \\
    \left|num_{D_R}\right|^2 =& 4k^2\Big(\psi'_2(-a)\psi_1(-a)-\psi_2(-a)\psi'_1(-a)\Big)^2. \label{eq:num2DR}
\end{align}
Since the time-independent Schr\"odinger equation does not contain the first derivative, the Wronskian is constant for all values of $x$ as per Abel's identity \cite[Eq.~(1.13.5)]{NIST:DLMF}. Therefore, $\left|num_{D_L}\right|^2 = \left|num_{D_R}\right|^2$, and as a result $T_L = T_R$, which completes the proof.


\acknowledgments

D.L. and D.I.B. are supported by by the W. M. Keck Foundation. D.I.B. is also supported by Army Research Office (ARO) (grant W911NF-19-1-0377, program manager Dr.~James Joseph, and cooperative agreement W911NF-21-2-0139). The research of J.R.W was carried out at the Jet Propulsion Laboratory, California Institute of Technology, under a contract with the National Aeronautics and Space Administration. The views and conclusions contained in this document are those of the authors and should not be interpreted as representing the official policies, either expressed or implied, of ARO or the U.S. Government. The U.S. Government is authorized to reproduce and distribute reprints for Government purposes notwithstanding any copyright notation herein. This project is supported by the German Space Agency (DLR) with funds provided by the Federal Ministry for Economic Affairs and Energy (BMWi) under Grant No. 50WM2254A (CAL-II) and 50WM2060 (CARIOQA).
D.S. gratefully acknowledges funding by the Federal Ministry of Education and Research (BMBF) through the funding program Photonics Research Germany under contract number 13N14875. D.S., E.M.R. and N.G. acknowledge support of the Deutsche Forschungsgemeinschaft (DFG, German Research Foundation) within the project A05, B07 and B09 of CRC 1227 (DQmat) and under Germany’s Excellence Strategy – EXC-2123 QuantumFrontiers – 390837967.

\bibliography{cited}

\begin{thebibliography}{59}%
\makeatletter
\providecommand \@ifxundefined [1]{%
 \@ifx{#1\undefined}
}%
\providecommand \@ifnum [1]{%
 \ifnum #1\expandafter \@firstoftwo
 \else \expandafter \@secondoftwo
 \fi
}%
\providecommand \@ifx [1]{%
 \ifx #1\expandafter \@firstoftwo
 \else \expandafter \@secondoftwo
 \fi
}%
\providecommand \natexlab [1]{#1}%
\providecommand \enquote  [1]{``#1''}%
\providecommand \bibnamefont  [1]{#1}%
\providecommand \bibfnamefont [1]{#1}%
\providecommand \citenamefont [1]{#1}%
\providecommand \href@noop [0]{\@secondoftwo}%
\providecommand \href [0]{\begingroup \@sanitize@url \@href}%
\providecommand \@href[1]{\@@startlink{#1}\@@href}%
\providecommand \@@href[1]{\endgroup#1\@@endlink}%
\providecommand \@sanitize@url [0]{\catcode `\\12\catcode `\$12\catcode
  `\&12\catcode `\#12\catcode `\^12\catcode `\_12\catcode `\%12\relax}%
\providecommand \@@startlink[1]{}%
\providecommand \@@endlink[0]{}%
\providecommand \url  [0]{\begingroup\@sanitize@url \@url }%
\providecommand \@url [1]{\endgroup\@href {#1}{\urlprefix }}%
\providecommand \urlprefix  [0]{URL }%
\providecommand \Eprint [0]{\href }%
\providecommand \doibase [0]{https://doi.org/}%
\providecommand \selectlanguage [0]{\@gobble}%
\providecommand \bibinfo  [0]{\@secondoftwo}%
\providecommand \bibfield  [0]{\@secondoftwo}%
\providecommand \translation [1]{[#1]}%
\providecommand \BibitemOpen [0]{}%
\providecommand \bibitemStop [0]{}%
\providecommand \bibitemNoStop [0]{.\EOS\space}%
\providecommand \EOS [0]{\spacefactor3000\relax}%
\providecommand \BibitemShut  [1]{\csname bibitem#1\endcsname}%
\let\auto@bib@innerbib\@empty
\bibitem [{\citenamefont {Balantekin}\ and\ \citenamefont
  {Takigawa}(1998)}]{balantekin1998quantum}%
  \BibitemOpen
  \bibfield  {author} {\bibinfo {author} {\bibfnamefont {A.}~\bibnamefont
  {Balantekin}}\ and\ \bibinfo {author} {\bibfnamefont {N.}~\bibnamefont
  {Takigawa}},\ }\bibfield  {title} {\bibinfo {title} {Quantum tunneling in
  nuclear fusion},\ }\href {https://doi.org/10.1103/RevModPhys.70.77}
  {\bibfield  {journal} {\bibinfo  {journal} {Rev. Modern Phys.}\ }\textbf
  {\bibinfo {volume} {70}},\ \bibinfo {pages} {77} (\bibinfo {year}
  {1998})}\BibitemShut {NoStop}%
\bibitem [{\citenamefont {Amini}\ \emph {et~al.}(2019)\citenamefont {Amini},
  \citenamefont {Biegert}, \citenamefont {Calegari}, \citenamefont
  {Chac{\'o}n}, \citenamefont {Ciappina}, \citenamefont {Dauphin},
  \citenamefont {Efimov}, \citenamefont {de~Morisson~Faria}, \citenamefont
  {Giergiel}, \citenamefont {Gniewek} \emph {et~al.}}]{amini2019symphony}%
  \BibitemOpen
  \bibfield  {author} {\bibinfo {author} {\bibfnamefont {K.}~\bibnamefont
  {Amini}}, \bibinfo {author} {\bibfnamefont {J.}~\bibnamefont {Biegert}},
  \bibinfo {author} {\bibfnamefont {F.}~\bibnamefont {Calegari}}, \bibinfo
  {author} {\bibfnamefont {A.}~\bibnamefont {Chac{\'o}n}}, \bibinfo {author}
  {\bibfnamefont {M.~F.}\ \bibnamefont {Ciappina}}, \bibinfo {author}
  {\bibfnamefont {A.}~\bibnamefont {Dauphin}}, \bibinfo {author} {\bibfnamefont
  {D.~K.}\ \bibnamefont {Efimov}}, \bibinfo {author} {\bibfnamefont {C.~F.}\
  \bibnamefont {de~Morisson~Faria}}, \bibinfo {author} {\bibfnamefont
  {K.}~\bibnamefont {Giergiel}}, \bibinfo {author} {\bibfnamefont
  {P.}~\bibnamefont {Gniewek}}, \emph {et~al.},\ }\bibfield  {title} {\bibinfo
  {title} {Symphony on strong field approximation},\ }\href@noop {} {\bibfield
  {journal} {\bibinfo  {journal} {Rep. Prog. Phys.}\ }\textbf {\bibinfo
  {volume} {82}},\ \bibinfo {pages} {116001} (\bibinfo {year}
  {2019})}\BibitemShut {NoStop}%
\bibitem [{\citenamefont {Liang}(2013)}]{shi2013quantum}%
  \BibitemOpen
  \bibfield  {author} {\bibinfo {author} {\bibfnamefont {S.-D.}\ \bibnamefont
  {Liang}},\ }\href@noop {} {\emph {\bibinfo {title} {Quantum tunneling and
  field electron emission theories}}}\ (\bibinfo  {publisher} {World
  Scientific},\ \bibinfo {year} {2013})\BibitemShut {NoStop}%
\bibitem [{\citenamefont {Smedarchina}\ \emph {et~al.}(2018)\citenamefont
  {Smedarchina}, \citenamefont {Siebrand},\ and\ \citenamefont
  {Fern{\'a}ndez-Ramos}}]{smedarchina2018entanglement}%
  \BibitemOpen
  \bibfield  {author} {\bibinfo {author} {\bibfnamefont {Z.}~\bibnamefont
  {Smedarchina}}, \bibinfo {author} {\bibfnamefont {W.}~\bibnamefont
  {Siebrand}},\ and\ \bibinfo {author} {\bibfnamefont {A.}~\bibnamefont
  {Fern{\'a}ndez-Ramos}},\ }\bibfield  {title} {\bibinfo {title} {Entanglement
  and co-tunneling of two equivalent protons in hydrogen bond pairs},\ }\href
  {https://doi.org/10.1063/1.5000681} {\bibfield  {journal} {\bibinfo
  {journal} {J. Chem. Phys.}\ }\textbf {\bibinfo {volume} {148}},\ \bibinfo
  {pages} {102307} (\bibinfo {year} {2018})}\BibitemShut {NoStop}%
\bibitem [{\citenamefont {Guzun}\ \emph {et~al.}(2013)\citenamefont {Guzun},
  \citenamefont {Mazur}, \citenamefont {Dorogan}, \citenamefont {Ware},
  \citenamefont {Marega~Jr}, \citenamefont {Tarasov}, \citenamefont {Lienau},\
  and\ \citenamefont {Salamo}}]{guzun2013effect}%
  \BibitemOpen
  \bibfield  {author} {\bibinfo {author} {\bibfnamefont {D.}~\bibnamefont
  {Guzun}}, \bibinfo {author} {\bibfnamefont {Y.~I.}\ \bibnamefont {Mazur}},
  \bibinfo {author} {\bibfnamefont {V.~G.}\ \bibnamefont {Dorogan}}, \bibinfo
  {author} {\bibfnamefont {M.~E.}\ \bibnamefont {Ware}}, \bibinfo {author}
  {\bibfnamefont {E.}~\bibnamefont {Marega~Jr}}, \bibinfo {author}
  {\bibfnamefont {G.~G.}\ \bibnamefont {Tarasov}}, \bibinfo {author}
  {\bibfnamefont {C.}~\bibnamefont {Lienau}},\ and\ \bibinfo {author}
  {\bibfnamefont {G.}~\bibnamefont {Salamo}},\ }\bibfield  {title} {\bibinfo
  {title} {Effect of resonant tunneling on exciton dynamics in coupled dot-well
  nanostructures},\ }\href {https://doi.org/10.1063/1.4801891} {\bibfield
  {journal} {\bibinfo  {journal} {J. App. Phys.}\ }\textbf {\bibinfo {volume}
  {113}},\ \bibinfo {pages} {154304} (\bibinfo {year} {2013})}\BibitemShut
  {NoStop}%
\bibitem [{\citenamefont {Albiez}\ \emph {et~al.}(2005)\citenamefont {Albiez},
  \citenamefont {Gati}, \citenamefont {F{\"o}lling}, \citenamefont {Hunsmann},
  \citenamefont {Cristiani},\ and\ \citenamefont
  {Oberthaler}}]{albiez2005direct}%
  \BibitemOpen
  \bibfield  {author} {\bibinfo {author} {\bibfnamefont {M.}~\bibnamefont
  {Albiez}}, \bibinfo {author} {\bibfnamefont {R.}~\bibnamefont {Gati}},
  \bibinfo {author} {\bibfnamefont {J.}~\bibnamefont {F{\"o}lling}}, \bibinfo
  {author} {\bibfnamefont {S.}~\bibnamefont {Hunsmann}}, \bibinfo {author}
  {\bibfnamefont {M.}~\bibnamefont {Cristiani}},\ and\ \bibinfo {author}
  {\bibfnamefont {M.~K.}\ \bibnamefont {Oberthaler}},\ }\bibfield  {title}
  {\bibinfo {title} {Direct observation of tunneling and nonlinear
  self-trapping in a single bosonic {Josephson} junction},\ }\href
  {https://doi.org/10.1103/PhysRevLett.95.010402} {\bibfield  {journal}
  {\bibinfo  {journal} {Phys. Rev. Lett.}\ }\textbf {\bibinfo {volume} {95}},\
  \bibinfo {pages} {010402} (\bibinfo {year} {2005})}\BibitemShut {NoStop}%
\bibitem [{\citenamefont {Bai}(2000)}]{bai2000scanning}%
  \BibitemOpen
  \bibfield  {author} {\bibinfo {author} {\bibfnamefont {C.}~\bibnamefont
  {Bai}},\ }\href@noop {} {\emph {\bibinfo {title} {Scanning tunneling
  microscopy and its application}}},\ Vol.~\bibinfo {volume} {32}\ (\bibinfo
  {publisher} {Springer Science \& Business Media},\ \bibinfo {year}
  {2000})\BibitemShut {NoStop}%
\bibitem [{\citenamefont {Arzano}\ \emph {et~al.}(2005)\citenamefont {Arzano},
  \citenamefont {Medved},\ and\ \citenamefont {Vagenas}}]{arzano2005hawking}%
  \BibitemOpen
  \bibfield  {author} {\bibinfo {author} {\bibfnamefont {M.}~\bibnamefont
  {Arzano}}, \bibinfo {author} {\bibfnamefont {A.~J.~M.}\ \bibnamefont
  {Medved}},\ and\ \bibinfo {author} {\bibfnamefont {E.~C.}\ \bibnamefont
  {Vagenas}},\ }\bibfield  {title} {\bibinfo {title} {Hawking radiation as
  tunneling through the quantum horizon},\ }\href
  {https://doi.org/10.1088/1126-6708/2005/09/037} {\bibfield  {journal}
  {\bibinfo  {journal} {JHEP}\ }\textbf {\bibinfo {volume} {2005}}\bibinfo
  {number} { (09)},\ \bibinfo {pages} {037}}\BibitemShut {NoStop}%
\bibitem [{\citenamefont {Lambert}\ \emph {et~al.}(2013)\citenamefont
  {Lambert}, \citenamefont {Chen}, \citenamefont {Cheng}, \citenamefont {Li},
  \citenamefont {Chen},\ and\ \citenamefont {Nori}}]{lambert2013quantum}%
  \BibitemOpen
\bibfield  {number} {  }\bibfield  {author} {\bibinfo {author} {\bibfnamefont
  {N.}~\bibnamefont {Lambert}}, \bibinfo {author} {\bibfnamefont {Y.-N.}\
  \bibnamefont {Chen}}, \bibinfo {author} {\bibfnamefont {Y.-C.}\ \bibnamefont
  {Cheng}}, \bibinfo {author} {\bibfnamefont {C.-M.}\ \bibnamefont {Li}},
  \bibinfo {author} {\bibfnamefont {G.-Y.}\ \bibnamefont {Chen}},\ and\
  \bibinfo {author} {\bibfnamefont {F.}~\bibnamefont {Nori}},\ }\bibfield
  {title} {\bibinfo {title} {Quantum biology},\ }\href@noop {} {\bibfield
  {journal} {\bibinfo  {journal} {Nature Physics}\ }\textbf {\bibinfo {volume}
  {9}},\ \bibinfo {pages} {10} (\bibinfo {year} {2013})}\BibitemShut {NoStop}%
\bibitem [{\citenamefont {Atkatz}(1994)}]{atkatz1994quantum}%
  \BibitemOpen
  \bibfield  {author} {\bibinfo {author} {\bibfnamefont {D.}~\bibnamefont
  {Atkatz}},\ }\bibfield  {title} {\bibinfo {title} {Quantum cosmology for
  pedestrians},\ }\href {https://doi.org/10.1119/1.17479} {\bibfield  {journal}
  {\bibinfo  {journal} {Am. J. Phys.}\ }\textbf {\bibinfo {volume} {62}},\
  \bibinfo {pages} {619} (\bibinfo {year} {1994})}\BibitemShut {NoStop}%
\bibitem [{\citenamefont {Razavy}(2013)}]{mohsen2013quantum}%
  \BibitemOpen
  \bibfield  {author} {\bibinfo {author} {\bibfnamefont {M.}~\bibnamefont
  {Razavy}},\ }\href@noop {} {\emph {\bibinfo {title} {Quantum Theory of
  Tunneling}}}\ (\bibinfo  {publisher} {World Scientific},\ \bibinfo {year}
  {2013})\BibitemShut {NoStop}%
\bibitem [{\citenamefont {Landau}\ and\ \citenamefont
  {Lifshitz}(1981)}]{landau2013quantum}%
  \BibitemOpen
  \bibfield  {author} {\bibinfo {author} {\bibfnamefont {L.~D.}\ \bibnamefont
  {Landau}}\ and\ \bibinfo {author} {\bibfnamefont {E.~M.}\ \bibnamefont
  {Lifshitz}},\ }\href {https://doi.org/10.1016/C2013-0-02793-4} {\emph
  {\bibinfo {title} {Quantum mechanics: non-relativistic theory}}},\
  Vol.~\bibinfo {volume} {3}\ (\bibinfo  {publisher} {Elsevier},\ \bibinfo
  {year} {1981})\BibitemShut {NoStop}%
\bibitem [{\citenamefont {Zakhariev}\ and\ \citenamefont
  {Sokolov}(1964)}]{zakhariev1964intensified}%
  \BibitemOpen
  \bibfield  {author} {\bibinfo {author} {\bibfnamefont {B.}~\bibnamefont
  {Zakhariev}}\ and\ \bibinfo {author} {\bibfnamefont {S.}~\bibnamefont
  {Sokolov}},\ }\bibfield  {title} {\bibinfo {title} {Intensified tunnel effect
  for complex particles},\ }\href@noop {} {\bibfield  {journal} {\bibinfo
  {journal} {Annalen der Physik}\ }\textbf {\bibinfo {volume} {469}},\ \bibinfo
  {pages} {229} (\bibinfo {year} {1964})}\BibitemShut {NoStop}%
\bibitem [{\citenamefont {Amirkhanov}\ and\ \citenamefont
  {Zakhariev}(1966)}]{Amirkhanov1966}%
  \BibitemOpen
  \bibfield  {author} {\bibinfo {author} {\bibfnamefont {I.}~\bibnamefont
  {Amirkhanov}}\ and\ \bibinfo {author} {\bibfnamefont {B.~N.}\ \bibnamefont
  {Zakhariev}},\ }\bibfield  {title} {\bibinfo {title} {Violation of barrier
  penetration symmetry for composite particles},\ }\href@noop {} {\bibfield
  {journal} {\bibinfo  {journal} {Sov. Phys. JETP}\ }\textbf {\bibinfo {volume}
  {22}},\ \bibinfo {pages} {764} (\bibinfo {year} {1966})}\BibitemShut
  {NoStop}%
\bibitem [{\citenamefont {Bondar}\ \emph {et~al.}(2010)\citenamefont {Bondar},
  \citenamefont {Liu},\ and\ \citenamefont {Ivanov}}]{bondar2010enhancement}%
  \BibitemOpen
  \bibfield  {author} {\bibinfo {author} {\bibfnamefont {D.~I.}\ \bibnamefont
  {Bondar}}, \bibinfo {author} {\bibfnamefont {W.-K.}\ \bibnamefont {Liu}},\
  and\ \bibinfo {author} {\bibfnamefont {M.~Y.}\ \bibnamefont {Ivanov}},\
  }\bibfield  {title} {\bibinfo {title} {Enhancement and suppression of
  tunneling by controlling symmetries of a potential barrier},\ }\href
  {https://doi.org/10.1103/PhysRevA.82.052112} {\bibfield  {journal} {\bibinfo
  {journal} {Phys. Rev. A}\ }\textbf {\bibinfo {volume} {82}},\ \bibinfo
  {pages} {052112} (\bibinfo {year} {2010})}\BibitemShut {NoStop}%
\bibitem [{\citenamefont {Potnis}\ \emph {et~al.}(2017)\citenamefont {Potnis},
  \citenamefont {Ramos}, \citenamefont {Maeda}, \citenamefont {Carr},\ and\
  \citenamefont {Steinberg}}]{potnis2017interaction}%
  \BibitemOpen
  \bibfield  {author} {\bibinfo {author} {\bibfnamefont {S.}~\bibnamefont
  {Potnis}}, \bibinfo {author} {\bibfnamefont {R.}~\bibnamefont {Ramos}},
  \bibinfo {author} {\bibfnamefont {K.}~\bibnamefont {Maeda}}, \bibinfo
  {author} {\bibfnamefont {L.~D.}\ \bibnamefont {Carr}},\ and\ \bibinfo
  {author} {\bibfnamefont {A.~M.}\ \bibnamefont {Steinberg}},\ }\bibfield
  {title} {\bibinfo {title} {Interaction-assisted quantum tunneling of a
  {Bose-Einstein} condensate out of a single trapping well},\ }\href
  {https://doi.org/10.1103/PhysRevLett.118.060402} {\bibfield  {journal}
  {\bibinfo  {journal} {Phys. Rev. Lett.}\ }\textbf {\bibinfo {volume} {118}},\
  \bibinfo {pages} {060402} (\bibinfo {year} {2017})}\BibitemShut {NoStop}%
\bibitem [{\citenamefont {Ramos}\ \emph {et~al.}(2020)\citenamefont {Ramos},
  \citenamefont {Spierings}, \citenamefont {Racicot},\ and\ \citenamefont
  {Steinberg}}]{Ramos2020}%
  \BibitemOpen
  \bibfield  {author} {\bibinfo {author} {\bibfnamefont {R.}~\bibnamefont
  {Ramos}}, \bibinfo {author} {\bibfnamefont {D.}~\bibnamefont {Spierings}},
  \bibinfo {author} {\bibfnamefont {I.}~\bibnamefont {Racicot}},\ and\ \bibinfo
  {author} {\bibfnamefont {A.~M.}\ \bibnamefont {Steinberg}},\ }\bibfield
  {title} {\bibinfo {title} {Measurement of the time spent by a tunnelling atom
  within the barrier region},\ }\href
  {https://doi.org/10.1038/s41586-020-2490-7} {\bibfield  {journal} {\bibinfo
  {journal} {Nature}\ }\textbf {\bibinfo {volume} {583}},\ \bibinfo {pages}
  {529} (\bibinfo {year} {2020})}\BibitemShut {NoStop}%
\bibitem [{\citenamefont {Smerzi}\ \emph {et~al.}(1997)\citenamefont {Smerzi},
  \citenamefont {Fantoni}, \citenamefont {Giovanazzi},\ and\ \citenamefont
  {Shenoy}}]{smerzi1997quantum}%
  \BibitemOpen
  \bibfield  {author} {\bibinfo {author} {\bibfnamefont {A.}~\bibnamefont
  {Smerzi}}, \bibinfo {author} {\bibfnamefont {S.}~\bibnamefont {Fantoni}},
  \bibinfo {author} {\bibfnamefont {S.}~\bibnamefont {Giovanazzi}},\ and\
  \bibinfo {author} {\bibfnamefont {S.}~\bibnamefont {Shenoy}},\ }\bibfield
  {title} {\bibinfo {title} {Quantum coherent atomic tunneling between two
  trapped {Bose-Einstein} condensates},\ }\href@noop {} {\bibfield  {journal}
  {\bibinfo  {journal} {Phys. Rev. Lett.}\ }\textbf {\bibinfo {volume} {79}},\
  \bibinfo {pages} {4950} (\bibinfo {year} {1997})}\BibitemShut {NoStop}%
\bibitem [{\citenamefont {Wu}\ and\ \citenamefont
  {Niu}(2000)}]{PhysRevA.61.023402}%
  \BibitemOpen
  \bibfield  {author} {\bibinfo {author} {\bibfnamefont {B.}~\bibnamefont
  {Wu}}\ and\ \bibinfo {author} {\bibfnamefont {Q.}~\bibnamefont {Niu}},\
  }\bibfield  {title} {\bibinfo {title} {Nonlinear {Landau-Zener} tunneling},\
  }\href {https://doi.org/10.1103/PhysRevA.61.023402} {\bibfield  {journal}
  {\bibinfo  {journal} {Phys. Rev. A}\ }\textbf {\bibinfo {volume} {61}},\
  \bibinfo {pages} {023402} (\bibinfo {year} {2000})}\BibitemShut {NoStop}%
\bibitem [{\citenamefont {Salasnich}\ \emph {et~al.}(2001)\citenamefont
  {Salasnich}, \citenamefont {Parola},\ and\ \citenamefont
  {Reatto}}]{Salasnich2001}%
  \BibitemOpen
  \bibfield  {author} {\bibinfo {author} {\bibfnamefont {L.}~\bibnamefont
  {Salasnich}}, \bibinfo {author} {\bibfnamefont {A.}~\bibnamefont {Parola}},\
  and\ \bibinfo {author} {\bibfnamefont {L.}~\bibnamefont {Reatto}},\
  }\bibfield  {title} {\bibinfo {title} {Pulsed macroscopic quantum tunneling
  of falling {Bose-Einstein} condensates},\ }\href
  {https://doi.org/10.1103/PhysRevA.64.023601} {\bibfield  {journal} {\bibinfo
  {journal} {Phys. Rev. A}\ }\textbf {\bibinfo {volume} {64}},\ \bibinfo
  {pages} {023601} (\bibinfo {year} {2001})}\BibitemShut {NoStop}%
\bibitem [{\citenamefont {Svidzinsky}\ and\ \citenamefont
  {Chui}(2003)}]{svidzinsky2003symmetric}%
  \BibitemOpen
  \bibfield  {author} {\bibinfo {author} {\bibfnamefont {A.~A.}\ \bibnamefont
  {Svidzinsky}}\ and\ \bibinfo {author} {\bibfnamefont {S.-T.}\ \bibnamefont
  {Chui}},\ }\bibfield  {title} {\bibinfo {title} {Symmetric-asymmetric
  transition in mixtures of {Bose-Einstein} condensates},\ }\href@noop {}
  {\bibfield  {journal} {\bibinfo  {journal} {Phys. Rev. A}\ }\textbf {\bibinfo
  {volume} {67}},\ \bibinfo {pages} {053608} (\bibinfo {year}
  {2003})}\BibitemShut {NoStop}%
\bibitem [{\citenamefont {Jona-Lasinio}\ \emph {et~al.}(2003)\citenamefont
  {Jona-Lasinio}, \citenamefont {Morsch}, \citenamefont {Cristiani},
  \citenamefont {Malossi}, \citenamefont {M{\"u}ller}, \citenamefont
  {Courtade}, \citenamefont {Anderlini},\ and\ \citenamefont
  {Arimondo}}]{jona2003asymmetric}%
  \BibitemOpen
  \bibfield  {author} {\bibinfo {author} {\bibfnamefont {M.}~\bibnamefont
  {Jona-Lasinio}}, \bibinfo {author} {\bibfnamefont {O.}~\bibnamefont
  {Morsch}}, \bibinfo {author} {\bibfnamefont {M.}~\bibnamefont {Cristiani}},
  \bibinfo {author} {\bibfnamefont {N.}~\bibnamefont {Malossi}}, \bibinfo
  {author} {\bibfnamefont {J.}~\bibnamefont {M{\"u}ller}}, \bibinfo {author}
  {\bibfnamefont {E.}~\bibnamefont {Courtade}}, \bibinfo {author}
  {\bibfnamefont {M.}~\bibnamefont {Anderlini}},\ and\ \bibinfo {author}
  {\bibfnamefont {E.}~\bibnamefont {Arimondo}},\ }\bibfield  {title} {\bibinfo
  {title} {Asymmetric {Landau-Zener} tunneling in a periodic potential},\
  }\href@noop {} {\bibfield  {journal} {\bibinfo  {journal} {Phys. Rev. Lett.}\
  }\textbf {\bibinfo {volume} {91}},\ \bibinfo {pages} {230406} (\bibinfo
  {year} {2003})}\BibitemShut {NoStop}%
\bibitem [{\citenamefont {Jona-Lasinio}\ \emph {et~al.}(2005)\citenamefont
  {Jona-Lasinio}, \citenamefont {Morsch}, \citenamefont {Cristiani},
  \citenamefont {Arimondo},\ and\ \citenamefont {Menotti}}]{jona2005nonlinear}%
  \BibitemOpen
  \bibfield  {author} {\bibinfo {author} {\bibfnamefont {M.}~\bibnamefont
  {Jona-Lasinio}}, \bibinfo {author} {\bibfnamefont {O.}~\bibnamefont
  {Morsch}}, \bibinfo {author} {\bibfnamefont {M.}~\bibnamefont {Cristiani}},
  \bibinfo {author} {\bibfnamefont {E.}~\bibnamefont {Arimondo}},\ and\
  \bibinfo {author} {\bibfnamefont {C.}~\bibnamefont {Menotti}},\ }\bibfield
  {title} {\bibinfo {title} {Nonlinear effects for {Bose-Einstein} condensates
  in optical lattices},\ }\href@noop {} {\bibfield  {journal} {\bibinfo
  {journal} {Las. Phys.}\ }\textbf {\bibinfo {volume} {15}},\ \bibinfo {pages}
  {1180} (\bibinfo {year} {2005})}\BibitemShut {NoStop}%
\bibitem [{\citenamefont {Carr}\ \emph {et~al.}(2005)\citenamefont {Carr},
  \citenamefont {Holland},\ and\ \citenamefont
  {Malomed}}]{carr2005macroscopic}%
  \BibitemOpen
  \bibfield  {author} {\bibinfo {author} {\bibfnamefont {L.}~\bibnamefont
  {Carr}}, \bibinfo {author} {\bibfnamefont {M.}~\bibnamefont {Holland}},\ and\
  \bibinfo {author} {\bibfnamefont {B.~A.}\ \bibnamefont {Malomed}},\
  }\bibfield  {title} {\bibinfo {title} {Macroscopic quantum tunnelling of
  {Bose-Einstein} condensates in a finite potential well},\ }\href@noop {}
  {\bibfield  {journal} {\bibinfo  {journal} {Journal of Physics B: Atomic,
  Molecular and Optical Physics}\ }\textbf {\bibinfo {volume} {38}},\ \bibinfo
  {pages} {3217} (\bibinfo {year} {2005})}\BibitemShut {NoStop}%
\bibitem [{\citenamefont {Dekel}\ \emph {et~al.}(2010)\citenamefont {Dekel},
  \citenamefont {Farberovich}, \citenamefont {Fleurov},\ and\ \citenamefont
  {Soffer}}]{PhysRevA.81.063638}%
  \BibitemOpen
  \bibfield  {author} {\bibinfo {author} {\bibfnamefont {G.}~\bibnamefont
  {Dekel}}, \bibinfo {author} {\bibfnamefont {V.}~\bibnamefont {Farberovich}},
  \bibinfo {author} {\bibfnamefont {V.}~\bibnamefont {Fleurov}},\ and\ \bibinfo
  {author} {\bibfnamefont {A.}~\bibnamefont {Soffer}},\ }\bibfield  {title}
  {\bibinfo {title} {Dynamics of macroscopic tunneling in elongated
  {Bose-Einstein} condensates},\ }\href
  {https://doi.org/10.1103/PhysRevA.81.063638} {\bibfield  {journal} {\bibinfo
  {journal} {Phys. Rev. A}\ }\textbf {\bibinfo {volume} {81}},\ \bibinfo
  {pages} {063638} (\bibinfo {year} {2010})}\BibitemShut {NoStop}%
\bibitem [{\citenamefont {Manju}\ \emph {et~al.}(2018)\citenamefont {Manju},
  \citenamefont {Hardman}, \citenamefont {Sooriyabandara}, \citenamefont
  {Wigley}, \citenamefont {Close}, \citenamefont {Robins}, \citenamefont
  {Hush},\ and\ \citenamefont {Szigeti}}]{manju2018quantum}%
  \BibitemOpen
  \bibfield  {author} {\bibinfo {author} {\bibfnamefont {P.}~\bibnamefont
  {Manju}}, \bibinfo {author} {\bibfnamefont {K.}~\bibnamefont {Hardman}},
  \bibinfo {author} {\bibfnamefont {M.}~\bibnamefont {Sooriyabandara}},
  \bibinfo {author} {\bibfnamefont {P.}~\bibnamefont {Wigley}}, \bibinfo
  {author} {\bibfnamefont {J.}~\bibnamefont {Close}}, \bibinfo {author}
  {\bibfnamefont {N.}~\bibnamefont {Robins}}, \bibinfo {author} {\bibfnamefont
  {M.}~\bibnamefont {Hush}},\ and\ \bibinfo {author} {\bibfnamefont
  {S.}~\bibnamefont {Szigeti}},\ }\bibfield  {title} {\bibinfo {title} {Quantum
  tunneling dynamics of an interacting {Bose-Einstein} condensate through a
  {Gaussian} barrier},\ }\href@noop {} {\bibfield  {journal} {\bibinfo
  {journal} {Phys. Rev. A}\ }\textbf {\bibinfo {volume} {98}},\ \bibinfo
  {pages} {053629} (\bibinfo {year} {2018})}\BibitemShut {NoStop}%
\bibitem [{\citenamefont {del Campo}\ \emph {et~al.}(2006)\citenamefont {del
  Campo}, \citenamefont {Delgado}, \citenamefont {Garc\'{\i}a-Calder\'on},
  \citenamefont {Muga},\ and\ \citenamefont {Raizen}}]{PhysRevA.74.013605}%
  \BibitemOpen
  \bibfield  {author} {\bibinfo {author} {\bibfnamefont {A.}~\bibnamefont {del
  Campo}}, \bibinfo {author} {\bibfnamefont {F.}~\bibnamefont {Delgado}},
  \bibinfo {author} {\bibfnamefont {G.}~\bibnamefont {Garc\'{\i}a-Calder\'on}},
  \bibinfo {author} {\bibfnamefont {J.~G.}\ \bibnamefont {Muga}},\ and\
  \bibinfo {author} {\bibfnamefont {M.~G.}\ \bibnamefont {Raizen}},\ }\bibfield
   {title} {\bibinfo {title} {Decay by tunneling of bosonic and fermionic
  {Tonks-Girardeau} gases},\ }\href
  {https://doi.org/10.1103/PhysRevA.74.013605} {\bibfield  {journal} {\bibinfo
  {journal} {Phys. Rev. A}\ }\textbf {\bibinfo {volume} {74}},\ \bibinfo
  {pages} {013605} (\bibinfo {year} {2006})}\BibitemShut {NoStop}%
\bibitem [{\citenamefont {Glick}\ and\ \citenamefont
  {Carr}(2011)}]{glick2011macroscopic}%
  \BibitemOpen
  \bibfield  {author} {\bibinfo {author} {\bibfnamefont {J.~A.}\ \bibnamefont
  {Glick}}\ and\ \bibinfo {author} {\bibfnamefont {L.~D.}\ \bibnamefont
  {Carr}},\ }\href@noop {} {\bibinfo {title} {Macroscopic quantum tunneling of
  solitons in {Bose-Einstein} condensates}} (\bibinfo {year} {2011}),\ \Eprint
  {https://arxiv.org/abs/1105.5164} {arXiv:1105.5164 [cond-mat.quant-gas]}
  \BibitemShut {NoStop}%
\bibitem [{\citenamefont {Lode}\ \emph {et~al.}(2014)\citenamefont {Lode},
  \citenamefont {Klaiman}, \citenamefont {Alon}, \citenamefont {Streltsov},\
  and\ \citenamefont {Cederbaum}}]{Lode2014}%
  \BibitemOpen
  \bibfield  {author} {\bibinfo {author} {\bibfnamefont {A.~U.~J.}\
  \bibnamefont {Lode}}, \bibinfo {author} {\bibfnamefont {S.}~\bibnamefont
  {Klaiman}}, \bibinfo {author} {\bibfnamefont {O.~E.}\ \bibnamefont {Alon}},
  \bibinfo {author} {\bibfnamefont {A.~I.}\ \bibnamefont {Streltsov}},\ and\
  \bibinfo {author} {\bibfnamefont {L.~S.}\ \bibnamefont {Cederbaum}},\
  }\bibfield  {title} {\bibinfo {title} {Controlling the velocities and the
  number of emitted particles in the tunneling to open space dynamics},\ }\href
  {https://doi.org/10.1103/PhysRevA.89.053620} {\bibfield  {journal} {\bibinfo
  {journal} {Phys. Rev. A}\ }\textbf {\bibinfo {volume} {89}},\ \bibinfo
  {pages} {053620} (\bibinfo {year} {2014})}\BibitemShut {NoStop}%
\bibitem [{\citenamefont {Alcala}\ \emph {et~al.}(2017)\citenamefont {Alcala},
  \citenamefont {Glick},\ and\ \citenamefont {Carr}}]{PhysRevLett.118.210403}%
  \BibitemOpen
  \bibfield  {author} {\bibinfo {author} {\bibfnamefont {D.~A.}\ \bibnamefont
  {Alcala}}, \bibinfo {author} {\bibfnamefont {J.~A.}\ \bibnamefont {Glick}},\
  and\ \bibinfo {author} {\bibfnamefont {L.~D.}\ \bibnamefont {Carr}},\
  }\bibfield  {title} {\bibinfo {title} {Entangled dynamics in macroscopic
  quantum tunneling of {Bose-Einstein} condensates},\ }\href
  {https://doi.org/10.1103/PhysRevLett.118.210403} {\bibfield  {journal}
  {\bibinfo  {journal} {Phys. Rev. Lett.}\ }\textbf {\bibinfo {volume} {118}},\
  \bibinfo {pages} {210403} (\bibinfo {year} {2017})}\BibitemShut {NoStop}%
\bibitem [{\citenamefont {Haldar}\ and\ \citenamefont
  {Alon}(2019)}]{Haldar2019}%
  \BibitemOpen
  \bibfield  {author} {\bibinfo {author} {\bibfnamefont {S.~K.}\ \bibnamefont
  {Haldar}}\ and\ \bibinfo {author} {\bibfnamefont {O.~E.}\ \bibnamefont
  {Alon}},\ }\bibfield  {title} {\bibinfo {title} {Many-body quantum dynamics
  of an asymmetric bosonic {Josephson} junction},\ }\href@noop {} {\bibfield
  {journal} {\bibinfo  {journal} {New J. Phys.}\ }\textbf {\bibinfo {volume}
  {21}},\ \bibinfo {pages} {103037} (\bibinfo {year} {2019})}\BibitemShut
  {NoStop}%
\bibitem [{\citenamefont {Pethick}\ and\ \citenamefont
  {Smith}(2008)}]{pethick2008bose}%
  \BibitemOpen
  \bibfield  {author} {\bibinfo {author} {\bibfnamefont {C.~J.}\ \bibnamefont
  {Pethick}}\ and\ \bibinfo {author} {\bibfnamefont {H.}~\bibnamefont
  {Smith}},\ }\href@noop {} {\emph {\bibinfo {title} {Bose--Einstein
  condensation in dilute gases}}}\ (\bibinfo  {publisher} {Cambridge University
  Press},\ \bibinfo {year} {2008})\BibitemShut {NoStop}%
\bibitem [{\citenamefont {Egorov}\ \emph {et~al.}(2013)\citenamefont {Egorov},
  \citenamefont {Opanchuk}, \citenamefont {Drummond}, \citenamefont {Hall},
  \citenamefont {Hannaford},\ and\ \citenamefont {Sidorov}}]{Egorov_2013}%
  \BibitemOpen
  \bibfield  {author} {\bibinfo {author} {\bibfnamefont {M.}~\bibnamefont
  {Egorov}}, \bibinfo {author} {\bibfnamefont {B.}~\bibnamefont {Opanchuk}},
  \bibinfo {author} {\bibfnamefont {P.}~\bibnamefont {Drummond}}, \bibinfo
  {author} {\bibfnamefont {B.}~\bibnamefont {Hall}}, \bibinfo {author}
  {\bibfnamefont {P.}~\bibnamefont {Hannaford}},\ and\ \bibinfo {author}
  {\bibfnamefont {A.}~\bibnamefont {Sidorov}},\ }\bibfield  {title} {\bibinfo
  {title} {Measurement of s-wave scattering lengths in a two-component
  {Bose-Einstein} condensate},\ }\href@noop {} {\bibfield  {journal} {\bibinfo
  {journal} {Phys. Rev. A}\ }\textbf {\bibinfo {volume} {87}},\ \bibinfo
  {pages} {053614} (\bibinfo {year} {2013})}\BibitemShut {NoStop}%
\bibitem [{\citenamefont {Lee}\ \emph {et~al.}(2007)\citenamefont {Lee},
  \citenamefont {Ostrovskaya},\ and\ \citenamefont {Kivshar}}]{Lee_2007}%
  \BibitemOpen
  \bibfield  {author} {\bibinfo {author} {\bibfnamefont {C.}~\bibnamefont
  {Lee}}, \bibinfo {author} {\bibfnamefont {E.~A.}\ \bibnamefont
  {Ostrovskaya}},\ and\ \bibinfo {author} {\bibfnamefont {Y.~S.}\ \bibnamefont
  {Kivshar}},\ }\bibfield  {title} {\bibinfo {title} {Nonlinearity-assisted
  quantum tunnelling in a matter-wave interferometer},\ }\href
  {https://doi.org/10.1088/0953-4075/40/21/010} {\bibfield  {journal} {\bibinfo
   {journal} {Journal of Physics B: Atomic, Molecular and Optical Physics}\
  }\textbf {\bibinfo {volume} {40}},\ \bibinfo {pages} {4235} (\bibinfo {year}
  {2007})}\BibitemShut {NoStop}%
\bibitem [{\citenamefont {Schlagheck}\ and\ \citenamefont
  {Wimberger}(2007)}]{Schlagheck2007}%
  \BibitemOpen
  \bibfield  {author} {\bibinfo {author} {\bibfnamefont {P.}~\bibnamefont
  {Schlagheck}}\ and\ \bibinfo {author} {\bibfnamefont {S.}~\bibnamefont
  {Wimberger}},\ }\bibfield  {title} {\bibinfo {title} {Nonexponential decay of
  {Bose–Einstein} condensates: a numerical study based on the complex scaling
  method},\ }\href {https://doi.org/10.1007/s00340-006-2511-8} {\bibfield
  {journal} {\bibinfo  {journal} {Applied Physics B}\ }\textbf {\bibinfo
  {volume} {86}},\ \bibinfo {pages} {385} (\bibinfo {year} {2007})}\BibitemShut
  {NoStop}%
\bibitem [{\citenamefont {Paul}\ \emph {et~al.}(2007)\citenamefont {Paul},
  \citenamefont {Hartung}, \citenamefont {Richter},\ and\ \citenamefont
  {Schlagheck}}]{paul2007}%
  \BibitemOpen
  \bibfield  {author} {\bibinfo {author} {\bibfnamefont {T.}~\bibnamefont
  {Paul}}, \bibinfo {author} {\bibfnamefont {M.}~\bibnamefont {Hartung}},
  \bibinfo {author} {\bibfnamefont {K.}~\bibnamefont {Richter}},\ and\ \bibinfo
  {author} {\bibfnamefont {P.}~\bibnamefont {Schlagheck}},\ }\bibfield  {title}
  {\bibinfo {title} {Nonlinear transport of {Bose-Einstein} condensates through
  mesoscopic waveguides},\ }\href {https://doi.org/10.1103/PhysRevA.76.063605}
  {\bibfield  {journal} {\bibinfo  {journal} {Phys. Rev. A}\ }\textbf {\bibinfo
  {volume} {76}},\ \bibinfo {pages} {063605} (\bibinfo {year}
  {2007})}\BibitemShut {NoStop}%
\bibitem [{\citenamefont {Zenesini}\ \emph {et~al.}(2008)\citenamefont
  {Zenesini}, \citenamefont {Sias}, \citenamefont {Lignier}, \citenamefont
  {Singh}, \citenamefont {Ciampini}, \citenamefont {Morsch}, \citenamefont
  {Mannella}, \citenamefont {Arimondo}, \citenamefont {Tomadin},\ and\
  \citenamefont {Wimberger}}]{Zenesini_2008}%
  \BibitemOpen
  \bibfield  {author} {\bibinfo {author} {\bibfnamefont {A.}~\bibnamefont
  {Zenesini}}, \bibinfo {author} {\bibfnamefont {C.}~\bibnamefont {Sias}},
  \bibinfo {author} {\bibfnamefont {H.}~\bibnamefont {Lignier}}, \bibinfo
  {author} {\bibfnamefont {Y.}~\bibnamefont {Singh}}, \bibinfo {author}
  {\bibfnamefont {D.}~\bibnamefont {Ciampini}}, \bibinfo {author}
  {\bibfnamefont {O.}~\bibnamefont {Morsch}}, \bibinfo {author} {\bibfnamefont
  {R.}~\bibnamefont {Mannella}}, \bibinfo {author} {\bibfnamefont
  {E.}~\bibnamefont {Arimondo}}, \bibinfo {author} {\bibfnamefont
  {A.}~\bibnamefont {Tomadin}},\ and\ \bibinfo {author} {\bibfnamefont
  {S.}~\bibnamefont {Wimberger}},\ }\bibfield  {title} {\bibinfo {title}
  {Resonant tunneling of {Bose–Einstein} condensates in optical lattices},\
  }\href {https://doi.org/10.1088/1367-2630/10/5/053038} {\bibfield  {journal}
  {\bibinfo  {journal} {New Journal of Physics}\ }\textbf {\bibinfo {volume}
  {10}},\ \bibinfo {pages} {053038} (\bibinfo {year} {2008})}\BibitemShut
  {NoStop}%
\bibitem [{\citenamefont {Paul}\ \emph {et~al.}(2005)\citenamefont {Paul},
  \citenamefont {Richter},\ and\ \citenamefont {Schlagheck}}]{paul2005}%
  \BibitemOpen
  \bibfield  {author} {\bibinfo {author} {\bibfnamefont {T.}~\bibnamefont
  {Paul}}, \bibinfo {author} {\bibfnamefont {K.}~\bibnamefont {Richter}},\ and\
  \bibinfo {author} {\bibfnamefont {P.}~\bibnamefont {Schlagheck}},\ }\bibfield
   {title} {\bibinfo {title} {Nonlinear resonant transport of bose-einstein
  condensates},\ }\href {https://doi.org/10.1103/PhysRevLett.94.020404}
  {\bibfield  {journal} {\bibinfo  {journal} {Phys. Rev. Lett.}\ }\textbf
  {\bibinfo {volume} {94}},\ \bibinfo {pages} {020404} (\bibinfo {year}
  {2005})}\BibitemShut {NoStop}%
\bibitem [{\citenamefont {Sias}\ \emph {et~al.}(2007)\citenamefont {Sias},
  \citenamefont {Zenesini}, \citenamefont {Lignier}, \citenamefont {Wimberger},
  \citenamefont {Ciampini}, \citenamefont {Morsch},\ and\ \citenamefont
  {Arimondo}}]{sias2007}%
  \BibitemOpen
  \bibfield  {author} {\bibinfo {author} {\bibfnamefont {C.}~\bibnamefont
  {Sias}}, \bibinfo {author} {\bibfnamefont {A.}~\bibnamefont {Zenesini}},
  \bibinfo {author} {\bibfnamefont {H.}~\bibnamefont {Lignier}}, \bibinfo
  {author} {\bibfnamefont {S.}~\bibnamefont {Wimberger}}, \bibinfo {author}
  {\bibfnamefont {D.}~\bibnamefont {Ciampini}}, \bibinfo {author}
  {\bibfnamefont {O.}~\bibnamefont {Morsch}},\ and\ \bibinfo {author}
  {\bibfnamefont {E.}~\bibnamefont {Arimondo}},\ }\bibfield  {title} {\bibinfo
  {title} {Resonantly enhanced tunneling of {Bose-Einstein} condensates in
  periodic potentials},\ }\href {https://doi.org/10.1103/PhysRevLett.98.120403}
  {\bibfield  {journal} {\bibinfo  {journal} {Phys. Rev. Lett.}\ }\textbf
  {\bibinfo {volume} {98}},\ \bibinfo {pages} {120403} (\bibinfo {year}
  {2007})}\BibitemShut {NoStop}%
\bibitem [{\citenamefont {Chin}\ \emph {et~al.}(2010)\citenamefont {Chin},
  \citenamefont {Grimm}, \citenamefont {Julienne},\ and\ \citenamefont
  {Tiesinga}}]{RevModPhys.82.1225}%
  \BibitemOpen
  \bibfield  {author} {\bibinfo {author} {\bibfnamefont {C.}~\bibnamefont
  {Chin}}, \bibinfo {author} {\bibfnamefont {R.}~\bibnamefont {Grimm}},
  \bibinfo {author} {\bibfnamefont {P.}~\bibnamefont {Julienne}},\ and\
  \bibinfo {author} {\bibfnamefont {E.}~\bibnamefont {Tiesinga}},\ }\bibfield
  {title} {\bibinfo {title} {Feshbach resonances in ultracold gases},\ }\href
  {https://doi.org/10.1103/RevModPhys.82.1225} {\bibfield  {journal} {\bibinfo
  {journal} {Rev. Mod. Phys.}\ }\textbf {\bibinfo {volume} {82}},\ \bibinfo
  {pages} {1225} (\bibinfo {year} {2010})}\BibitemShut {NoStop}%
\bibitem [{\citenamefont {Alon}\ \emph {et~al.}(2008)\citenamefont {Alon},
  \citenamefont {Streltsov},\ and\ \citenamefont
  {Cederbaum}}]{AlonStreltsov2008}%
  \BibitemOpen
  \bibfield  {author} {\bibinfo {author} {\bibfnamefont {O.~E.}\ \bibnamefont
  {Alon}}, \bibinfo {author} {\bibfnamefont {A.~I.}\ \bibnamefont
  {Streltsov}},\ and\ \bibinfo {author} {\bibfnamefont {L.~S.}\ \bibnamefont
  {Cederbaum}},\ }\bibfield  {title} {\bibinfo {title} {Multiconfigurational
  time-dependent hartree method for bosons: Many-body dynamics of bosonic
  systems},\ }\href {https://doi.org/10.1103/PhysRevA.77.033613} {\bibfield
  {journal} {\bibinfo  {journal} {Phys. Rev. A}\ }\textbf {\bibinfo {volume}
  {77}},\ \bibinfo {pages} {033613} (\bibinfo {year} {2008})}\BibitemShut
  {NoStop}%
\bibitem [{Note1()}]{Note1}%
  \BibitemOpen
  \bibinfo {note} {In Ref.~\cite {Haldar2019}, the notion of survival
  probability is used, which is just 1 minus the tunneling
  probability.}\BibitemShut {Stop}%
\bibitem [{\citenamefont {Ryu}\ and\ \citenamefont
  {Boshier}(2015)}]{ryu2015integrated}%
  \BibitemOpen
  \bibfield  {author} {\bibinfo {author} {\bibfnamefont {C.}~\bibnamefont
  {Ryu}}\ and\ \bibinfo {author} {\bibfnamefont {M.~G.}\ \bibnamefont
  {Boshier}},\ }\bibfield  {title} {\bibinfo {title} {Integrated coherent
  matter wave circuits},\ }\href@noop {} {\bibfield  {journal} {\bibinfo
  {journal} {New J. Phys.}\ }\textbf {\bibinfo {volume} {17}},\ \bibinfo
  {pages} {092002} (\bibinfo {year} {2015})}\BibitemShut {NoStop}%
\bibitem [{\citenamefont {Roy}\ \emph {et~al.}(2016)\citenamefont {Roy},
  \citenamefont {Green}, \citenamefont {Bowler},\ and\ \citenamefont
  {Gupta}}]{Roy2016}%
  \BibitemOpen
  \bibfield  {author} {\bibinfo {author} {\bibfnamefont {R.}~\bibnamefont
  {Roy}}, \bibinfo {author} {\bibfnamefont {A.}~\bibnamefont {Green}}, \bibinfo
  {author} {\bibfnamefont {R.}~\bibnamefont {Bowler}},\ and\ \bibinfo {author}
  {\bibfnamefont {S.}~\bibnamefont {Gupta}},\ }\bibfield  {title} {\bibinfo
  {title} {Rapid cooling to quantum degeneracy in dynamically shaped atom
  traps},\ }\href {https://doi.org/10.1103/PhysRevA.93.043403} {\bibfield
  {journal} {\bibinfo  {journal} {Phys. Rev. A}\ }\textbf {\bibinfo {volume}
  {93}},\ \bibinfo {pages} {043403} (\bibinfo {year} {2016})}\BibitemShut
  {NoStop}%
\bibitem [{\citenamefont {Albers}\ \emph {et~al.}(2021)\citenamefont {Albers},
  \citenamefont {Corgier}, \citenamefont {Herbst}, \citenamefont {Rajagopalan},
  \citenamefont {Schubert}, \citenamefont {Vogt}, \citenamefont {Woltmann},
  \citenamefont {L\"ammerzahl}, \citenamefont {Herrmann}, \citenamefont
  {Charron}, \citenamefont {Ertmer}, \citenamefont {Rasel}, \citenamefont
  {Gaaloul},\ and\ \citenamefont {Schlippert}}]{Albers2021alloptical}%
  \BibitemOpen
  \bibfield  {author} {\bibinfo {author} {\bibfnamefont {H.}~\bibnamefont
  {Albers}}, \bibinfo {author} {\bibfnamefont {R.}~\bibnamefont {Corgier}},
  \bibinfo {author} {\bibfnamefont {A.}~\bibnamefont {Herbst}}, \bibinfo
  {author} {\bibfnamefont {A.}~\bibnamefont {Rajagopalan}}, \bibinfo {author}
  {\bibfnamefont {C.}~\bibnamefont {Schubert}}, \bibinfo {author}
  {\bibfnamefont {C.}~\bibnamefont {Vogt}}, \bibinfo {author} {\bibfnamefont
  {M.}~\bibnamefont {Woltmann}}, \bibinfo {author} {\bibfnamefont
  {C.}~\bibnamefont {L\"ammerzahl}}, \bibinfo {author} {\bibfnamefont
  {S.}~\bibnamefont {Herrmann}}, \bibinfo {author} {\bibfnamefont
  {E.}~\bibnamefont {Charron}}, \bibinfo {author} {\bibfnamefont
  {W.}~\bibnamefont {Ertmer}}, \bibinfo {author} {\bibfnamefont {E.~M.}\
  \bibnamefont {Rasel}}, \bibinfo {author} {\bibfnamefont {N.}~\bibnamefont
  {Gaaloul}},\ and\ \bibinfo {author} {\bibfnamefont {D.}~\bibnamefont
  {Schlippert}},\ }\href@noop {} {\bibinfo {title} {All-optical matter-wave
  lens using time-averaged potentials}} (\bibinfo {year} {2021}),\ \Eprint
  {https://arxiv.org/abs/2109.08608} {arXiv:2109.08608 [physics.atom-ph]}
  \BibitemShut {NoStop}%
\bibitem [{\citenamefont {Raghavan}\ \emph {et~al.}(1999)\citenamefont
  {Raghavan}, \citenamefont {Smerzi}, \citenamefont {Fantoni},\ and\
  \citenamefont {Shenoy}}]{raghavan}%
  \BibitemOpen
  \bibfield  {author} {\bibinfo {author} {\bibfnamefont {S.}~\bibnamefont
  {Raghavan}}, \bibinfo {author} {\bibfnamefont {A.}~\bibnamefont {Smerzi}},
  \bibinfo {author} {\bibfnamefont {S.}~\bibnamefont {Fantoni}},\ and\ \bibinfo
  {author} {\bibfnamefont {S.~R.}\ \bibnamefont {Shenoy}},\ }\bibfield  {title}
  {\bibinfo {title} {Coherent oscillations between two weakly coupled
  bose-einstein condensates: {Josephson} effects, $\ensuremath{\pi}$
  oscillations, and macroscopic quantum self-trapping},\ }\href
  {https://doi.org/10.1103/PhysRevA.59.620} {\bibfield  {journal} {\bibinfo
  {journal} {Phys. Rev. A}\ }\textbf {\bibinfo {volume} {59}},\ \bibinfo
  {pages} {620} (\bibinfo {year} {1999})}\BibitemShut {NoStop}%
\bibitem [{\citenamefont {Cataldo}\ and\ \citenamefont
  {Jezek}(2014)}]{cataldojezek}%
  \BibitemOpen
  \bibfield  {author} {\bibinfo {author} {\bibfnamefont {H.~M.}\ \bibnamefont
  {Cataldo}}\ and\ \bibinfo {author} {\bibfnamefont {D.~M.}\ \bibnamefont
  {Jezek}},\ }\bibfield  {title} {\bibinfo {title} {Dynamics in asymmetric
  double-well condensates},\ }\href
  {https://doi.org/10.1103/PhysRevA.90.043610} {\bibfield  {journal} {\bibinfo
  {journal} {Phys. Rev. A}\ }\textbf {\bibinfo {volume} {90}},\ \bibinfo
  {pages} {043610} (\bibinfo {year} {2014})}\BibitemShut {NoStop}%
\bibitem [{\citenamefont {Ananikian}\ and\ \citenamefont
  {Bergeman}(2006)}]{ananikian}%
  \BibitemOpen
  \bibfield  {author} {\bibinfo {author} {\bibfnamefont {D.}~\bibnamefont
  {Ananikian}}\ and\ \bibinfo {author} {\bibfnamefont {T.}~\bibnamefont
  {Bergeman}},\ }\bibfield  {title} {\bibinfo {title} {Gross-pitaevskii
  equation for bose particles in a double-well potential: Two-mode models and
  beyond},\ }\href {https://doi.org/10.1103/PhysRevA.73.013604} {\bibfield
  {journal} {\bibinfo  {journal} {Phys. Rev. A}\ }\textbf {\bibinfo {volume}
  {73}},\ \bibinfo {pages} {013604} (\bibinfo {year} {2006})}\BibitemShut
  {NoStop}%
\bibitem [{\citenamefont {Jia}\ \emph {et~al.}(2008)\citenamefont {Jia},
  \citenamefont {Li},\ and\ \citenamefont {Liang}}]{jia}%
  \BibitemOpen
  \bibfield  {author} {\bibinfo {author} {\bibfnamefont {X.}~\bibnamefont
  {Jia}}, \bibinfo {author} {\bibfnamefont {W.}~\bibnamefont {Li}},\ and\
  \bibinfo {author} {\bibfnamefont {J.~Q.}\ \bibnamefont {Liang}},\ }\bibfield
  {title} {\bibinfo {title} {Nonlinear correction to the boson
  {Josephson-junction} model},\ }\href
  {https://doi.org/10.1103/PhysRevA.78.023613} {\bibfield  {journal} {\bibinfo
  {journal} {Phys. Rev. A}\ }\textbf {\bibinfo {volume} {78}},\ \bibinfo
  {pages} {023613} (\bibinfo {year} {2008})}\BibitemShut {NoStop}%
\bibitem [{\citenamefont {Daley}(2015)}]{daley2015towards}%
  \BibitemOpen
  \bibfield  {author} {\bibinfo {author} {\bibfnamefont {A.~J.}\ \bibnamefont
  {Daley}},\ }\bibfield  {title} {\bibinfo {title} {Towards an atomtronic
  diode},\ }\href@noop {} {\bibfield  {journal} {\bibinfo  {journal} {Physics}\
  }\textbf {\bibinfo {volume} {8}},\ \bibinfo {pages} {72} (\bibinfo {year}
  {2015})}\BibitemShut {NoStop}%
\bibitem [{\citenamefont {Manju}\ \emph {et~al.}(2020)\citenamefont {Manju},
  \citenamefont {Hardman}, \citenamefont {Wigley}, \citenamefont {Close},
  \citenamefont {Robins},\ and\ \citenamefont {Szigeti}}]{manju2020atomic}%
  \BibitemOpen
  \bibfield  {author} {\bibinfo {author} {\bibfnamefont {P.}~\bibnamefont
  {Manju}}, \bibinfo {author} {\bibfnamefont {K.~S.}\ \bibnamefont {Hardman}},
  \bibinfo {author} {\bibfnamefont {P.~B.}\ \bibnamefont {Wigley}}, \bibinfo
  {author} {\bibfnamefont {J.~D.}\ \bibnamefont {Close}}, \bibinfo {author}
  {\bibfnamefont {N.~P.}\ \bibnamefont {Robins}},\ and\ \bibinfo {author}
  {\bibfnamefont {S.~S.}\ \bibnamefont {Szigeti}},\ }\bibfield  {title}
  {\bibinfo {title} {An atomic {Fabry-Perot} interferometer using a pulsed
  interacting {Bose-Einstein} condensate},\ }\href@noop {} {\bibfield
  {journal} {\bibinfo  {journal} {Scientific Reports}\ }\textbf {\bibinfo
  {volume} {10}},\ \bibinfo {pages} {1} (\bibinfo {year} {2020})}\BibitemShut
  {NoStop}%
\bibitem [{\citenamefont {Schach}\ \emph {et~al.}(2022)\citenamefont {Schach},
  \citenamefont {Friedrich}, \citenamefont {Williams}, \citenamefont
  {Schleich},\ and\ \citenamefont {Giese}}]{schach_tunneling_2022}%
  \BibitemOpen
  \bibfield  {author} {\bibinfo {author} {\bibfnamefont {P.}~\bibnamefont
  {Schach}}, \bibinfo {author} {\bibfnamefont {A.}~\bibnamefont {Friedrich}},
  \bibinfo {author} {\bibfnamefont {J.~R.}\ \bibnamefont {Williams}}, \bibinfo
  {author} {\bibfnamefont {W.~P.}\ \bibnamefont {Schleich}},\ and\ \bibinfo
  {author} {\bibfnamefont {E.}~\bibnamefont {Giese}},\ }\bibfield  {title}
  {\bibinfo {title} {Tunneling gravimetry},\ }\href
  {https://doi.org/10.1140/epjqt/s40507-022-00140-3} {\bibfield  {journal}
  {\bibinfo  {journal} {EPJ Quantum Technology}\ }\textbf {\bibinfo {volume}
  {9}},\ \bibinfo {pages} {20} (\bibinfo {year} {2022})}\BibitemShut {NoStop}%
\bibitem [{\citenamefont {Galitski}\ \emph {et~al.}(2019)\citenamefont
  {Galitski}, \citenamefont {Juzeliūnas},\ and\ \citenamefont
  {Spielman}}]{galitski_artificial_2019}%
  \BibitemOpen
  \bibfield  {author} {\bibinfo {author} {\bibfnamefont {V.}~\bibnamefont
  {Galitski}}, \bibinfo {author} {\bibfnamefont {G.}~\bibnamefont
  {Juzeliūnas}},\ and\ \bibinfo {author} {\bibfnamefont {I.~B.}\ \bibnamefont
  {Spielman}},\ }\bibfield  {title} {\bibinfo {title} {Artificial gauge fields
  with ultracold atoms},\ }\href {https://doi.org/10.1063/PT.3.4111} {\bibfield
   {journal} {\bibinfo  {journal} {Physics Today}\ }\textbf {\bibinfo {volume}
  {72}},\ \bibinfo {pages} {38} (\bibinfo {year} {2019})}\BibitemShut {NoStop}%
\bibitem [{\citenamefont {Aidelsburger}\ \emph {et~al.}(2018)\citenamefont
  {Aidelsburger}, \citenamefont {Nascimbene},\ and\ \citenamefont
  {Goldman}}]{aidelsburger_artificial_2018}%
  \BibitemOpen
  \bibfield  {author} {\bibinfo {author} {\bibfnamefont {M.}~\bibnamefont
  {Aidelsburger}}, \bibinfo {author} {\bibfnamefont {S.}~\bibnamefont
  {Nascimbene}},\ and\ \bibinfo {author} {\bibfnamefont {N.}~\bibnamefont
  {Goldman}},\ }\bibfield  {title} {\bibinfo {title} {Artificial gauge fields
  in materials and engineered systems},\ }\href
  {https://doi.org/10.1016/j.crhy.2018.03.002} {\bibfield  {journal} {\bibinfo
  {journal} {Comptes Rendus Physique}\ }\textbf {\bibinfo {volume} {19}},\
  \bibinfo {pages} {394} (\bibinfo {year} {2018})}\BibitemShut {NoStop}%
\bibitem [{\citenamefont {Juli{\'a}-D{\'\i}az}\ \emph
  {et~al.}(2010)\citenamefont {Juli{\'a}-D{\'\i}az}, \citenamefont
  {Martorell},\ and\ \citenamefont {Polls}}]{julia2010bose}%
  \BibitemOpen
  \bibfield  {author} {\bibinfo {author} {\bibfnamefont {B.}~\bibnamefont
  {Juli{\'a}-D{\'\i}az}}, \bibinfo {author} {\bibfnamefont {J.}~\bibnamefont
  {Martorell}},\ and\ \bibinfo {author} {\bibfnamefont {A.}~\bibnamefont
  {Polls}},\ }\bibfield  {title} {\bibinfo {title} {{Bose-Einstein} condensates
  on slightly asymmetric double-well potentials},\ }\href@noop {} {\bibfield
  {journal} {\bibinfo  {journal} {Phys. Rev. A}\ }\textbf {\bibinfo {volume}
  {81}},\ \bibinfo {pages} {063625} (\bibinfo {year} {2010})}\BibitemShut
  {NoStop}%
\bibitem [{\citenamefont {Fl{\"u}gge}(2012)}]{flugge2012practical}%
  \BibitemOpen
  \bibfield  {author} {\bibinfo {author} {\bibfnamefont {S.}~\bibnamefont
  {Fl{\"u}gge}},\ }\href@noop {} {\emph {\bibinfo {title} {Practical quantum
  mechanics}}}\ (\bibinfo  {publisher} {Springer Science \& Business Media},\
  \bibinfo {year} {2012})\BibitemShut {NoStop}%
\bibitem [{\citenamefont {Shegelski}\ and\ \citenamefont
  {Sample}(2020)}]{shegelski_equal_2020}%
  \BibitemOpen
  \bibfield  {author} {\bibinfo {author} {\bibfnamefont {M.~R.~A.}\
  \bibnamefont {Shegelski}}\ and\ \bibinfo {author} {\bibfnamefont
  {C.}~\bibnamefont {Sample}},\ }\bibfield  {title} {\bibinfo {title} {Equal
  reflection and transmission probabilities},\ }\href
  {https://doi.org/10.1088/1361-6404/ab7c37} {\bibfield  {journal} {\bibinfo
  {journal} {Eur. J. Phys.}\ }\textbf {\bibinfo {volume} {41}},\ \bibinfo
  {pages} {035405} (\bibinfo {year} {2020})}\BibitemShut {NoStop}%
\bibitem [{\citenamefont {Tannor}(2007)}]{tannor2007introduction}%
  \BibitemOpen
  \bibfield  {author} {\bibinfo {author} {\bibfnamefont {D.~J.}\ \bibnamefont
  {Tannor}},\ }\href@noop {} {\emph {\bibinfo {title} {Introduction to quantum
  mechanics: a time-dependent perspective}}}\ (\bibinfo {year}
  {2007})\BibitemShut {NoStop}%
\bibitem [{{\relax DLMF}()}]{NIST:DLMF}%
  \BibitemOpen
  {\relax DLMF},\ \href {http://dlmf.nist.gov/} {\bibinfo {title} {{\it NIST
  Digital Library of Mathematical Functions}}},\ \bibinfo {howpublished}
  {http://dlmf.nist.gov/, Release 1.0.26 of 2020-03-15},\ \bibinfo {note}
  {{F.}~W.~J. Olver, A.~B. {Olde Daalhuis}, D.~W. Lozier, B.~I. Schneider,
  R.~F. Boisvert, C.~W. Clark, B.~R. Miller, B.~V. Saunders, H.~S. Cohl, and
  M.~A. McClain, eds.}\BibitemShut {Stop}%
\end{thebibliography}%

\end{document}